\documentclass{article}

\usepackage{graphicx}
\usepackage{amsmath}
\usepackage{amsfonts}
\usepackage{amssymb}
\usepackage{verbatim}

\renewcommand{\arraystretch}{1.3}
\catcode`\@=11
\def\marginnote#1{}

\newcount\hour
\newcount\minute
\newtoks\amorpm
\hour=\time\divide\hour by60 \minute=\time{\multiply\hour by60
\global\advance\minute by-\hour}
\edef\standardtime{{\ifnum\hour<12 \global\amorpm={am}%
        \else\global\amorpm={pm}\advance\hour by-12 \fi
        \ifnum\hour=0 \hour=12 \fi
        \number\hour:\ifnum\minute<10 0\fi\number\minute\the\amorpm}}
\edef\militarytime{\number\hour:\ifnum\minute<10 0\fi\number\minute}

\def\draftlabel#1{{\@bsphack\if@filesw {\let\thepage\relax
      \xdef\@gtempa{\write\@auxout{\string
          \newlabel{#1}{{\@currentlabel}{\thepage}}}}}\@gtempa \if@nobreak
    \ifvmode\nobreak\fi\fi\fi\@esphack} \gdef\@eqnlabel{#1}}
    \def\@eqnlabel{}
\def\@vacuum{}
\def\draftmarginnote#1{\marginpar{\raggedright\scriptsize\tt#1}}

\def\draft{
  \oddsidemargin -.5truein
  \def\@oddfoot{\footnotesize \sl preliminary draft \hfil
    \rm\thepage\hfil\sl\today\quad\militarytime}
  \let\@evenfoot\@oddfoot \overfullrule 3pt
    \let\label=\draftlabel
    \let\marginnote=\draftmarginnote
  \def\@eqnnum{(\theequation)\rlap{\kern\marginparsep\tt\@eqnlabel}%
    \global\let\@eqnlabel\@vacuum}

  }
\makeatletter
\newdimen\normalarrayskip              % skip between lines
\newdimen\minarrayskip                 % minimal skip between lines
\normalarrayskip\baselineskip \minarrayskip\jot
\newif\ifold             \oldtrue            \def\new{\oldfalse}
\def\arraymode{\ifold\relax\else\displaystyle\fi} % mode of array entries
\def\eqnumphantom{\phantom{(\theequation)}}     % right phantom in eqnarray
\def\@arrayskip{\ifold\baselineskip\z@\lineskip\z@
     \else
     \baselineskip\minarrayskip\lineskip2\minarrayskip\fi}
\def\@arrayclassz{\ifcase \@lastchclass \@acolampacol \or
\@ampacol \or \or \or \@addamp \or
   \@acolampacol \or \@firstampfalse \@acol \fi
\edef\@preamble{\@preamble
  \ifcase \@chnum
     \hfil$\relax\arraymode\@sharp$\hfil
     \or $\relax\arraymode\@sharp$\hfil
     \or \hfil$\relax\arraymode\@sharp$\fi}}
\def\@array[#1]#2{\setbox\@arstrutbox=\hbox{\vrule
     height\arraystretch \ht\strutbox
     depth\arraystretch \dp\strutbox
     width\z@}\@mkpream{#2}\edef\@preamble{\halign
\noexpand\@halignto
\bgroup \tabskip\z@ \@arstrut \@preamble \tabskip\z@ \cr}%
\let\@startpbox\@@startpbox \let\@endpbox\@@endpbox
  \if #1t\vtop \else \if#1b\vbox \else \vcenter \fi\fi
  \bgroup \let\par\relax
  \let\@sharp##\let\protect\relax
  \@arrayskip\@preamble}
\def\eqnarray{\stepcounter{equation}%
              \let\@currentlabel=\theequation
              \global\@eqnswtrue
              \global\@eqcnt\z@
              \tabskip\@centering
              \let\\=\@eqncr

 \halign to \displaywidth\bgroup
    \eqnumphantom\@eqnsel\hskip\@centering
    $\displaystyle \tabskip\z@ {##}$%
    \global\@eqcnt\@ne \hskip 2\arraycolsep
         $\displaystyle\arraymode{##}$\hfil
    \global\@eqcnt\tw@ \hskip 2\arraycolsep
         $\displaystyle\tabskip\z@{##}$\hfil
         \tabskip\@centering
    &{##}\tabskip\z@\cr}
\newfont{\hr}{msbm10}
\newfont{\ams}{msam10}
\hoffset= - 0.9in        % switch off for draft style
\def\beq{\begin{equation}}
\def\eeq{\end{equation}}
\def\ba{\beq\new\begin{array}{c}}
\def\ea{\end{array}\eeq}
\def\be{\ba}
\def\ee{\ea}

\def\N2{${\cal N}=2$}

\def\1N{${\cal N}=1$}
\def\4N{${\cal N}=4$}
\def\nn{\nonumber}

\def\p{\partial}

\newdimen\linethick  \linethick=0.4pt
\newdimen\hboxitspace    \hboxitspace=5pt
\newdimen\vboxitspace    \vboxitspace=5pt

\def\fr#1{%
\beq\new \vcenter{ \hrule height\linethick
          \hbox{\vrule width\linethick
                \kern\hboxitspace
                \vbox{\kern\vboxitspace
                      \hbox{$\begin{array}{c}\displaystyle#1
         \end{array}$}%
                      \kern\vboxitspace}%
                \kern\hboxitspace
                \vrule width\linethick}%
          \hrule height\linethick}%
\eeq}
\renewcommand{\tt}[1][mer]{\hbox{\tiny{#1}}}

\newcommand{\Tr}{\mathop{\rm Tr}\nolimits}

\def\p{\partial}
\def\tr{{\rm tr}\,}
\def\Tr{{\rm Tr}\,}

\def\rp{[p\,]}

\def\l[{\phantom.[}

\def\l[{\phantom.[}
\def\rp{[p\,]}
\def\2rp{[2p\,]}
\def\nnn{\\ }
\def\N{{\cal N}}

\title{{\bf Eigenvalue hypothesis for Racah matrices and\\
HOMFLY polynomials for 3-strand knots \\in any symmetric and
antisymmetric representations} \vspace{.5cm}}
\author{{\bf H.Itoyama}\footnote{ {\small {\it
Department of Mathematics and Physics, Osaka City University} and
{\it Osaka City University Advanced Mathematical Institute (OCAMI),
Osaka, Japan}}; itoyama@sci.osaka-cu.ac.jp}, \ {\bf
A.Mironov}\footnote{ {\small {\it Lebedev Physics Institute} and
{\it ITEP, Moscow, Russia}}; mironov@itep.ru; mironov@lpi.ru}, \
{\bf A.Morozov}\thanks{{\small {\it ITEP, Moscow, Russia}};
morozov@itep.ru}, \ {\bf And.Morozov}\thanks{{\small {\it Moscow
State University} and {\it ITEP, Moscow, Russia}};
Andrey.Morozov@itep.ru}\date{ }}

\begin{document}

\setcounter{footnote}{3}

\setcounter{tocdepth}{3}

\maketitle

\vspace{-7.cm}

\begin{center}
\hfill FIAN/TD-24/12\\
\hfill ITEP/TH-46/12\\
\hfill OCU-PHYS-375
\end{center}

\vspace{4.cm}

\begin{abstract}
Character expansion expresses extended HOMFLY polynomials through
traces of products of finite dimensional ${\cal R}$- and Racah
mixing matrices. We conjecture that the mixing matrices are
expressed entirely in terms of the eigenvalues of the corresponding
${\cal R}$-matrices. Even a weaker (and, perhaps, more reliable)
version of this conjecture is sufficient to explicitly calculate
HOMFLY polynomials for all the 3-strand braids in arbitrary
(anti)symmetric representations. We list the examples of so obtained
polynomials for $R=[3]$ and $R=[4]$, and they are in accordance with
the known answers for torus and figure-eight knots, as well as for
the colored special and Jones polynomials. This provides an indirect
evidence in support of our conjecture.
\end{abstract}

\bigskip

\bigskip

\section{Introduction}

Knot polynomials \cite{knotp} attract a new attention these days, because they
provide a non-trivial generalization of conformal blocks and possess
a vast variety of non-trivial interrelations, which deserve
investigation and understanding (see \cite{MMapols} for a recent
review and references).

Nowadays there are two main approaches to knot polynomials,
implied by consideration of Chern-Simons theory \cite{CS} in two
different gauges.
In the holomorphic gauge $A_{\bar z}=0$, the $3d$ functional
integral reduces to the one-dimensional Kontsevich integral,
expanded in chord diagrams with polylogarithmic coefficients
and closely related to the Vassiliev invariants \cite{VK}.
In the temporal gauge $A_0=0$, the $3d$ functional integral
reduces to an ordered product of quantum ${\cal R}$-matrices
standing at the vertices of the knot diagram, obtained
by projection of the knot onto 2-dimensional plane \cite{TR,inds,MSm}.
This representation is at the moment the most convenient
tool for study of the knot polynomials.

In \cite{DMMSS,MMMI} we proposed to use this representation
for construction of a character expansion of the HOMFLY \cite{HOMFLY}
and superpolynomials \cite{sp},
promoting them to {\it extended} knot polynomials,
depending on infinitely many time-variables (like $\tau$-functions do).
These quantities are no longer topological, only braid invariants,
yet instead are easily studied by a variety of powerful
matrix model methods.
In particular, one can obtain generic explicit formulas in terms
of a finite number of finite-dimensional matrices
for arbitrary number $m$ of strands in the braid.
Such a braid is parameterized by a sequence of integers
\be
{\cal B} = \Big\{\,a_{11},\ldots,a_{1,m-1}\,|\,a_{21},\ldots,a_{2,m-1}\,|
\ \ldots\ |
\,a_{n1},\ldots,a_{n,m-1}\Big\} \
\stackrel{m=3}{\longrightarrow}\
\Big\{\,a_1,b_1\,|\,a_2,b_2\,|\ \ldots\ |\,a_n,b_n\Big\}
\ee
and the corresponding HOMFLY polynomial is equal to
\be
{\cal H}_R^{\cal B} =
\sum_{Q\,\vdash\, m|R|} C_{RQ}^{\cal B} S_Q \ =\
%q^{-4\varkappa_R}A^{-|R|}???
\sum_{Q\,\vdash\, m|R|} \left\{\Tr \prod_{i=1}^n
\left(\prod_{j=1}^{m-1} {\cal R}_Q^{a_{ij}}U_Q^{j,j+1}\right)\right\}\, S_Q
%\ \longrightarrow
\\
\stackrel{m=3}{\longrightarrow}\
%q^{-4\varkappa_R}A^{-|R|}???
\sum_{Q\,\vdash\, 3|R|} \left\{\Tr \Big( {\cal R}_Q^{a_1} U_Q {\cal R}_Q^{b_1}
U_Q^\dagger\  {\cal R}_Q^{a_2} U_Q {\cal R}_Q^{b_2}
U_Q^\dagger\ \ldots\ {\cal R}_Q^{a_n} U_Q {\cal R}_Q^{b_n}
U_Q^\dagger\Big)\right\}\, S_Q
\label{extHexpan}
\ee
%This formula is written for $m=3$, which is the case that we are
%going to study in the present paper.
This formula represents the HOMFLY polynomial as a linear combination
of the Schur polynomials $S_Q$ in the irreducible representations $Q$
appearing in the product
\be
R^{\otimes m} = \oplus_{Q\,\vdash m|R|}\  {\cal M}_R^Q\otimes\, Q
\label{Rmexpan}
\ee
and they are described by the Young diagrams of size $m|R|$:
$m$ times bigger than the size of $R$.

Eq.(\ref{extHexpan}) looks like a direct generalization of the celebrated
Rosso-Jones formula \cite{RJ} for the torus knots,
only the coefficients $C_{RQ}$ in front of the Schur polynomials $S_Q$
are in general represented as traces of the ${\cal R}$-matrices
and Racah mixing matrices $U$ over the spaces of intertwining operators ${\cal M}_R^Q$,
while for the torus knots there is
a much simpler Adams rule \cite{RJ} to determine these coefficients.

If $S_Q=S_Q\{p\}$ are the Schur polynomials of arbitrary time variables
(actually they depend on the first $Q$ times), one has an {\it extended}
HOMFLY. It is this space of all time variables where the Adams rule
\be
S_R\{p_{mk}\} = \widehat{{\rm Ad}}_m S_R\{p_k\}
= \sum_{Q\,\vdash\,m|R|} c_{RQ}^{{\rm torus}}S_R\{p_k\}
\label{Adrule}
\ee
is applicable (and $C_{RQ}^{{\rm torus}}\sim q^{2n/m\varkappa_Q}c_{RQ}^{{\rm torus}}$, see (\ref{kappa})).
Only on this space the character decomposition is unambiguously defined
for $m>3$ and $|R|>1$.
The standard topologically invariant HOMFLY
polynomials (Wilson-loop averages in Chern-Simons theory)
appear on the {\it topological locus} in the time-variable space (a kind of a Miwa transform with
finitely many Miwa variables):
\be
p_k = p_k^*\equiv \frac{A^k-A^{-k}}{q^k-q^{-k}} \ \stackrel{\ \ A=q^N}{=}\ \
\frac{[Nk]_q}{[k]_q}
\label{tolo}
\ee
and are normalized as
\be
H_R^{\cal K}(A,q) = \Big(q^{-4\varkappa_R}A^{-|R|}\Big)^{\#(B)}
{\cal H}_R^{\cal B}\{p^*\}
\label{tiHOMFLY}
\ee
$\#(B)$ being the number of intersections in the braid. The quantum numbers are defined here as
$[x]_q={x^q-x^{-q}\over q-q^{-1}}$.

For ${\cal R}$-matrices in (\ref{extHexpan}) there is a simple general
expression: they act on any irrep $Q$ as diagonal matrices of the size
$\,{\rm dim}({\cal M}_R^Q)$ with the eigenvalues $r^Q_\alpha$,
all taken
(with appropriately chosen signs)
from the set of the exponentiated symmetric group characters,
eigenvalues $\varkappa_T$ of the cut-and-join operator \cite{caj}
\be\label{kappa}
\hat W_{[2]} = \frac{1}{2}
\sum_{a,b} \left((a+b)p_ap\,_b \frac{\p}{\p p_{a+b}} +
abp_{a+b}\frac{\p^2}{\p p_a\p p\,_b}\right), \\
\hat W_{[2]} S_T\{p\} = \varkappa_T S_T\{p\}, \\
\varkappa_T = \varphi_{[2]}(T) = \sum_{(ab)\in T} (a-b) =
\frac{1}{2}\sum_a t_a(t_a-2a-1)
\ee
where $t_a$ are the lengths of the lines of Young diagram $T$.

The relevant $T$ in the case of (\ref{extHexpan}) are the ones which
appear in the expansion of a product of just {\it two} representations $R$:
\be
R\otimes R = \oplus_T\ {\cal M}_R^T \otimes\, T
\ee
To see which eigenvalues contribute to ${\cal R}_Q$,
one should look at the representation product tree \cite{MMMII}
\be
\begin{array}{ccccc}
(R\otimes R)\otimes R &=&
R\otimes R \otimes R &=& R\otimes (R\otimes R)\\
\downarrow &&&&\downarrow \\
\sum_T {\cal M}_R^T \otimes (T\otimes R) &&&&
\sum_T {\cal M}_R^T \otimes (R\otimes T)\\
\downarrow &&&&\downarrow \\
\sum_Q {\cal M}_R^Q \otimes Q &&&&
\sum_T {\cal M}_R^Q \otimes Q\\
\end{array}
%\nn\\
%\sum_Q \ \longleftarrow \ \sum T\otimes R \ \longleftarrow \
%(R\otimes R)\otimes R =
%R\otimes R \otimes R = R\otimes (R\otimes R)\ \longrightarrow \
%\sum R\otimes T' \ \longrightarrow \sum Q'
\label{twodeco}
\ee
and pick up representations $T$  which contribute at a given $Q$.

The orthogonal Racah mixing matrix $U_Q$ describes the rotation in the space
${\cal M}_R^Q$, corresponding to the transformation between
the left and right decompositions in (\ref{twodeco}).
For $m>3$ there are $m-1$ different decompositions and thus $m-2$
different mixing matrices, the last one, appearing in (\ref{extHexpan}), is
\be
U^{m-1,m}_Q = \left(\prod_{j=1}^{m-2} U^{j,j+1}_Q\right)^\dagger
\ee

To have a complete description of HOMFLY polynomials one just needs
to know the mixing matrices. For $m=3$ (three-strand braids) these
are just the standard Racah coefficients, the problem being that
they are well known \cite{sl2Racah} only for the $SU_q(2)$
group,which is far not sufficient for our purposes of constructing
the HOMFLY polynomial at arbitrary $A=q^N$, i.e. for arbitrary
$SU(N)$. In fact, in order to build a comprehensive and interesting
theory of knot polynomials on the base of (\ref{extHexpan}), one
needs some clever description of generic $U_Q$, revealing their
structure, not just concrete expressions. Some steps have been
already made in this direction in
\cite{MMMI,MMMII,IMMMIII,IMMM,AMMM}, see also \cite{inds} for a
fruitful parallel development. In particular, in \cite{An,AMMM} the
general structure of mixing matrices is described for arbitrary $m$,
but only for the fundamental representation $R=\Box$. The present
paper is a step in "orthogonal" direction: for $|R|>1$, but only for
$m=3$, when there is just one mixing matrix. Finally a synthesis of
both developments is needed.

What we do in the present paper, we formulate a conjecture about the
form of the mixing matrix.
Basically, we suggest that {\bf $U_Q$ depends only on the eigenvalues of
the corresponding ${\cal R}_Q$}.
If this was the case, then one could take the known $SU_q(2)$ formulas,
express them through eigenvalues and then use the same expression
in all other cases just by substituting the other eigenvalue sets,
relevant in particular situations.
However, so far we managed to get such expressions only for
${\rm dim}({\cal M}_R^Q) = 2,3,4,5$ (for mixing matrices of the
sizes up to $5$), what is sufficient only for calculations
in representations $[2],[3],[4],[5]$.
Hence, it is not quite clear at the moment if the conjecture is true in general.

For calculations in symmetric representations $R=S^r=[r]$,
however, a weaker form of the conjecture is sufficient,
since the relevant eigenvalue sets are somewhat special.
Accepting the conjecture in this weaker form,
one looses a test
(the very possibility to express the Racah matrix
through the eigenvalues of associated ${\cal R}_Q$),
but instead one gets a tool for calculating ${\cal H}_{S^r}^{\cal B}$
in {\it arbitrary} symmetric representation
(antisymmetric representations $\Lambda^r$ are then obtained by
the application of the duality transform
${\cal H}_{R'}(A,q) = {\cal H}_R (A,-1/q)$ for the transposed Young diagram $R'$).

All this makes our calculation {\it not} a derivation
from the first principles, and justification comes from
the possibility to reproduce the known answers in particular
examples.
For such examples we use the torus knots
(the Adams rule), the figure eight ($4_1$) knot \cite{IMMM}
and the colored special \cite{DMMSS,Zhu}
and Jones \cite{katlas} polynomials.
For $m=3$ and $R=[3],[4]$ the tests are positive,
so our conjecture, at least in its weak form (i.e. for the
special sets of the ${\cal R}_Q$-eigenvalues) is justified.
%However, the crucial test requires an examination of higher $r\ge 5$.
Also of crucial important is a synthesis with \cite{AMMM}
and similar calculations for at least $m=4$ (where a number
of other examples are known from \cite{GS} and \cite{twist}).
The question of primary importance would be a similar
calculation for non-symmetric representation $R=[21]$ and,
better, for non-hook $R=[22]$ and also non-symmetric $R=[32]$,
where non-trivial new effects are expected from the study
of the special (super)polynomials \cite{DMMSS,AntMor} and of the
Alexander polynomials \cite{IMMM,SerM}.

In what follows, in ss.\ref{conj1}-\ref{conj}
we formulate our conjectures and in further sections provide
associated (hypothetical) formulas for the mixing matrices.
% up to the size $5$.
Our main result is formulated  in s.\ref{ans}.
%and illustrated in the tables in Appendix ???.
In the tables we provide explicit expressions for colored
([1],[2],[3] and [4]) HOMFLY polynomials for the first few
$m=3$-strand knots.
Making use of eq.(\ref{extHexpan}) and equations for the mixing matrices of ss.4-8,
one can obtain analogous results for {\it any} given $3$-strand braid
${\cal B} = \{a_1,b_1|a_2,b_2|\ldots\}$ by a half minute MAPLE or
Mathematica calculation.

\section{The fundamental representation $R=[1]$: a hint of conjecture
\label{conj1}}

To make use of (\ref{extHexpan}), one first of all needs to know
the decomposition (\ref{Rmexpan}).
For a $3$-strand knot and the fundamental representation $R=[1]$
it is just
\be
\l[1]\otimes[1]\otimes[1] = [3] + 2\cdot[21] + [111]
\ee
and there is a single non-trivial $2\times 2$ mixing matrix $U_{[21]}$.
Eq.(\ref{extHexpan}) in this case is just
\be
{\cal H}^{\,a_1,b_1|\ldots}_{[1]} = q^{a_1+b_1+\ldots}S_3 + (-1/q)^{a_1+b_1+\ldots}S_{111} +
\label{3str1}
\ee
\vspace{-0.2cm}
$$
+ \left\{\Tr_{2\times 2}\left(
\underbrace{\left(\begin{array}{cc} q & \\ & -1/q \end{array}\right)^{a_1}}_{{\cal R}_{[21]}}
\underbrace{
\frac{1}{[2]_q}\!\left(\begin{array}{cc} 1  & \sqrt{[3]_q} \\
-\sqrt{[3]_q}  & 1  \end{array}\right)}_{{U}_{[21]}}
\underbrace{\left(\begin{array}{cc} q & \\ & -1/q \end{array}\right)^{b_1}}_{{\cal R}_{[21]}}
\underbrace{\frac{1}{[2]_q}\!\left(\begin{array}{cc} 1  & -\sqrt{[3]_q} \\
\sqrt{[3]_q}  & 1  \end{array}\right)}_{{U}^\dagger_{[21]}}\ldots\right)
\right\}
S_{[21]}
$$
Here we used the fact that the two ${\cal R}$ matrix eigenvalues, associated with
$T=[2]$ and $T=[11]$ in the {\it pair} decomposition $[1]\otimes [1] = [2]+[11]$,
are $\xi_1=q^{\varkappa_{[2]}}=q$ and $\xi_2=-q^{\varkappa_{[11]}}=-1/q$ respectively,
and we borrowed the $2\times 2$ mixing matrix from \cite{MMMII}.
From now on we often omit the index $q$, whenever it can not cause a confusion
(with the similarly denoted Young diagrams for symmetric representations).
We also use a convenient notation $\{x\} = x-x^{-1}$.
The Schur polynomials ($SU(N)$ characters) in this particular case are
\be
S_3 = {1\over 3}p_3 + p_2p_1 + \frac{1}{6}p_1^3 \ \stackrel{\ \ p=p^*}{=}\
\frac{ \{A\}\{Aq\}\{Aq^2\} }{ \{q\}\{q^2\}\{q^3\} }, \\
S_{21} = -{1\over 3}p_3 + \frac{1}{3}p_1^3 \ \stackrel{\ \ p=p^*}{=}\
\frac{ \{Aq^{-1}\}\{A\}\{Aq\} }{ \{q\}^2\{q^3\} }, \\
S_{111} = {1\over 3}p_3 + {1\over 2}p_2p_1 + \frac{1}{6}p_1^3 \ \stackrel{\ \ p=p^*}{=}\
\frac{ \{Aq^{-2}\}\{Aq^{-1}\}\{A\}\}}{\{q\}\{q^2\}\{q^3\}}
\ee
This formula is rigorously proved and known to reproduce all the 3-strand
HOMFLY polynomials in the fundamental representation \cite{MMMII}.

\bigskip

However, let us reverse the logic.
Imagine that the group theory calculation in \cite{MMMII}
has not been done, and we do not know the answer for $U_{[21]}$
from the first principles, which is actually the case in more complicated situations.
Can there be an alternative, indirect way to obtain $U$?
One can attempt to use three pieces of knowledge, which are always available:
formula (\ref{3str1}), the Rosso-Jones formula (\ref{Adrule})
for the torus knots
and the eigenvalues of the matrix ${\cal R}_{[21]}$.
It is immediate to see that this is indeed enough,
at least, in this particular case.

Indeed, since
\be
S_1\{p_{3k}\} = p_3 = S_{[3]}\{p_k\} - S_{[21]}\{p_k\} + S_{[111]}\{p_k\},  \\
S_1^3\{p_k\} = p_1^3 = S_{[3]}\{p_k\} + 2S_{[21]}\{p_k\} + S_{[111]}\{p_k\}
\ee
the Adams rule (\ref{Adrule}) and its counterpart for links imply
that the coefficients $C_{[1],[21]}$ should have definite values
$-1$ and $2$ for the $3$-strand torus knots and links respectively.
At the same time, since
for the torus knots/links $[3,n]$ all $a_1=b_1=\ldots=a_n=b_n=1$,
from (\ref{3str1}) the coefficient in front of $S_{21}$ is equal to
$\ \Tr \Big({\cal R}_{[21]}U_{[21]}{\cal R}_{21}U_{[21]}^\dagger\Big)^n$.
Comparing these two statements, one obtains:
\be
\Tr \Big({\cal R}_{[21]}U_{[21]}{\cal R}_{21}U_{[21]}^\dagger\Big)^n
=   \left\{ \begin{array}{cccc}
-1 & {\rm for} & n = 3k\pm 1 & ({\rm knots}) \\
2 & {\rm for} & n = 3k  & ({\rm links})
\end{array}\right.
\label{toco3}
\ee
and this implies that the two eigenvalues of the product
${\cal R}_{[21]}U_{[21]}{\cal R}_{[21]}U_{[21]}^\dagger$
are $e^{\pm 2\pi i/3}$.
Given ${\cal R}_{[21]}$, this is actually enough to obtain $U_{[21]}$.
%We perform this calculation in a slightly more general form,
%not to repeat it again in what follows.

It is clear that if the $U$-matrix is unambiguously obtained
from such a reasoning, it will depend only on the matrix ${\cal R}$,
and since ${\cal R}$ is diagonal  in this basis, $U$ will depend only on
the eigenvalues of ${\cal R}$. Let us denote this eigenvalues
via $\xi_1$ and $\xi_2$.
At the same time, an orthogonal $2\times 2$ matrix $U$ depends on
a single parameter: mixing angle, and we denote its sine
and cosine through $c$ and $s$ respectively; of course, $c^2+s^2=1$.
Finally, if $\xi_1\xi_2 \neq -1$, the r.h.s. of
(\ref{toco3}) should be multiplied by $(-\xi_1\xi_2)^n$
so that the characteristic equation for ${\cal U}$ is
\be
\det_{2\times 2} \Big({\cal R}U{\cal R}U^\dagger - \lambda\cdot I\Big)
= \left(\lambda +\xi_1\xi_2 e^{2\pi i/3}\right)
\left(\lambda +\xi_1\xi_2 e^{-2\pi i/3}\right)
= \lambda^2 - \xi_1\xi_2 \lambda + (\xi_1\xi_2)^2
\ee
Thus, we see that
\be
\Tr \left\{\left(\begin{array}{cc} \xi_1 & 0 \\ 0 & \xi_2 \end{array}\right)
\left(\begin{array}{cc} c & s \\ -s & c \end{array}\right)
\left(\begin{array}{cc} \xi_1 & 0 \\ 0 & \xi_2 \end{array}\right)
\left(\begin{array}{cc} c & -s \\ s & c \end{array}\right)\right\}
%= \left(\begin{array}{cc}   &   \\  &  \end{array}\right)
= (\xi_1^2+\xi_2^2)c^2 + 2\xi_1\xi_2 s^2 =
(\xi_1-\xi_2)^2c^2 + 2\xi_1\xi_2
\ee
is equal to $\xi_1\xi_2$, i.e. that
\be
c = \frac{\sqrt{-\xi_1\xi_2}}{\xi_1-\xi_2},\ \ \ \ \
s = \frac{\sqrt{\xi_1^2-\xi_1\xi_2+\xi_2^2}}{\xi_1-\xi_2}
\ \ \ \ \Longrightarrow \ \ \ \ U =
\left(\begin{array}{cc} \frac{\sqrt{-\xi_1\xi_2}}{\xi_1-\xi_2} &
\frac{\sqrt{\xi_1^2-\xi_1\xi_2+\xi_2^2}}{\xi_1-\xi_2} \\ \\
-\frac{\sqrt{\xi_1^2-\xi_1\xi_2+\xi_2^2}}{\xi_1-\xi_2}
& \frac{\sqrt{-\xi_1\xi_2}}{\xi_1-\xi_2}\end{array}\right)
\label{U2ev}
\ee
Moreover, one can substitute $\xi_1$ and $\xi_2$ in this formulas
by their normalized counterparts
\be
\tilde \xi_\alpha = \frac{\xi_\alpha}{\left(\pm\prod_{\beta=1}^\N \xi_\beta \right)^{1/\N}}
\ee
where $\N = {\rm dim}({\cal M}_R^Q)$ is the size of the mixing matrix $U$
(in our current example $\N=2$), and the sign in front of the product
drops out of the answer for $U$. It is however convenient to adjust it
so that no roots of unity appear in intermediate formulas
(in our case it deserves choosing minus sign).
Substituting $\xi_1=q$ and $\xi_2=-1/q$, one obtains
\be
{\cal R}_{[21]} = \left(\begin{array}{cc} q & 0 \\ 0 & -\frac{1}{q}\end{array}\right)
\ \ \ \Longrightarrow \ \ \
U_{[21]} = \left(\begin{array}{cc} \frac{1}{[2]} & \frac{\sqrt{[3]}}{[2]} \\ \\
-\frac{\sqrt{[3]}}{[2]} & \frac{1}{[2]}\end{array}\right)
\ee
We shall see below that formula (\ref{U2ev}) works nicely not only for
$Q=[21]$, but in many (all?) other situations when the size of the mixing
matrix $U_Q$ is $\N=2$.

\section{Eigenvalue conjectures: strong and weak forms
\label{conj}}

Now we are ready to formulate our main conjectures:

\bigskip

{\bf 1. The mixing matrix $U_Q$ depends only on the eigenvalues $\xi_\alpha$
$\ (\alpha = 1,\ldots, \N)\ $ of the
corresponding ${\cal R}$-matrix ${\cal R}_Q$.}

{\bf 2. It actually depends on the normalized eigenvalues $\tilde\xi_\alpha$.}

\bigskip

This is the strong form of our conjectures.
The weak form (w) implies that this is true not for {\it arbitrary}
sets  $\{\xi_\alpha\}$, but only for those, which actually appear
as eigenvalues of ${\cal R}$-matrices, i.e.
for $\xi_\alpha = \pm q^{\varkappa_{T(\alpha)}}$.

An even weaker form (ww) is that
this is true only when eigenvalues of
${\cal R}_Q$ are made from $\pm q^{\varkappa_T}$,
where the Young diagrams $T$ have no more than two lines.
Since $T$ are representations contributing to decomposition
of $R\otimes R$, this actually means that conjectures are
restricted to $R$ which are pure symmetric representations
(described by the single line diagram $R=[r]$).
Of course, answers for the duality-related antisymmetric representations
are also available in this case.

The weakening refers actually only to applicability of the conjectures.
Since so far we looked only at a restricted set of examples,
we actually tested them only in the weakest form.
At the same time, we found explicit expressions for the $U$-matrices
through {\it arbitrary} sets of eigenvalues $\{\xi_\alpha\}$
only for $\N=2,3,4,5$; thus it is not clear, if the strongest
form of the conjecture is at all viable.

Still, if true, the conjectures are very powerful.
The crucial point is that for a very special set of eigenvalues,
that is, in the case of $SU_q(2)$ Racah coefficients
the $U$-matrices are fully known.
It is, however, highly non-trivial to rewrite them in
terms of the corresponding eigenvalues, and the "minimal"
expression of such a type, if exists at all, looks to be unique:
eq.(\ref{U2ev}) for $\N=2$ is a perfect example.
Once it is found, one has the complete knowledge of the arbitrary $U$-matrices,
{\it de facto} of generic Racah matrices for  $SU_q(\infty)$.
It is still unclear for us if the things can work this way for $\N>5$.

However, conjecture  ${\bf 2}$ implies that, whenever the two sets of normalized
eigenvalues $\{\tilde\xi_\alpha\}$ are the same, the $U$-matrices coincide.
It turns out that for all representations $Q \in [r]^{\otimes 3}$,
i.e. appearing in the decomposition of three symmetric representations,
the normalized eigenvalues are the ones, appearing in the known
$SU_q(2)$ Racah series (of course, the eigenvalues themselves are
different, but normalization makes them the same!)
This allows one to solve completely the problem for arbitrary symmetric
(and antisymmetric) representations.

\section{Representation $[2]$
\label{rep2}}

This case is studied in detail in \cite{IMMMIII}.
We use the mixing matrices, found in that paper,
to illustrate our conjectures.
The ${\cal R}$-matrix eigenvalues are defined in the channel
$[2]\otimes [2] = [4] + [31] + [22]$ and they are
\be
\begin{array}{ccccc}
\l[4]: && q^{\varkappa_{[64]}} &=& q^{6}   \\
\l[31]: && -q^{\varkappa_{[31]}} &=& -q^2   \\
\l[22]: && q^{\varkappa_{[22]}} &=& 1 \\
\end{array}
\label{R2ev}
\ee
%\subsection{Decomposition of $[2]^{\times 3}$ and generic formula}
Since
\be
\l[2]\otimes[2]\otimes[2] = \Big([4]+[31]+[22]\Big)\otimes [2] = \\
= \Big([6] + [51] + [42]\Big) + \Big([51]+[42]+[411]+[33]+[321]\Big)
+ \Big([42]+[321]+[222]\Big)
=\\
= [6]+[411]+ [33]+[222]  + 2\Big([51]+[321]\Big) + 3\cdot[42]
\ee
in this example we have two $2\times 2$ mixing matrices and one $3\times 3$,
which in this case corresponds to the two line Young diagram and can be directly found
from the $SU_q(2)$ Racah coefficients.

The mixing matrices are \cite{IMMMIII}:
\be
{\cal R}_{[51]} = \left(\begin{array}{cc} q^{\varkappa_{[4]}} & 0 \\
0 & -q^{\varkappa_{[31]}}\end{array}\right)
= \left(\begin{array}{cc} q^6 & 0 \\ 0 & - q^2\end{array}\right)
\ \ \ \ \ \Longrightarrow \ \ \ \ \
U_{[51]} = \frac{1}{[2]_{q^2}}\left(\begin{array}{cc} 1 & \sqrt{[3]_{q^2}} \\
-\sqrt{[3]_{q^2}} & 1 \end{array}\right)
\ee
\be
{\cal R}_{[321]} = \left(\begin{array}{cc} -q^{\varkappa_{[31]}} & 0 \\
0 & q^{\varkappa_{[22]}}\end{array}\right)
= \left(\begin{array}{cc} -q^2 & 0 \\ 0 & 1\end{array}\right)
\ \ \ \ \ \Longrightarrow \ \ \ \ \
U_{[321]} = \frac{1}{[2]_q}\left(\begin{array}{cc} 1 & \sqrt{[3]_q} \\
-\sqrt{[3]}_q & 1\end{array}\right)
\ee
They are deduced in \cite{IMMMIII} by the same trick:
comparison with the Rosso-Jones formula for the torus knots.
The only difference from s.\ref{conj1} is that now $\xi_1\xi_2\neq 1$.
It is easy to check that these matrices are  given by rule (\ref{U2ev}).

The $3\times 3$ mixing matrix is found in \cite{IMMMIII} in a rather sophisticated
form, see eqs.(51) and (52) of that paper, but the final answer is rather simple:
\be
{\cal R}_{[42]} = \left(\!\begin{array}{ccc} q^{\varkappa_{[4]}} & 0 & 0 \\
0 & -q^{\varkappa_{[31]}} & 0 \\ 0 & 0 & q^{\varkappa_{[22]}} \end{array}\right)
= \left(\begin{array}{ccc} q^6 & 0 & 0\\ 0 & - q^2 & 0 \\ 0&0& 1\end{array}\right)
\ \ \Longrightarrow \ \
U_{[42]} = \left(\!\begin{array}{ccc}  \frac{[2]}{[3][4]}
& -\frac{[2]}{[4]}\sqrt{\frac{[5]}{[3]}} &-\frac{\sqrt{[5]}}{[3]}\\
\frac{[2]}{[4]}\sqrt{\frac{[5]}{[3]}} & -\frac{[6]}{[3][4]}& \frac{1}{\sqrt{[3]}} \\
-\frac{\sqrt{[5]}}{[3]} &-\frac{1}{\sqrt{[3]}}& \frac{1}{[3]}
\end{array}\right)
\label{U42}
\ee
The question is whether it follows from some generic eigenvalue formula, a counterpart
of (\ref{U2ev}).
The answer is affirmative, see eq.(\ref{Uxxx}) below.

\section{Representation $[3]$
\label{rep3}}

%\subsection{Decomposition of $[3]^{\times 3}$ and generic formula}

The $R$-matrix eigenvalues in the four different channels in
$[3]\otimes[3] = [6]+[51]+[42]+[33]$ are:
\be
\begin{array}{ccccc}
\l[6]: && q^{\varkappa_{[6]}} &=& q^{15}   \\
\l[51]: && -q^{\varkappa_{[51]}} &=& -q^9   \\
\l[42]: && q^{\varkappa_{[42]}} &=& q^5 \\
\l[33]: && -q^{\varkappa_{[33]}} &=& -q^3
\end{array}
\label{R3ev}
\ee
Since
\be
\l[3]\times[3]\times[3] = \Big([6]+[51]+[42]+[33]]\Big)\times [3] =  \\
= ([9]+[81]+[72]+[63]) + ([81]+[72]+[711]+[63]+[621]+[54]+[531]) +  \\
+ ([72]+[63]+[621]+[54]+[531]+[522]+[441]+[432]) + ([63]+[531]+[432]+[333]) =
\\
= [9]+ [711]+[522]+[441]+[333] + 2\cdot\Big([81]+[621]+[54]+[432]\Big)
+ 3\cdot\Big([72]+[531]\Big) + 4\cdot[63]
\ee
eq.(\ref{extHexpan}) implies:
$$
{\cal H}^{\,a_1,b_1|\ldots}_{[3]} =
q^{15(a_1+b_1+\ldots)} S_{[9]} + (-q)^{9(a_1+b_1+\ldots)} S_{[711]}
+ q^{5(a_1+b_1+\ldots)} \Big(S_{[522]} + S_{[441]}\Big)
+ (-q)^{3(a_1+b_1+\ldots)} S_{[333]} +
$$
\vspace{-0.4cm}
\be
+  \tr_{2\times 2} \left\{\left(\begin{array}{cc} q^{15} & \\ & -q^9
\end{array}\right)^{a_1}
U_{15,9}
%\left(\begin{array}{cc} c_{15,9]} & s_{15,9} \\ -s_{15,9} & c_{15,9}\end{array}\right)
\left(\begin{array}{cc} q^{15} & \\ & -q^9 \end{array}\right)^{b_1}
U_{15,9}^\dagger
%\left(\begin{array}{cc} c_{15,9} & -s_{15,9} \\ s_{15,9} & c_{15,9} \end{array}\right)
\ \ldots\
\right\}  S_{[81]} +  \\
+ \tr_{2\times 2}\left\{ \left(\begin{array}{cc} -q^{9} & \\ & q^5
\end{array}\right)^{a_1}
U_{9,5}
%\left(\begin{array}{cc} c_{9,5} & s_{9,5} \\ -s_{9,5} & c_{9,5}\end{array}\right)
\left(\begin{array}{cc} -q^{9} & \\ & q^5 \end{array}\right)^{b_1}
U_{9,5}^\dagger
%\left(\begin{array}{cc} c_{9,5} & -s_{9,5} \\ s_{9,5} & c_{9,5}\end{array}\right)
\ \ldots\
\right\} \Big(S_{[621]} + S_{[54]}\Big) +  \\
+ \tr_{2\times 2} \left\{\left(\begin{array}{cc} q^{5} & \\ & -q^3
\end{array}\right)^{a_1}
U_{5,3}
%\left(\begin{array}{cc} c_{5,3} & s_{5,3} \\ -s_{5,3} & c_{5,3}\end{array}\right)
\left(\begin{array}{cc} q^{5} & \\ & -q^3 \end{array}\right)^{b_1}
U_{5,3}^\dagger
%\left(\begin{array}{cc} c_{5,3} & -s_{5,3} \\ s_{5,3} & c_{5,3}\end{array}\right)
\ \ldots\
\right\} S_{[432]} +  \\
+ \tr_{3\times 3}\left\{ \left(\begin{array}{ccc} q^{15} && \\ & -q^9 & \\
&&q^5 \end{array}\right)^{a_1}
U_{15,9,5}
%\left(\begin{array}{ccc} c_{5,3} & s_{5,3} \\ -s_{5,3} & c_{5,3}\end{array}\right)
\left(\begin{array}{ccc} q^{15} & &\\ & -q^9&\\&&q^5 \end{array}\right)^{b_1}
U_{15,9,5}^\dagger
%\left(\begin{array}{ccc} c_{5,3} & -s_{5,3} \\ s_{5,3} &c_{5,3} \end{array}\right)
\ \ldots\
\right\} S_{[72]} +  \\
+ \tr_{3\times 3}\left\{ \left(\begin{array}{ccc} -q^{9} && \\ & q^5 & \\
&&-q^3 \end{array}\right)^{a_1}
U_{9,5,3}
%\left(\begin{array}{ccc} c_{5,3} & s_{5,3} \\ -s_{5,3} & c_{5,3}\end{array}\right)
\left(\begin{array}{ccc} -q^{9} & &\\ & q^5&\\&&-q^{3}
\end{array}\right)^{b_1}
U_{9,5,3}^\dagger
%\left(\begin{array}{ccc} c_{5,3} & -s_{5,3} \\ s_{5,3} &c_{5,3} \end{array}\right)
\ \ldots\
\right\} S_{[531]} +  \\
+\tr_{4\times 4}  \left\{\left(\begin{array}{cccc} q^{15} &&& \\ & -q^9 && \\
&&q^5&\\&&&-q^3 \end{array}\right)^{a_1}
U_{15,9,5,3}
%\left(\begin{array}{ccc} c_{5,3} & s_{5,3} \\ -s_{5,3} & c_{5,3}\end{array}\right)
\left(\begin{array}{cccc} q^{15} & &&\\ & -q^9&&\\&&q^5&\\&&&-q^3
\end{array}\right)^{b_1}
U_{15,9,5,3}^\dagger
%\left(\begin{array}{ccc} c_{5,3} & -s_{5,3} \\ s_{5,3} &c_{5,3} \end{array}\right)
\ \ldots\
\right\} S_{[63]}
\label{H3-3}
\ee
This time we labeled the mixing matrices $U_Q$ by the eigenvalues of the corresponding
${\cal R}_Q$, in accordance with our conjectures.
These conjectures are also used, when the contributions of $Q=[621]$ and $Q=[54]$
are unified: the ${\cal R}$-matrices are the same and we assume that the same
is true for the mixing matrices.

Moreover, conjecture {\bf 2} implies that
\be
U_{[621]} = U_{[54]} = U_{9,5} \ \stackrel{\bf 2}{=}\  U_{6,2} = U_{[51]}  \\
U_{[432]} = U_{5,3}  \ \stackrel{\bf 2}{=}\  U_{2,0} = U_{[321]}
\ \stackrel{\bf 2}{=}\ U_{1,-1} = U_{[21]}  \\
U_{[531]} = U_{9,5,3}  \ \stackrel{\bf 2}{=}\  U_{6,2,0} = U_{[42]}  \\
\ee
i.e. these mixing matrices are given by  expressions,
already known from the study of $R=[1]$ and $R=[2]$.
The "new" mixing matrices are
\be
U_{[81]} = U_{15,9}    \\
U_{[72]} = U_{15,9,5}    \\
U_{[63]} = U_{15,9,5,3}
\label{newma3}
\ee
So, what is our way to define these matrices?

Our suggestion is to make use of the known Racah coefficients for the
$SU_q(2)$ algebra.
They involve only the Young diagrams with at most two lines,
but under our conjectures this appears sufficient to define the
mixing matrices like (\ref{newma3}).

\section{Racah coefficients for $SU_q(2)$}

The Racah matrix (i.e. basically the $6j$-symbols) $u^T_{T'}$, or rather
$\ u\!\left[\!\!\begin{array}{c|c}RRR&T\\Q&T'\!\!\end{array}\right]$,
describes the ordering dependence of the product of three representations:

\begin{picture}(300,100)(-120,-45)
\put(0,0){\line(-1,1){30}}
\put(0,0){\line(1,1){30}}
\put(-15,15){\line(1,1){15}}
\put(0,0){\line(0,-1){20}}
\put(-40,27){\mbox{$R$}}
\put(7,27){\mbox{$R$}}
\put(35,27){\mbox{$R$}}
\put(-10,-25){\mbox{Q}}
\put(-20,0){\mbox{$T$}}
\put(200,0){\line(-1,1){30}}
\put(200,0){\line(1,1){30}}
\put(215,15){\line(-1,1){15}}
\put(200,0){\line(0,-1){20}}
\put(160,27){\mbox{$R$}}
\put(187,27){\mbox{$R$}}
\put(235,27){\mbox{$R$}}
\put(190,-25){\mbox{Q}}
\put(210,0){\mbox{$T'$}}
\put(85,0){\mbox{$= \ \ \ \sum_{T'}\ \  u^T_{T'} $}}
\end{picture}

\noindent
This matrix is exactly what we call the mixing matrix $U_{RQ}$ above
and below.

The crucial feature of representation theory is that the fusion properties
of representations, in particular, the entries $u^T_{T'}$ of the Racah matrix,
do not depend on the rank of the algebra $SU_q(N)$: they are the same for
all $N$. What differs, are the non-vanishing representations:
only those with no more than $N$ lines in the Young diagram can be seen
for a given $N$.
This means that representation theory of $SU_q(2)$, the only one which
is sufficiently developed allows one to extract
$\ u\!\left[\!\!\begin{array}{c|c}RRR&T\\Q&T'\!\!\end{array}\right]$
only when $R,Q,T,T'$ are all one- or two-line Young diagrams.
Our conjectures in s.\ref{conj} are designed to extend these available
quantities to other $R,Q,T,T'$ without performing a tedious
representation theory analysis for higher rank groups.

In this section we briefly remind some well-known $SU_q(2)$ formulas
and demonstrate how they can be used, if our conjectures are true.

\bigskip

When $R$ is a one-line diagram with $p$ boxes (we use $p$ instead
of $r$, because in our applications they are not always the same),
i.e. a symmetric representation $R=[p]$, then $T$ and $T'$ are of the
form $[2p-j,j]$, while $Q$ can be a one-, two- or three-line diagram.
Three-line $Q$'s are not seen at the $SU_q(2)$ level,
where one can consider only $Q$ of restricted form, $Q=[3p-k,k]$.
Then $0\leq j,j'\leq k$ for $T$ and $T'$.
In other words,
%The knowledge of these coefficients allows one to construct
%particular mixing matrices: between the $\N$
%copies of the 2-column representation $[3p-(\N-1),\N-1]$
%in the product
\be
\rp^{\otimes 3} =
[3p\,] + 2\cdot [3p-1,1] + 3\cdot [3p-2,2] + \ldots =\\
= \sum_{k=0}^{p} (k+1)[3p-k,k]
+ {\rm other \ Young\ diagrams\ with}\ \geq 2\ {\rm lines}
\label{p3expan}
\ee
and the elements of the $(k+1)\times (k+1)$ Racah matrix are labeled as
$u^{[2p-j,j]}_{[2p-j\,',j\,']},\ \ 0\leq j,j\,'\leq k$.
In the remaining part of this section we
list these matrices and express their entries
through the eigenvalues of the corresponding ${\cal R}$-matrices.
The $2$-line representations with $0\leq k\leq p$, which
are explicitly mentioned in (\ref{p3expan}) are distinguished by
simplicity of these eigenvalues: they are equal to
%The corresponding ${\cal R}$ matrix has peculiar eigenvalues
\be
(-)^j\cdot q^{\varkappa_{[2p-j,j]}}\ =\ (-)^j\cdot q^{2p^2-(2j+1)p+j(j-1)}
\ee
Thus, the $\N\times\N$ Racah matrix is what we called
$U_{\varkappa_{[2p]},\varkappa_{[2p-1,1]},\ldots, \varkappa_{[2p-k,k]}}$
with $k+1=\N$.
Our {\bf conjecture 2} implies that this is the same as
$U_{c-p,\,c-3p,\,\ldots,\,c-(2k+1)p+k(k-1)}$ with arbitrary $c$.

Note that there are more two-line contributions with $k>p$ in (\ref{p3expan}),
with other multiplicities.
They are also under control of the $SU_q(2)$ representation theory.

\subsection{$2\times 2$}

The $SU_q(2)$ Racah matrix of size $2\times 2$ is
\be\label{U2p}
%U^{(2)}[p] =
U(2|p\,)\ =\
\left(
\begin{array}{cc}
\frac{[p\,]}{[2p\,]} & \epsilon\frac{\sqrt{[p\,][3p\,]}}{[2p\,]} \\ \\
-\epsilon\frac{\sqrt{[p\,][3p\,]}}{[2p\,]} & \frac{[p\,]}{[2p\,]}
\end{array}\right)
\ee
and is obtained for the fusion of $\rp^{\otimes 3}\to [3p-1]$.
The sign $\epsilon=\pm 1$ of the square root in the off-diagonal element is arbitrary, they depend on the
choice of normalization of the basis.
In what follows we put $\epsilon=1$ in order to match the standard definition of the Racah coefficients.
If expressed through the ${\cal R}$-matrix eigenvalues,
$\xi_1 = q^{\varkappa_{[2p]}}$ and $\xi_2 = -q^{\varkappa_{[2p-1,1]}}$
i.e. $\tilde\xi_1 = q^p$ and $\tilde\xi_2=-q^{-p} =-\tilde\xi_1^{-1}$,
this $U(2|p\,)$ becomes:
\be
%U(2|p\,)
U(2|p\,)\ =\ U[\tilde\xi_1,\tilde\xi_2]\ =\ \frac{1}{\tilde\xi_1-\tilde\xi_2}
\left(
\begin{array}{cc} 1 & \sqrt{\tilde\xi_1^2+1+\tilde\xi_2^2} \\
-\sqrt{\tilde\xi_1^2+1+\tilde\xi_2^2} & 1 \end{array}\right)
\label{Uxx}
\ee
and our weak conjecture is:
\be
\boxed{
U_{c-p,\ c-3p} = U\big[q^p,-q^{-p}\big] = U(2|p\,)
}
\ee
with an arbitrary $c$.
In particular, the $2\times 2$ mixing matrices, which we encountered
in ss.\ref{rep2},\ref{rep3} above, are:
\be
\begin{array}{ccccc}
U_{[432]} = U_{[321]} &=& U_{5,3} = U_{1,-1} = U_{4,2} &=& U(2|p=1),   \\
U_{[51]} = U_{[621]} = U_{[54]} &=& U_{9,5} = U_{6,2} &=& U(2|p=2),  \\
U_{[81]} &=& U_{15,9} &=& U(2|p=3) \\
\ldots
\end{array}
\ee

\subsection{$3\times 3$}

The $3\times 3$ Racah matrix is

{\footnotesize
\be
U(3|p\,)\ = \
\left(\begin{array}{ccc}
\frac{[p-1][p]}{[2p-1][2p]} &
\frac{\epsilon_1[p]}{[2p]}\sqrt{\frac{[2]\cdot[p-1]\cdot[3p-1]}{[2p-2][2p-1]}}
&\frac{\epsilon_2}{[2p-1]}
\sqrt{\frac{[p-1][p\,]\cdot[3p-2][3p-1]}{[2p-2]][2p\,]}} \\ \\
-\frac{\epsilon_1[p]}{[2p]}\sqrt{\frac{[2]\cdot[p-1]\cdot[3p-1]}{[2p-2][2p-1]}}
&-\frac{[p-1][p\,]\cdot[4p-2]}{[2p-2][2p-1][2p]}
&\frac{\epsilon_1\epsilon_2[p-1]}{[2p-2]}\sqrt{\frac{[2][p\,]\cdot[3p-2]}{[2p-1][2p\,]}}\\ \\
\frac{\epsilon_2}{[2p-1]}\sqrt{\frac{[p-1][p\,]\cdot[3p-2][3p-1]}{[2p-2]][2p\,]}}
& -\frac{\epsilon_1\epsilon_2[p-1]}{[2p-2]}\sqrt{\frac{[2][p\,]\cdot[3p-2]}{[2p-1][2p\,]}}
& \frac{[p-1][p\,]}{[2p-2][2p-1]}
\end{array}\right)
\label{U3p}
\ee}
\hspace{-.3cm}
Now there are {\it two} independent square roots and two arbitrary signs
$\epsilon_{1,2}=\pm 1$.
Like eq. (\ref{U42}), we accept  the sign convention
$\epsilon_1 = \epsilon_2 = -1$.
Note also that the transposition reverses signs
of elements at the odd diagonals, and does not change those at the even diagonals.
For a general mixing matrix of size $(k+1)$ there is a sign ambiguity such that
$k$ sign factors remain arbitrary, $\epsilon_1,\ldots,\epsilon_k=\pm 1$:
$U_{ij} \sim \epsilon_{i-1}\epsilon_{j-1}U_{ij}$, while the Racah matrix keeps to be orthogonal.
In what follows we omit these arbitrary sign factors:
they do not affect the answers for the HOMFLY polynomials either.

Being expressed in terms of the ${\cal R}$-matrix eigenvalues
$\xi_1 = q^{\varkappa_{[2p]}}$, $\xi_2 = -q^{\varkappa_{[2p-1,1]}}$,
$\xi_3 = q^{\varkappa_{[2p-2,2]}}$ and
%$\tilde\xi_1 = q^{2p-\frac{2}{3}}$,
%$\tilde\xi_2 = - -q^{-\frac{2}{3}}$,
%$\tilde\xi_3 = q^{-\left(2p-\frac{4}{3}\right)}$,
\be
\tilde\xi_1 = q^{2p-\frac{2}{3}}\\
\tilde\xi_2 = -q^{-\frac{2}{3}}\\
\tilde\xi_3 = q^{-\left(2p-\frac{4}{3}\right)}
\ee
the $3\times 3$ Racah matrix (\ref{U3p}) becomes:
\be
U(3|p\,)\ =\ U[\tilde\xi_1,\tilde\xi_2,\tilde\xi_3]\ =\
\left(\begin{array}{ccc}
-{\tilde\xi_1(\tilde\xi_2+\tilde\xi_3)\over\tilde\xi_{12}\tilde\xi_{13}}
& -\frac{1}{\tilde\xi_{12}}\sqrt{\frac{(\tilde\xi_1^3+1)(\tilde\xi_2^3+1)}
{\tilde\xi_1\tilde\xi_2\tilde\xi_{13}\tilde\xi_{23}}}
& -\frac{1}{\tilde\xi_{13}}\sqrt{\frac{(\tilde\xi_1^3+1)(\tilde\xi_3^3+1)}
{\tilde\xi_1\tilde\xi_3\tilde\xi_{12}\tilde\xi_{32}}}\\
\frac{1}{\tilde\xi_{12}}\sqrt{\frac{(\tilde\xi_1^3+1)(\tilde\xi_2^3+1)}
{\tilde\xi_1\tilde\xi_2\tilde\xi_{13}\tilde\xi_{23}}}
&{\tilde\xi_2(\tilde\xi_1+\tilde\xi_3)\over\tilde\xi_{12}\tilde\xi_{23}}
&\frac{1}{\tilde\xi_{23}}\sqrt{\frac{(\tilde\xi_2^3+1)(\tilde\xi_3^3+1)}
{\tilde\xi_2\tilde\xi_3\tilde\xi_{12}\tilde\xi_{13}}}\\
-\frac{1}{\tilde\xi_{13}}\sqrt{\frac{(\tilde\xi_1^3+1)(\tilde\xi_3^3+1)}
{\tilde\xi_1\tilde\xi_3\tilde\xi_{12}\tilde\xi_{32}}}
& -\frac{1}{\tilde\xi_{23}}\sqrt{\frac{(\tilde\xi_2^3+1)(\tilde\xi_3^3+1)}
{\tilde\xi_2\tilde\xi_3\tilde\xi_{12}\tilde\xi_{13}}}
& -{\tilde\xi_3(\tilde\xi_1+\tilde\xi_2)\over\tilde\xi_{13}\tilde\xi_{23}}
\end{array}\right)
\label{Uxxx}
\ee
As usual, $\tilde\xi_{ij} = \tilde\xi_i-\tilde\xi_j$.
Note that $\tilde\xi_2$ is defined with the minus sign.

Our weak conjecture is
\be
\boxed{
U_{c-p,\ c-3p,\ c-5p+2} = U\big[q^{2p},-1,q^{2-2p}\big] = U(3|p\,)
}
\ee
with arbitrary $c$.

This implies that
\be
%??? = U(3|p=1),  \\
U_{[531]} =  U_{[42]}  = U_{9,5,3}= U_{6,2,0} = U(3|p=2),  \\
U_{[72]} = U_{15,9,5} = U(3|p=3),  \\
\ldots
\ee

In particular,
\be
  U_{[72]} = U(3|p=3) =
\left(\begin{array}{ccc}
\frac{[2][3]}{[5][6]} & -\frac{[3]}{[6]}\sqrt{\frac{[2]\cdot[2]\cdot[8]}{[4][5]}}
& -\frac{1}{[5]}\sqrt{\frac{[2][3\,]\cdot[7][8]}{[4]][6\,]}} \\ \\
 \frac{[3]}{[6]}\sqrt{\frac{[2]\cdot[2]\cdot[8]}{[4][5]}} &-\frac{[2][3\,]\cdot[10]}{[4][5][6]}
&\frac{[2]}{[4]}\sqrt{\frac{[2][3\,]\cdot[7]}{[5][6\,]}}\\ \\
 - \frac{1}{[5]}\sqrt{\frac{[2][3\,]\cdot[7][8]}{[4]][6\,]}}  & -\frac{[2]}{[4]}\sqrt{\frac{[2][3\,]\cdot[7]}{[5][6\,]}}
        & \frac{[2][3\,]}{[4][5]}
\end{array}\right)
\ee

\subsection{$4\times 4$}

The $4\times 4$ Racah matrix is
\be
U(4|p\,) =
\label{U4p}
\ee

\bigskip

{\footnotesize
\centerline{
$
%\!\!\!\!\!\!\!\!\!\!\!\!\!\!\!\!\!\!\!\!\!\!\!\!\!\!\!\!\!\!\!\!\!\!\!\!\!\!\!\!\!
 \left(\begin{array}{cccc}
[2p\,]& [2p-1,1] & [2p-2,2] & [2p-3,3]\\
\hline \\
\frac{[p-2][p-1][p]}{[2p-2][2p-1][2p]}
& \frac{[p-1][p]}{[2p-2][2p]}\sqrt{\frac{[p-2][3][3p-2]}{[2p-1][2p-3]}}
& \frac{[p]}{[2p-1][2p-2]}\sqrt{\frac{[3][p-2][p-1][3p-3][3p-2]}{[2p][2p-4]}}
&
\frac{1}{[2p-2]}\sqrt{\frac{[p-2][p-1][p][3p-4][3p-3][3p-2]}{[2p-4][2p-3][2p-1][2p]}}
\\ \\
& \frac{[p-1][p]\Big([3p-2]+[3p-4]-[p]\Big)}{[2p-3][2p-2][2p]}
&
\frac{[p-2]\Big([2][p-1]-[3p-2]\Big)}{[2p-2]}\sqrt{\frac{[p-1][3p-3]}{[2p-4][2p-3][2p-1][2p]}}
&
\frac{[p-2]}{[2p-2][2p-3]}\sqrt{\frac{[p]}{[2p]}\frac{[3][p-1][3p-3][3p-4]}{[2p-4]}}\\
\\
&
& \frac{[p-2][p-1]\Big([2][3p-3]-[p-2]\Big)}{[2p-1][2p-2][2p-4]}
& \frac{[p-2][p-1]}{[2p-4][2p-2]}\sqrt{\frac{[p][3][3p-4]}{[2p-3][2p-1]}}
\\ \\
&&&\frac{[p-2][p-1][p]}{[2p-4][2p-3][2p-2]}
\end{array}\right)
$
}}

\bigskip

At this moment we just note that the last matrix we need to complete
the calculation in the $[3]^{\otimes 3}$ case is
\be
\begin{array}{c}
U_{63} = \\ U_{15,9,5,3} = \\ U(4|p=3) =
\end{array}
%U_{63} = U_{15,9,5,3} = U(4|p=3) =
%U_{[63]} =
\left(\begin{array}{cccc}
%[2p\,]& [2p-1,1] & [2p-2,2] & [2p-3,3]\\
%\hline \\
 -\frac{[2]\cdot[3]}{[4]\cdot[5]\cdot[6]}
& \frac{[2]\cdot[3]}{[4]\cdot[6]}\sqrt{\frac{[7]}{[5]}}
& -\frac{[3]}{[5]\cdot[4]}\sqrt{[3]\cdot[7]}
&
 \frac{1}{[4]}\sqrt{[7]}
\\ \\
 - \frac{[2]\cdot[3]}{[4]\cdot[6]}\sqrt{\frac{[7]}{[5]}}  &
  \frac{[2]\cdot \Big([7]+[5]-[3]\Big)}{[4]\cdot[6]}
&
-\frac{\Big(-[2]\cdot[2] + [7]\Big)}{[4]}\frac{1}{\sqrt{[3]\cdot[5]}}
&
-\frac{1}{[4]}\sqrt{[5]}\\\\
-\frac{[3]}{[5]\cdot[4]}\sqrt{[3]\cdot[7]}
&\frac{\Big(-[2]\cdot[2] + [7]\Big)}{[4]} \frac{1}{\sqrt{[3]\cdot[5]}}
& \frac{\Big([2]\cdot[6]-1\Big)}{[5]\cdot[4]}
& \frac{1}{[4]}\sqrt{[3]}
\\ \\
-\frac{1}{[4]}\sqrt{[7]}  & -\frac{1}{[4]}\sqrt{[5]}   & -\frac{1}{[4]}\sqrt{[3]} &-\frac{1}{[4]}
\end{array}\right)\nn
\ee

%\be
%%U_{[63]} =
%\left(\begin{array}{cccc}
%%[2p\,]& [2p-1,1] & [2p-2,2] & [2p-3,3]\\
%%\hline \\
% -\frac{[2]\cdot[3]}{[4]\cdot[5]\cdot[6]}
%& sgn_{12} \frac{[2]\cdot[3]}{[4]\cdot[6]}\sqrt{\frac{[7]}{[5]}}
%&  sgn_{12} sgn_{23}\frac{[3]}{[5]\cdot[4]}\sqrt{[3]\cdot[7]}
%&
% -sgn_{12} sgn_{23}sgn_{34}\frac{1}{[4]}\sqrt{[7]}
%\\ \\
% -sgn_{12} \frac{[2]\cdot[3]}{[4]\cdot[6]}\sqrt{\frac{[7]}{[5]}}  &
%  \frac{[2]\cdot \Big([7]+[5]-[3]\Big)}{[4]\cdot[6]}
%&
%sgn_{23} \frac{\Big(-[2]\cdot[2] + [7]\Big)}{[4]}\frac{1}{\sqrt{[3]\cdot[5]}}
%&
%sgn_{23}sgn_{34} \frac{1}{[4]}\sqrt{[5]}\\\\
%- sgn_{12}sgn_{23} \frac{[3]}{[5]\cdot[4]}\sqrt{[3]\cdot[7]}
%&-sgn_{23} \frac{\Big(-[2]\cdot[2] + [7]\Big)}{[4]} \frac{1}{\sqrt{[3]\cdot[5]}}
%& \frac{\Big([2]\cdot[6]-1\Big)}{[5]\cdot[4]}
%& sgn_{34} \frac{1}{[4]}\sqrt{[3]}
%\\ \\
%sgn_{12}sgn_{23}sgn_{34}\frac{1}{[4]}\sqrt{[7]}  & sgn_{23} sgn_{34} \frac{1}{[4]}\sqrt{[5]}
% & -sgn_{34}\frac{1}{[4]}\sqrt{[3]} &-\frac{1}{[4]}
%\end{array}\right)
%\ee
%
%  We can choose $sgn_{12} = \pm$, $sgn_{23} = \pm$ and $sgn_{34} = \pm$ independently.
%  The choice made here  corresponds to
%\be
%sgn_{12} = - sgn_{23} = sgn_{34} = 1
%\ee
% As is easily seen, there are prominent patterns in the matrix elements of the general
%  $N\times N$ mixing? or Racah matrices with regard to their (anti)symmtricity and sings.
%  The off-diagonal matrix elements lying nearest  the diagoal are antisymmtric, while those
%   lying  nextnerest the diagonal are symmetric. This alternating pattern persists.
%   The sign factors are controled by those contained in the nearest  off-diagonals.

We return to the eigenvalue description of this formula in s.\ref{evcond} below,
where a more condensed notation is introduced.

\section{Precise formulation of eigenvalue conjecture}

\subsection{Weak form}

The two mixing matrices, one of them being $SU_q(2)$ are the same provided the following condition is satisfied.
Associate with $Q$ a set of integers:

$Q \longrightarrow \{r| j_1,\ldots,j_2\}$, such that $j_1<j_2$ ($k=j_2-j_1$, $p=r-j_1$). Then

$$
\ u\!\left[\!\!\begin{array}{c|c}\l[r],[r],[r]&j_1+j\\Q&j_1+j'\!\!\end{array}\right]
= \ u\!\left[\!\!\begin{array}{c|c}\l[p\,],[p\,],[p\,] &j \\ \l[3p-k] &j'\!\!\end{array}\right]
$$

\begin{picture}(300,100)(-120,-45)
\put(0,0){\line(-1,1){30}}
\put(0,0){\line(1,1){30}}
\put(-15,15){\line(1,1){15}}
\put(0,0){\line(0,-1){20}}
\put(-40,27){\mbox{$[r]$}}
\put(7,27){\mbox{$[r]$}}
\put(35,27){\mbox{$[r]$}}
\put(-10,-30){\mbox{[3r-l-m,l,m]}}
\put(-50,0){\mbox{$[2r-j,j]$}}
\put(200,0){\line(-1,1){30}}
\put(200,0){\line(1,1){30}}
\put(215,15){\line(-1,1){15}}
\put(200,0){\line(0,-1){20}}
\put(160,27){\mbox{$[r]$}}
\put(187,27){\mbox{$[r]$}}
\put(235,27){\mbox{$[r]$}}
\put(190,-30){\mbox{[3r-l-m,l,m]}}
\put(210,0){\mbox{$[2r-j\,',j\,']$}}
\put(85,0){\mbox{$= \ \ \ \sum_{j\,'}\ \  u^{j}_{j'} $}}
\end{picture}

\begin{picture}(300,100)(-120,-45)
\put(0,0){\line(-1,1){30}}
\put(0,0){\line(1,1){30}}
\put(-15,15){\line(1,1){15}}
\put(0,0){\line(0,-1){20}}
\put(-40,27){\mbox{$[p]$}}
\put(7,27){\mbox{$[p]$}}
\put(35,27){\mbox{$[p]$}}
\put(-10,-30){\mbox{[3p-k,k]}}
\put(-50,0){\mbox{$[2p-j,j]$}}
\put(200,0){\line(-1,1){30}}
\put(200,0){\line(1,1){30}}
\put(215,15){\line(-1,1){15}}
\put(200,0){\line(0,-1){20}}
\put(160,27){\mbox{$[p]$}}
\put(187,27){\mbox{$[p]$}}
\put(235,27){\mbox{$[p]$}}
\put(190,-30){\mbox{[3p-k,k]}}
\put(210,0){\mbox{$[2p-j\,',j\,']$}}
\put(85,0){\mbox{$= \ \ \ \sum_{j\,'}\ \  u^j_{j\,'} $}}
\end{picture}

\be
\boxed{
U_{\varkappa_{[2r-j_1,j_1]},\varkappa_{[2r-j_1-1,j_1+1]},\ldots,
\varkappa_{[2r-j_2,j_2]}} \ =\
U[j_2-j_1+1|p=r-j_1]
}
\ee

In this form it is applicable to calculation of $[r]^{\otimes 3}$.

If true, this substitutes the honest calculation of the Racah matrix for just  the
$SU_q(3)$ case, since for higher groups there will emerge Young diagrams with more than three lines.

\subsection{Strong form}

In its strong form the conjecture can be applied for the Racah matrix for $R^{\otimes 3}$
with arbitrary $R$, in particular, it can be checked for non-trivial representations like $[2,1]^{\otimes 3}$.
Within this framework one suffices to know the size of the corresponding Racah matrix and the ${\cal R}$-matrix
eigenvalues, and use formulas like (\ref{Uxx}), (\ref{Uxxx}) and so on.
If true, this substitutes full $SU_q(\infty)$ Racah calculus for three coinciding representations.

\subsection{Checks}

The conjecture has to be further examined with various checks. Apart from testing answers for the colored
HOMFLY polynomials obtained with help of the conjecture, one may check immediately the Racah matrices. A few evident tests are:

\begin{itemize}

\item One may check other 2-line contributions to (\ref{p3expan}).

\item The Racah matrices have to satisfy various consistency conditions like the pentagon identity.

\item One can compare the eigenvalues of the matrix ${\cal R}_{Q}U_{Q}{\cal R}_{Q}U_{Q}^\dagger$ with those
known from the torus knot/link results obtained by applying the Rosso-Jones formula.

\end{itemize}

\section{Racah matrices for higher $\N$
\label{evcond}}

While for the purposes of the present paper, for evaluation of the 3-strand HOMFLY
polynomials in symmetric and antisymmetric representations
sufficient is the weak form of our conjecture, and thus just the $SU_q(2)$
Racah matrices $U(\N|p)$ are needed,
the strong form, if correct, would open a way for study of other
representations and multi-strand braids.
But for it to work one should at least attempt to re-express $U(\N|p)$
through associated ${\cal R}$-matrix eigenvalues, as we did above for $\N=2$ and $3$.
In this section we present the {\it answers} for the next two sizes, $\N=4$ and $5$.
However, for $\N>5$ we did not succeed yet, thus, we have no any direct evidence that
such formulas do exist at all at the moment. If they do not exist at $\N>5$, this would imply immediate
failure of our conjecture in its strong form.

\subsection{Mixing matrices through ${\cal R}$-matrix eigenvalues}

The results for $\N = 2,3,4,5$ can be presented in a universal form and are summarized in the following table.
The {\it squares}(!) of all the off-diagonal elements $i\neq j$ can be presented as
\be
%\!\!\!\!\!\!\!\!\!\!\!\!\!\!\!\!\!\!\!\!\!\!\!\!\!\!\!\!\!\!\!\!\!\!\!\!
U_{ij}^2 = \frac{
%????\tilde\xi_i\tilde\xi_j????
1}{\tilde\xi_{ij}^2 \prod_{k\neq i,j}^{\cal N} \tilde\xi_{ik}\tilde\xi_{jk}}
\ \cdot
\label{offdiag}
%\ee
%
%\centerline{
%$
%\cdot
\left(
\begin{array}{ccccc}
{\cal N}=2: &&
%\bar Q_{i}^{(2,3)} =  \bar Q_{j}^{(2,3)}
%= \sqrt{ \bar Q_{i}^{(2,3)} \bar Q_{j}^{(2,3)}}
&&  =\tilde\xi_1^2-1+\tilde\xi_1^{-2} =\tilde\xi_2^2-1+\tilde\xi_2^{-2}\\ \\
{\cal N}=3: &&
%(\tilde\xi_i\tilde\xi_j)^{1/2}Q_{i}^{(3,2)}Q_{j}^{(3,2)}
&&  =-(\tilde\xi_i^3+1)(\tilde\xi_j^3+1)(\tilde\xi_i\tilde\xi_j)^{-1} = \\
&&&& = -(\tilde\xi_i\tilde\xi_j)^{1/2}
(\tilde\xi_i^{3/2}+\tilde\xi_i^{-3/2})(\tilde\xi_j^{3/2}+\tilde\xi_j^{-3/2}) \\ \\
{\cal N}=4: &&
%??? \bar Q_i^{(2,2)}\bar Q_j^{(2,2)}
%\sqrt{\prod_{k\neq i,j}^4 \bar Q_{ik}^{(1,3)}\bar Q_{jk}^{(1,3)}}
&& \sim (\tilde\xi_i^2-1)(\tilde\xi_j^2-1)
\prod_{k\neq i,j}\left(\tilde\xi_i\tilde\xi_k - 1 + (\tilde\xi_i\tilde\xi_k)^{-1}\right) \\
&&&& = (\tilde\xi_i^2-1)(\tilde\xi_j^2-1)
\prod_{k\neq i,j}\left(\tilde\xi_j\tilde\xi_k - 1 + (\tilde\xi_j\tilde\xi_k)^{-1}\right)\\  \\
{\cal N}=5: &&
%(\tilde\xi_i\tilde\xi_j)Q^{(1,3)}_iQ^{(1,3)}_j
%\prod_{k\neq i,j} Q_{ik}^{(1,2)}Q_{jk}^{(1,2)}
&& (\tilde\xi_i\tilde\xi_j)(\tilde\xi_i+1+\tilde\xi_i^{-1})
(\tilde\xi_j+1+\tilde\xi_j^{-1})\cdot\\
&&&&\cdot
\prod_{k\neq i,j}(\tilde\xi_i\tilde\xi_k+1)(\tilde\xi_j\tilde\xi_k+1)\\ \\
{\cal N}=6: && &&?  \\  \\
%{\cal N}=7: && \\  \\
\end{array}
\right)
%$
%}
\ee

\bigskip

\noindent
The signs of the elements $U_{ij}$ are regulated by the rule
\be
\sigma U = U^\dagger \sigma
\label{symU1}
\ee
where $\sigma$ is a diagonal signature (alternating) matrix.
There is still an inessential sign ambiguity, described above,
in a comment after eq.(\ref{U3p}).

The diagonal elements can be restored from the orthogonality of the Racah matrix,
$U_{ii}^2 = 1- \sum_{j\neq i} U_{ij}^2$ and they are full squares
of factorized expressions (no square in this case!):
\be
U_{ii}=\frac{%???\tilde\xi_i???
1}{\prod_{k\ne i}\tilde\xi_{ik}}\cdot\left(
\begin{array}{ccc}
{\cal N}=2: &&    \sqrt{-1} \\ \\
{\cal N}=3: &&   -\tilde\xi_i\left(\sum_{k\neq i} \tilde\xi_k\right) \\ \\
{\cal N}=4: && \tilde\xi_i\Big( \tilde\xi_i (\sum_{j\ne k\ne i} \tilde\xi_j\tilde\xi_k)
-\sum_{k\ne i}\tilde\xi_k\Big) \\  \\
{\cal N}=5: &&  \tilde\xi_i\Big((\tilde\xi_i+1)(1+\sum_{k\ne i}(\tilde\xi_k+1/\tilde\xi_k)
+(\sum_{k\ne j\ne i}\frac{1}{\tilde\xi_k\xi_j})\Big)\\ \\
{\cal N}=6: &&  ? \\  \\
%{\cal N}=7: && \\  \\
\end{array}
\right)
\label{diag}
\ee

These formulas provide the eigenvalue description of the $SU_q(2)$ Racah
matrices $U(\N|p)$, which we now list for $\N\leq 5$ (and first few elements for $\N=6$ and arbitrary $\N$),
and our strong conjecture is that (\ref{offdiag}) and (\ref{diag})
describe the Racah matrices for {\it arbitrary} sets of ${\cal R}$-matrix
eigenvalues, yet for the {\it honest} evaluation of such matrices the
full representation theory for $SU_q(\infty)$ should be developed.

The formulas for $U(\N|p)$ are accompanied with the corresponding ($SU_q(2)$-related)
values of the normalized eigenvalues $\tilde\xi_i$.

\subsection{$SU_q(2)$ Racah matrices: ${\cal N}=2$}

For the $2\times 2$ Racah matrix see (\ref{U2p}).
According to our {\it strong} conjecture it is a {\it particular} case
of (\ref{Uxx}) for
$\left(\prod_{a=0}^1 \xi_a\right)^{-1/2} = q^{-2p^2+2p}$,
\be
\tilde\xi_0 = q^{p}\\
\tilde\xi_1 = -q^{-p}
\ee

\subsection{$SU_q(2)$ Racah matrices: ${\cal N}=3$}

For the $3\times 3$ Racah matrix see (\ref{U3p}).
According to our {\it strong} conjecture it is a {\it particular} case
of (\ref{Uxxx}) for
$\left(\prod_{a=0}^2 \xi_a\right)^{-1/3} = q^{-2p^2+3p-\frac{2}{3}}$,
\be
\tilde\xi_0 = q^{2p-\frac{2}{3}}\\
\tilde\xi_1 = -q^{-\frac{2}{3}}\\
\tilde\xi_2 = q^{-\left(2p-\frac{4}{3}\right)}
\ee

\subsection{$SU_q(2)$ Racah matrices: ${\cal N}=4$}

For the $4\times 4$ Racah matrix see (\ref{U4p}).

$\left(\prod_{a=0}^3 \xi_a\right)^{-1/4} = q^{-2p^2+4p-2}$,
\be
\tilde\xi_0 = q^{3p-2}\\
\tilde\xi_1 = -q^{p-2}\\
\tilde\xi_2 = q^{-p}\\
\tilde\xi_3 = -q^{-(3p-4)}
\ee

\subsection{$SU_q(2)$ Racah matrices: $5\times 5$}

$\left(\prod_{a=0}^4 \xi_a\right)^{-1/5} = q^{-2p^2+5p-4}$,
\be
\tilde\xi_0 = q^{4p-4}\\
\tilde\xi_1 = -q^{2p-4}\\
\tilde\xi_2 = q^{-2}\\
\tilde\xi_3 = -q^{-(2p-2)}\\
\tilde\xi_4 = q^{-(4p-8)}
\ee

\bigskip

\be
U_{00} = \frac{[p-3][p-2][p-1]\rp}{[2p-3][2p-2][2p-1]\2rp}, \nnn\nnn
U_{01} = \frac{[p-2][p-1]\rp}{[2p-3][2p-2]\2rp}
\sqrt{\frac{[4][p-3]\cdot[3p-3]}{[2p-4][2p-1]}},\nnn \nnn
U_{02} = \frac{[p-1]\rp}{[2p-2][2p-1]}
\sqrt{\frac{[3][4]\cdot[p-3][p-2]\cdot[3p-4][3p-3]}{[2]\cdot[2p-5][2p-4][2p-3]\2rp}},
\nnn \nnn
U_{03} = \frac{\rp}{[2p-3][2p-2]}\sqrt{\frac{[4]\cdot[p-3][p-2][p-1]\cdot[3p-5][3p-4][3p-3]}
{[2p-6][2p-4][2p-1]\2rp}},\nnn\nnn
U_{04} = \frac{1}{[2p-3]}
\sqrt{\frac{[p-3][p-2][p-1]\rp\cdot[3p-6][3p-5][3p-4][3p-3]}
{[2p-6][2p-5][2p-4][2p-2][2p-1]\2rp}}, \nnn\nnn\nnn
U_{11} = \frac{[p-2][p-1]\Big([p-3]^2-[3][p-1][3p-3]\Big)}{[2p-4][2p-3][2p-2]\2rp},\nnn\nnn
U_{12} =
%???\frac{[p-1]\Big([p-3]-[3p-3]\Big)}{[2p-4][2p-2]}
-\frac{[p-2]^{3/2}[p-1][p\,]\cdot[4p-6]}{[2p-4][2p-3][2p-2]}
\sqrt{\frac{[2][3]
%\cdot[p-2]^3
\cdot[3p-4]}{[2p-5][2p-3][2p-1]\2rp}},
\ee
\be
%\nnn\nnn
U_{13} = \frac{[p-3]\Big([3][p-1]-[3p-3]\Big)}{[2p-4][2p-3][2p-2]}
\sqrt{\frac{[p-2][p-1]\cdot[3p-5][3p-4]}{[2p-6]\2rp}},\ \ \ \ -\epsilon^2[p-2][p-1][p]\nnn\nnn
U_{14} = \frac{[p-3]}{[2p-4][2p-3]}
\sqrt{\frac{[4]\cdot[p-2][p-1]\rp\cdot[3p-6][3p-5][3p-4]}{[2p-6][2p-5][2p-2]\2rp}}, \nnn\nnn\nnn
U_{22} = \frac{[p-2]\Big([p-3]^2[p-2]-[2]^2[p-2]^2[3p-4]+[p-3][3p-4][3p-3]\Big)}
{[2p-5][2p-4][2p-2][2p-1]}, \nnn\nnn
U_{23} =
%???\frac{[p-3][p-2]\Big([p-2]-[3p-4]\Big)}{[2p-4][2p-2]}
-\frac{[p-3][p-2][p-1]^{3/2}\cdot[4p-6]}{[2p-4][2p-3][2p-2]}
\sqrt{\frac{[2][3]
%\cdot[p-1]
\cdot[3p-5]}{[2p-6][2p-5][2p-3][2p-1]}},\nnn\nnn
U_{24} = \frac{[p-3][p-2]}{[2p-5][2p-4]}\sqrt{\frac{[3][4]\cdot[p-1]\rp\cdot[3p-6][3p-5]}
{[2]\cdot[2p-6][2p-3][2p-2][2p-1]}},\nnn\nnn\nnn
U_{33} = \frac{[p-3][p-2][p-1]\Big([p-3]-[3][3p-5]\Big)}{[2p-6][2p-4][2p-3][2p-2]},\nnn\nnn
U_{34} = \frac{[p-3][p-2][p-1]}{[2p-6][2p-4][2p-3]}\sqrt{\frac{[4]\cdot\rp\cdot[3p-6]}{[2p-5][2p-2]}},
\nnn\nnn\nnn
U_{44} = \frac{[p-3][p-2][p-1]\rp}{[2p-6][2p-5][2p-4][2p-3]}\nn
\ee

\subsection{$SU_q(2)$ Racah matrices: $6\times 6$}

$\left(\prod_{a=0}^5 \xi_a\right)^{-1/6} = q^{-2p^2+6p-\frac{20}{3}}$,
\be
\tilde\xi_0 = q^{5p-\frac{20}{3}}\\
\tilde\xi_1 = -q^{3p-\frac{20}{3}}\\
\tilde\xi_2 = q^{p-\frac{14}{3}}\\
\tilde\xi_3 = -q^{-\left(p+\frac{2}{3}\right)}\\
\tilde\xi_4 = q^{-\left(3p-\frac{16}{3}\right)}\\
\tilde\xi_5 = -q^{-\left(5p-\frac{40}{3}\right)}
\ee

\be
U_{00} = \frac{[p-4][p-3][p-2][p-1]\rp}{[2p-4][2p-3][2p-2][2p-1]\2rp}, \nnn\nnn
U_{01} = \frac{[p-3][p-2][p-1]\rp}{[2p-4][2p-3][2p-2]\2rp}
\sqrt{\frac{[5][p-4]\cdot[3p-4]}{[2p-5][2p-1]}},\nnn \nnn
U_{02} = \frac{[p-2][p-1]\rp}{[2p-4][2p-2][2p-1]}
\sqrt{\frac{[4][5]\cdot[p-4][p-3]\cdot[3p-5][3p-4]}{[2]\cdot[2p-6][2p-5][2p-3]\2rp}},
\nnn \nnn
\ldots
\ee

%\subsection{$Sl_q(2)$ Racah matrices: ${\cal N}=7$}
%
%????
%
%
%$\left(\prod_{a=0}^6 \xi_a\right)^{-1/7} = q^{-2p^2+7p-10}$
%\be
%\tilde\xi_0 = q^{6p-10}\nn\\
%\tilde\xi_1 = -q^{4p-10}\nn\\
%\tilde\xi_2 = q^{2p-8}\nn\\
%\tilde\xi_3 = -q^{-4}\nn\\
%\tilde\xi_4 = q^{-(2p-2)}\nn\\
%\tilde\xi_5 = -q^{-(4p-10)}\nn\\
%\tilde\xi_6 = -q^{-(6p-20)}
%\ee

\subsection{$SU_q(2)$ Racah matrices: ${\cal N}\times {\cal N}$}

\be
U_{00} = \prod_{i=0}^{{\cal N}-2} \frac{[p-i]}{[2p-i]}, \nnn\nnn
U_{01} = \frac{\prod_{i=0}^{{\cal N}-3} [p-i]}{\2rp \prod_{i=2}^{{\cal N}-2}[2p-i]}
\sqrt{\frac{[{\cal N}-1]\cdot[p-({\cal N}-2)]\cdot[3p-{\cal N}-2)]}
{[2p-({\cal N}-1)][2p-1]}}
\nnn\nnn
U_{02} = %\frac{[p-1]\rp}{[2p-2][2p-1]}
\sqrt{\frac{[{\cal N}-2][{\cal N}-1]}{[2]}}
\left(\prod_{i=0}^{{\cal N}-4} [p-i] \right)
\sqrt{\frac{[p-({\cal N}-2)][p-({\cal N}-3)]\cdot[2p-3]
\cdot[3p-({\cal N}-1)][3p-({\cal N}-2)]}
{\prod_{i=0}^{{\cal N}}[2p-i] \prod_{i=1}^{{\cal N}-2}[2p-i] }}=\nnn
= \frac{\prod_{i=0}^{{\cal N}-4} [p-i]}{\prod_{i=1}^{{\cal N}-2}[2p-i]}
\sqrt{\frac{[{\cal N}-2][{\cal N}-1]\cdot [p-({\cal N}-2)][p-({\cal N}-3)]\cdot[2p-3]
\cdot[3p-({\cal N}-1)][3p-({\cal N}-2)]}{[2]\cdot[2p-{\cal N}][2p-({\cal N}-1)]\2rp}},
\nnn\nnn
\ldots
\ee

\section{ Summary:
Arbitrary 3-strand HOMFLY polynomial for arbitrary (anti)symmetric
representations}

\subsection{The answer
\label{ans}}

Now we are ready to propose a general answer for
colored HOMFLY polynomial for the case of the $3$-strand braid and
(anti)symmetric representation \ \ $R=S^r=[r]$ \ \
($R=\Lambda^r=[1^r]$):
\be
\boxed{ {\cal
H}_{[r]}^{(a_1b_1|a_2b_2|\ldots)} \{q|p_k\} = \sum_{Q\,\vdash\, 3r}
\left\{\Tr_Q \Big( {\cal R}_Q^{a_1}U_Q {\cal R}_Q^{b_1}U_Q^\dagger \
{\cal R}_Q^{a_2}U_Q {\cal R}_Q^{b_2}U_Q^\dagger\ \ldots
\Big)\right\} \, S_Q\{p_k\} } \label{ansH} \ee \be \boxed{ {\cal
H}_{[1^r]}^{(a_1b_1|a_2b_2|\ldots)} \{q|p_k\} = {\cal
H}_{[r]}^{(a_1b_1|a_2b_2|\ldots)} \{-q^{-1}|(-1)^kp_k\}}
\ee
\be
H_{[r]}(q|A) = {\cal H}_{[r]}^{(a_1b_1|a_2b_2|\ldots)} \{q|p_k=p_k^*\}, \\
H_{[1^r]}(q|A) = {\cal H}_{[1^r]}^{(a_1b_1|a_2b_2|\ldots)}
\{q|p_k=p_k^*\} = H_{[r]}(q^{-1}|A^{-1})
\ee
The sum in (\ref{ansH})
is over all the three-line Young diagrams $Q=[lmn]$ with $l\geq
m\geq n$, $l+m+n=3r$, and it remains to explain what are ${\cal
R}_{[lmn]}$ and $U_{[lmn]}$. Representations $\ Q=[lmn]\ $ enters
the decomposition of the representation product
\be
[r]^{\otimes 3} =
\Big([r]^{\otimes 2}\Big)\otimes [r]\ =\ \Big(\oplus_{j=0}^{\,r}\,
[2r-j,j]\Big)\otimes [r]\ = \sum_{\stackrel{l\geq m\geq
n}{l+m+n=3r}} \N_{[lmn]}\, [lmn]
\ee
with multiplicity $\N_{[lmn]}$.
For a given $Q$ only $j\in [j_Q,J_Q]$ contributes into the above sum (i.e.
$j_Q\leq j\leq J_Q$), and $\N_Q = J_Q-j_Q+1$. The matrix ${\cal R}_Q$ is the
$\N_Q\times\N_Q$ diagonal matrix with the entries (eigenvalues) $\
(-)^j\cdot q^{\varkappa_{[2r-j,j]}}$:
\be
\boxed{ {\cal R}_Q = {\rm
diag} \left\{ (-)^j \cdot q^{ 2r^2 - (2j+1)r+j(j-1) } \right\}, \
\ \ \ \ \ j_Q\leq j \leq J_Q }
\ee
The matrix $U_Q$ of the same size is
conjecturally the same as the $SU_q(2)$ Racah matrix $U[k+1|p]$
with $k,p$ made from $r, j_q, J_Q$:
\be
\boxed{ U_Q =
U\Big[\N_Q=J_Q-j_Q+1\ \Big|\ p= r - j_Q \Big] }
\ee
i.e.
$$
\boxed{
U_Q = U\left[\begin{array}{ccc|c} [r]&[r]&[r] & j \\ &Q&& j'
\end{array}\right]
= U\left[\begin{array}{ccc|c} [r-j_Q]&[r-j_Q]&[r-j_Q] & j \\
&[2r-J_Q-j_Q] && j'
\end{array}\right] = U\Big[\N_Q\Big|r-j_Q\Big]
}
$$
where the Racah matrix at the l.h.s. is for a
3-line $Q$, i.e. for the $SU_q(3)$ representation theory, while the Racah
matrix on the r.h.s. is for a $2$-line representation, for which
the $SU_q(2)$ theory is sufficient.

To complete the description of the answer (\ref{ansH}) it remains to
describe $U[k+1|p]$ and $j_Q,J_Q$.

\subsection{$SU_q(2)$ Racah matrices}
These matrices are well known \cite{sl2Racah} and widely used in the
physical literature \cite{R2}. The standard formula is
$$
U\left[\begin{array}{ccc|c} p_1&p_2&p_3 & j \\ &q&
&j'\end{array}\right] = \sqrt{[2j+1][2j'+1]}(-1)^{p_1+p_2-p_3-q-j}
\sqrt{{[p_2-p_1+j]![p_1-p_2+j]![p_1+p_2-j]!\over [p_1+p_2+j+1]!}}\times
$$
$$
\times\sqrt{{[q-p_3+j]![p_3-q+j]![p_3+q-j]!\over [p_3+q+j+1]!}}\times\sqrt{
{[q-p_1+j']![p_1-q+j']![p_1+q-j']!\over [p_1+q+j'+1]!}}\times
$$
$$
\times\sqrt{{[p_2-p_3+j']![p_3-p_2+j']![p_2+p_3-j']!\over [p_2+p_3+j'+1]!}}\times
$$
$$
\times
\sum_{k=0}^{p_1+p_2+p_3+q}(-1)^k{[k+1]!\over [k-p_1-p_2-j]![k-p_3-q-j]![k-p_1-q-j']!}\times
$$
$$
\times{1\over[k-p_2-p_3-j']![p_1+p_2+p_3+q-k]!
[p_1+p_3+j+j'-k]![p_2+q+j+j'-k]!}
$$
However, for $p_1=p_2=p_3=p$ this
formula is considerably simplified: as already stated in the previous section, the
sums actually turn into products. Such factorized expressions are
much more convenient, both for calculational purposes and
conceptually.

\subsection{Selection rules for the cubes of symmetric representations}

The diagram $[lmn] \in [2r-j,j]\otimes [r]$, provided \be l\geq 2r-j
\geq m \geq j\geq n \ee and, obviously, $0\leq j\leq 2r-j$. They
obviously follow from the picture, one should only remember that,
increasing the line number $a$ one should not get to the right of
the original line number $a-1$ i.e. should not cross the extended
vertical lines in the picture. In other words, no two added boxes
should be at the same column.

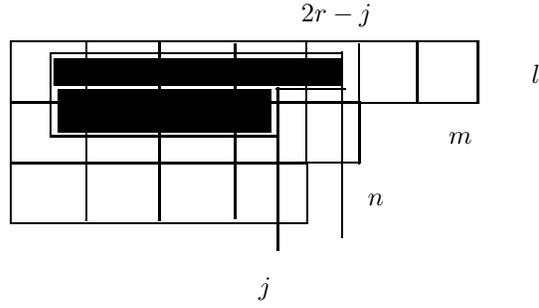
\unitlength 1mm % = 2.845pt
\linethickness{0.4pt}
\ifx\plotpoint\undefined\newsavebox{\plotpoint}\fi % GNUPLOT compatibility
\begin{picture}(127.5,162.25)(0,0)
\put(57.75,150.75){\framebox(62,8)[cc]{}}
\put(57.75,142.74){\framebox(46.25,8)[cc]{}}
\put(57.75,134.75){\framebox(39.25,7.75)[cc]{}}
\put(97,150.75){\line(0,-1){8}}
\put(111.75,159){\line(0,-1){8.25}}
\put(104,158.5){\line(0,-1){16}}
\put(97,158.75){\line(0,-1){7.75}}
\put(77.5,158.75){\line(0,-1){23.75}}
\put(87.5,158.5){\line(0,-1){23.25}}
\put(67.75,158.75){\line(0,-1){23.75}}
\put(63,154.5){\line(0,-1){8.25}}
\put(63,146.25){\line(1,0){30.25}}
\put(93,152.5){\line(1,0){9.25}}
\put(101.75,157.25){\line(-1,0){38.75}}
\put(63,157.25){\line(0,-1){3}}
\put(93.25,152.75){\line(0,-1){21.75}}
\put(101.75,157.5){\line(0,-1){24.75}}
\put(127.5,154.5){\makebox(0,0)[cc]{$l$}}
\put(117.5,146){\makebox(0,0)[cc]{$m$}}
\put(106.25,137.75){\makebox(0,0)[cc]{$n$}}
\put(101,162.25){\makebox(0,0)[cc]{$2r-j$}}
\put(91.5,126){\makebox(0,0)[cc]{$j$}}
\put(63.5,153){\rule{38\unitlength}{3.5\unitlength}}
\put(64,146.75){\rule{28.25\unitlength}{5.75\unitlength}}
\end{picture}

\vspace{-13cm}

\noindent
These five inequalities imply:
\be
j_{[lmn]} = {\rm
max}\Big(2r-l,n,0\Big) \ \leq j\ \leq\ {\rm min}(m,r,2r-m)=J_{[lmn]}
\ee

\bigskip

%\be
%\begin{array}{|c|cc|c| c | c|c|c|c|}
%r=2 &&&& \ \ \ \ \ \ \ \ \ \ \ \  & r=3 &&& \\
%&&&&&&&&\\
%Q & j_Q & J_Q & \N_Q && Q & j_Q & J_Q & \N_Q\\
%\hline
%6 & 0 & 0 & 1 &&  9 & 0 & 0 & 1 \\
%51 & 0 & 1 & 2 && 81 & 0 & 1 & 2\\
%42 & 0 & 2 & 3&& 72 & 0 & 2 & 3\\
%411 & 1 & 1 & 1 && 711 & 1 & 1 & 1\\
%33 & 1 & 1(2) & 1 && 63 & 0 & 3 & 4 \\
%321 & 1 & 2 & 2 && 621 & 1 & 2 & 2\\
%&&&&& 54 & 1 & 2(3) & 2 \\
%&&&&& 531 & 1 & 3 & 3\\
%&&&&& 522 & 2 & 2 & 1\\
%&&&&& 441 & 2 & 2(3) & 1 \\
%&&&&& 432 & 2 & 3 & 2\\
%&&&&& 333 & 3 & 3 & 1
%\end{array}
%\ee
$$
\begin{array}{|c|cc|c| c | c|c|c|c| c | c|c|c|c|}
\hline
r=2 &&&& \ \ \ \ \ \ \ \ \ \ \ \  & r=3 &&&& \ \ \ \ \ \ \ \ \ \ \ \  & r=4 &&& \\
&&&&&&&&&&&&&\\
Q & j_Q & J_Q & \N_Q && Q & j_Q & J_Q & \N_Q&& Q & j_Q & J_Q & \N_Q\\
&&&&&&&&&&&&&\\
\hline
&&&&&&&&&&&&&\\
\l[6] & 0 & 0 & 1 &&  [9] & 0 & 0 & 1 && [12,0] & 0&0&1\\
\l[51] & 0 & 1 & 2 && [81] & 0 & 1 & 2 && [11,1] & 0 & 1 & 2\\
\l[42] & 0 & 2 & 3&& [72] & 0 & 2 & 3 && [10,2] & 0&2&3\\
\l[411] & 1 & 1 & 1 && [711] & 1 & 1 & 1 && [10,1,1] & 1 & 1 & 1\\
\l[33] & 1 & 1  & 1 && [63] & 0 & 3 & 4 && [93] & 0 & 3 & 4\\
\l[321] & 1 & 2 & 2 && [621] & 1 & 2 & 2 && [921] & 1&2&2\\
&&&&& [54] & 1 & 2  & 2&& [84]&0&4&5 \\
&&&&& [531] & 1 & 3 & 3&& [831] & 1&3&3\\
&&&&& [522] & 2 & 2 & 1&& [822]&2&2&1\\
&&&&& & & & && [75] & 1 & 3  & 3 \\
&&&&& [441] & 2 & 2  & 1&&[741]&1&4&4\\
&&&&& [432] & 2 & 3 & 2 && [732] &2&3&2 \\
&&&&& & & & && [66]& 2 & 2  & 1 \\
&&&&& & & & && [651]& 2 & 3  & 2 \\
&&&&& & & & && [642]& 2 & 4 & 3 \\
&&&&& [333] & 3 & 3 & 1 && [633] & 3 & 3  & 1 \\
&&&&&&&&&& [552] & 3 & 3  & 1\\
&&&&&&&&&& [543] & 3 & 4 & 2\\
&&&&&&&&&& [444] & 4&4&1\\
\hline
\end{array}
$$

\section{Conclusion}

The main result of this paper is a direct algorithm to calculate explicit formulas for
the HOMFLY polynomials of 3-strand knots in symmetric and
antisymmetric representations with a list of illustrative examples.
In order to derive these formulas within Turaev-Reshetikhin formalism \cite{TR}
one needs the Racah coefficients for $SU_q(3)$,
which are not easily available in the literature.
Instead we made a conjecture that the Racah matrices
for the quantum groups of different ranks are directly related:
they depend only on the eigenvalues of the corresponding ${\cal R}$-matrices,
(which are powers of the symmetric group characters).
If this conjecture is true in its strong form,
one can evaluate the HOMFLY polynomials in arbitrary representations.
In this paper only the weak form of the conjecture is tested,
sufficient for all symmetric representations.
The antisymmetric representations are then immediately provided
by the duality symmetry (see, e.g., \cite{DMMSS}).
We checked that the answers following from this conjecture
reproduce the colored HOMFLY polynomials  from \cite{IMMM}
and the colored Jones polynomials from \cite{katlas}.
Testing the strong form of the conjecture and evaluation of the
HOMFLY polynomials in non-trivial multi-line and multi-column
representations remains an open task for the future work.

In our conjecture we used only the first non-trivial symmetric group character $\phi_{[2]}(T)$, since the
${\cal R}$-matrix eigenvalues are expressed through this. A role of higher symmetric group characters
$\phi_R(T)$ has to be clarified yet.

\section*{Acknowledgements}

Our work is partly supported by Ministry of Education and Science of
the Russian Federation under contract 8207, by NSh-3349.2012.2,
by RFBR grants 10-02-00509 (A.Mir.), 10-02-00499 (A.Mor.), 12-02-31078-young\_a (And.Mor.) and
by joint grants 11-02-90453-Ukr, 12-02-91000-ANF,
11-01-92612-Royal Society.
The research of H.I.
is supported in part by the Grant-in-Aid for Scientific Research
(23540316)
from the Ministry of Education, Science and Culture, Japan.
Support from JSPS/RFBR bilateral collaboration "Synthesis of integrabilities
 arising from gauge-string duality" (FY2010-2011: 12-02-92108-Yaf-a) is gratefully appreciated.

\newpage

\section*{Tables. Some answers for $R=[1],[2],[3],[4]$}

The tables list expressions for the {\it reduced} colored HOMFLY
polynomials
\be
h_R^{{\cal K}}(q|A) = \frac{H_R^{\cal K}\{q|p_k^*\}}{S_R^*}
\ee
for all the $3$-strand knots with up to 8 crossings from
the Rolfsen tables in \cite{katlas}.
We do not give here the answers for the antisymmetric representation, i.e. $[1,1],\ [1,1,1]$ and $[1,1,1,1]$,
because these results are given by the same formulas as for the corresponding symmetrical ones with the change of variables
$q\rightarrow -1/q$.

\be
\boxed{H_{[1^k]}(A|q)=H_{[k]}\left(A\Big|-\frac{1}{q}\right)}
\ee

From all these obtained expressions for the colored HOMFLY polynomials one immediately obtains

\begin{itemize}
\item the colored Jones polynomials by putting $A=q^2$
\item the special polynomials
\be
{\mathfrak{H}}_R^{\cal K}(A) = \lim_{q\rightarrow 1} \frac{H_R^{\cal
K}(q,A)}{S_R^*(q,A)}
\label{speHdef}
\ee
which celebrate the property \cite{DMMSS,Zhu}
\be
{\mathfrak{H}}^{\cal K}_R(A) = \Big({\mathfrak{H}}_{[1]}^{\cal
K}(A)\Big)^{|R|}
\label{spepro}
\ee
\end{itemize}

We do not list here the corresponding Jones and special polynomials to avoid
increasing the volume in vain.

Having the results for these first colored HOMFLY polynomials, it is natural to generate the polynomials
in arbitrary symmetric representation $S^r$ (and antisymmetric $\Lambda^r$) following the method
of \cite{IMMM}. These polynomials  for the twisted knots and recurrent relations between them
have been recently found in
\cite{IMMM,FGS1,twist,FGS2}. Since their list far does not cover all the 3-strand knots, we do not include
them in our tables.

\subsection*{\fbox{Knot $3_1$}\hspace{1cm} {\large $(a_1b_1|a_2b_2) = (-1,-1|-1,-1)$}\hspace{1cm} $K_1$-twist knot}

\begin{footnotesize}

$$
h_{[1]} = A^4\Big((q^{2}+q^{-2})A^{-2}-1\Big)
$$

$$
h_{[2]} = A^{8}q^{16}\Big((q^{-8}+q^{-12}+q^{-14}+q^{-20})A^{-4}+(-q^{-8}-q^{-10}-q^{-14}-q^{-16})A^{-2}+q^{-10}
\Big)
$$

$$
\begin{array}{c}
h_{[3]} =\ \ \
A^{6}\left(q^{18}+q^{14}+q^{12}+q^{10}+q^{6}+q^{4}+q^{2}+q^{-6}\right)+
\\
+A^{8}\left(-q^{20}-q^{18}-q^{16}-q^{14}-2q^{12}-2q^{10}-q^{8}-q^{4}-q^{2}-1\right)+
\\
+A^{10}\left(q^{20}+q^{18}+q^{16}+q^{12}+q^{10}+q^{8}\right)-A^{12}q^{18}
\end{array}
$$

$$
\begin{array}{c}
h_{[4]} =\ \ \ A^{8}\left(q^{32}+q^{28}+q^{26}+q^{24}+q^{22}+q^{20}+q^{18}+2q^{16}+q^{14}+q^{12}+q^{8}+q^{6}+q^{4}+q^{2}+q^{-8}\right)+
\\
+A^{10}\left(-q^{36}-q^{34}-q^{32}-2q^{30}-2q^{28}-3q^{26}-3q^{24}-2q^{22}-2q^{20}-2q^{18}-3q^{16}-3q^{14}-2q^{12}-q^{10}-q^{6}-q^{4}-q^{2}-1\right)+
\\
+A^{12}\left(q^{38}+q^{36}+2q^{34}+q^{32}+2q^{30}+2q^{28}+3q^{26}+3q^{24}+2q^{22}+q^{20}+q^{18}+q^{16}+2q^{14}+q^{12}+q^{10}\right)+
\\
+A^{14}\left(-q^{38}-q^{36}-q^{34}-q^{32}-q^{28}-q^{26}-q^{24}-q^{22}\right)+A^{16}q^{36}
\end{array}
$$

\end{footnotesize}

\subsection*{\fbox{Knot $4_1$}\hspace{1cm} {\large $(a_1b_1|a_2b_2) = (1,-1|1,-1)$}\hspace{1cm} $K_{-1}$-twist knot}

\begin{footnotesize}

$$
h_{[1]} = A^{-2}+(-q^{2}+1-q^{-2})+A^{2}
$$

$$
h_{[2]} = q^{-4}A^{-4}+(-q^{2}+q^{-2}-q^{-4}-q^{-6})A^{-2}+(q^{6}-q^{4}+3-q^{-4}+q^{-6})+(-q^{6}-q^{4}+q^{2}-q^{-2})A^{2}+q^{4}A^{4}
\Big)
$$

$$
\begin{array}{c}
h_{[3]} =\ \ \
A^{-6}q^{-12}+A^{-4}\left(-q^{-2}+q^{-8}-q^{-10}-q^{-12}-q^{-14}\right)+
\\
+A^{-2}\left(q^{6}-q^{2}+2q^{-2}+3q^{-4}-q^{-8}+q^{-12}+q^{-14}\right)+\left(-q^{12}+q^{10}-2q^{6}-3q^{4}+3-3q^{-4}-2q^{-6}+q^{-10}-q^{-12}\right)+
\\
+A^{2}\left(q^{14}+q^{12}-q^{8}+3q^{4}+2q^{2}-q^{-2}+q^{-6}\right)+A^{4}\left(-q^{14}-q^{12}-q^{10}+q^{8}-q^{2}\right)+A^{6}q^{12}
\end{array}
$$

$$
\begin{array}{c}
h_{[4]} =\ \ \ A^{-8}q^{-24}+A^{-6}\left(-q^{-10}+q^{-18}-q^{-20}-q^{-22}-q^{-24}-q^{-26}\right)+
\\
+A^{-4}\left(q^{2}-q^{-4}+2q^{-8}+2q^{-10}+3q^{-12}-q^{-16}+2q^{-22}+q^{-24}+q^{-26}\right)+
\\
+A^{-2}\left(-q^{12}+q^{8}-q^{4}-4q^{2}-3+2q^{-4}+3q^{-6}-2q^{-8}-4q^{-10}-4q^{-12}-2q^{-14}+q^{-16}-q^{-22}-q^{-24}\right)+
\\
+\left(q^{20}-q^{18}+2q^{14}+2q^{12}+3q^{10}-2q^{8}-3q^{6}+3q^{2}+9+3q^{-2}-3q^{-6}-2q^{-8}+3q^{-10}+2q^{-12}+2q^{-14}-q^{-18}+q^{-20}\right)+
\\
+A^{2}\left(-q^{24}-q^{22}+q^{16}-2q^{14}-4q^{12}-4q^{10}-2q^{8}+3q^{6}+2q^{4}-3-4q^{-2}-q^{-4}+q^{-8}-q^{-12}\right)+
\\
+A^{4}\left(q^{26}+q^{24}+2q^{22}-q^{16}+3q^{12}+2q^{10}+2q^{8}-q^{4}+q^{-2}\right)+
\\
+A^{6}\left(-q^{26}-q^{24}-q^{22}-q^{20}+q^{18}-q^{10}\right)+A^{8}q^{24}
\end{array}
$$

\end{footnotesize}

\subsection*{\fbox{Knot $5_2$}\hspace{1cm} {\large $(a_1b_1|a_2b_2) = (1,-1|1,3)$}\hspace{1cm} $K_2$-twist knot}

\begin{footnotesize}

$$
h_{[1]} = A^{-4}\Big((q^{2}-1+q^{-2})A^{-2}+(q^{2}-1+q^{-2})-A^{2}\Big)
$$

$$
\begin{array}{c}
h_{[2]} =\ \ \ q^{-16}A^{-8}\Big((q^{-8}-q^{-10}+2q^{-14}-q^{-16}-q^{-18}+q^{-20})A^{-4}+
\\
+(q^{-6}-q^{-8}-q^{-10}+3q^{-12}-2q^{-16}+q^{-18}+q^{-20})A^{-2}+
\\
+(q^{-4}-q^{-6}-q^{-8}+2q^{-10}-2q^{-14})+(-q^{-4}-q^{-6}+q^{-8}-q^{-12})A^{2}+q^{-6}A^{4}
\Big)
\Big)
\end{array}
$$

$$
\begin{array}{c}
h_{[3]} =\ \ \
-A^{-18}q^{-30}+A^{-16}\left(q^{-20}-q^{-26}+q^{-28}+q^{-30}+q^{-32}\right)+
\\
+A^{-14}\left(q^{-14}+2q^{-16}-q^{-20}-2q^{-22}+q^{-24}+q^{-26}-q^{-30}-q^{-32}\right)+
\\
+A^{-12}\left(-q^{-4}-q^{-6}-2q^{-8}+q^{-10}+2q^{-12}-3q^{-16}-3q^{-18}+q^{-20}+2q^{-22}-q^{-26}-q^{-28}+q^{-30}\right)+
\\
+A^{-10}\left(q^{2}-q^{-2}-3q^{-4}+q^{-6}+4q^{-8}+2q^{-10}-3q^{-12}-4q^{-14}+q^{-16}+3q^{-18}-q^{-22}-q^{-24}+q^{-26}\right)+
\\
+A^{-8}\left(q^{6}+q^{4}-3-q^{-2}+3q^{-4}+4q^{-6}-q^{-8}-4q^{-10}+3q^{-14}+q^{-16}-q^{-18}-q^{-20}+q^{-22}\right)+
\\
+A^{-6}\left(q^{6}-q^{4}-q^{2}+2q^{-2}+q^{-4}-2q^{-6}-q^{-8}+q^{-10}+q^{-12}-q^{-16}+q^{-18}\right)
\end{array}
$$

$$
\begin{array}{c}
h_{[4]} =\ \ \ A^{-24}q^{-60}+A^{-22}\left(-q^{-46}+q^{-54}-q^{-56}-q^{-58}-q^{-60}-q^{-62}\right)+
\\
+A^{-20}\left(-q^{-36}-q^{-38}-2q^{-40}+q^{-44}+q^{-46}+2q^{-48}-q^{-50}-q^{-52}+2q^{-58}+q^{-60}+q^{-62}\right)+
\\
+A^{-18}\big(q^{-22}+q^{-24}+2q^{-26}+2q^{-28}-q^{-30}-q^{-32}-2q^{-34}+q^{-36}+
\\
+4q^{-38}+3q^{-40}+3q^{-42}-2q^{-44}-2q^{-46}-q^{-48}+2q^{-52}-q^{-58}-q^{-60}\big)+
\\
+A^{-16}\big(-q^{-12}-q^{-14}+q^{-18}+3q^{-20}+3q^{-22}-2q^{-24}-4q^{-26}-6q^{-28}-q^{-30}+
\\
+4q^{-32}+4q^{-34}+3q^{-36}-4q^{-38}-4q^{-40}-2q^{-42}+3q^{-46}-q^{-52}-q^{-54}+q^{-56}\big)+
\\
+A^{-14}\big(-q^{-6}-3q^{-8}-2q^{-10}-q^{-12}+4q^{-14}+6q^{-16}-4q^{-20}-10q^{-22}-4q^{-24}+
\\
+4q^{-26}+6q^{-28}+6q^{-30}-4q^{-32}-5q^{-34}-3q^{-36}+4q^{-40}-q^{-46}-q^{-48}+q^{-50}\big)+
\\
+A^{-12}\big(q^{4}+q^{2}+1-2q^{-2}-3q^{-4}-3q^{-6}+2q^{-8}+8q^{-10}+5q^{-12}-9q^{-16}-8q^{-18}+q^{-20}+
\\
+6q^{-22}+9q^{-24}-q^{-26}-5q^{-28}-4q^{-30}-q^{-32}+4q^{-34}+q^{-36}-q^{-40}-q^{-42}+q^{-44}\big)+
\\
+A^{-10}\big(q^{8}+q^{6}-q^{2}-4-q^{-2}+3q^{-4}+5q^{-6}+5q^{-8}-4q^{-10}-6q^{-12}-4q^{-14}+2q^{-16}+
\\
+8q^{-18}+2q^{-20}-q^{-22}-4q^{-24}-2q^{-26}+3q^{-28}+q^{-30}+q^{-32}-q^{-34}-q^{-36}+q^{-38}\big)+
\\
+A^{-8}\big(q^{8}-q^{6}-q^{4}+3q^{-2}-q^{-6}-2q^{-8}-2q^{-10}+3q^{-12}+q^{-14}+q^{-16}-q^{-18}-2q^{-20}+q^{-22}+q^{-26}-q^{-30}+q^{-32}\big)
\end{array}
$$

\end{footnotesize}

\subsection*{\fbox{Knot $6_2$}\hspace{1cm} {\large $(a_1b_1|a_2b_2) = (1,-1|1,-3)$}}

\begin{footnotesize}

$$
h_{[1]} = A^2\Big((q^{2}+q^{-2})A^{-2}+(-q^{4}+q^{2}-2+q^{-2}-q^{-4})+(q^{2}-1+q^{-2})A^{2}
$$

$$
\begin{array}{c}
h_{[2]} =\ \ \ A^{4}q^8\Big((q^{-4}+q^{-8}+q^{-10}+q^{-16})A^{-4}+(-q^{2}-2q^{-4}-3q^{-10}-q^{-12}+q^{-14}-q^{-16}-q^{-18})A^{-2}+
\\
+(q^{6}-q^{4}+q^{2}+2-2q^{-2}+2q^{-4}+4q^{-6}-3q^{-8}+4q^{-12}-q^{-16}+q^{-18})+
\\
+(-q^{6}+q^{2}-3+3q^{-4}-3q^{-6}-2q^{-8}+2q^{-10}-q^{-14})A^{2}+(q^{4}-q^{2}+2q^{-2}-q^{-4}-q^{-6}+q^{-8})A^{4}
\Big)
\end{array}
$$

$$
\begin{array}{c}
h_{[3]} =\ \ \ \big(q^{6}+q^{2}+1+q^{-2}+q^{-6}+q^{-8}+q^{-10}+q^{-18}\big)+
\\
+A^{2}\big(-q^{16}-q^{12}-2q^{8}-2q^{4}-2q^{2}-4-q^{-2}-q^{-4}-q^{-6}-4q^{-8}-2q^{-10}-q^{-12}+q^{-14}-q^{-16}-q^{-18}-q^{-20}\big)+
\\
+A^{4}\big(q^{24}+q^{18}+q^{16}+q^{14}+q^{12}+4q^{10}+3q^{8}+5q^{2}+7+3q^{-2}-
\\
-2q^{-4}+4q^{-8}+5q^{-10}+q^{-12}-q^{-14}+q^{-18}+q^{-20}\big)+
\\
+A^{6}\big(-q^{30}+q^{28}-q^{26}-q^{24}-q^{22}-3q^{18}-3q^{16}+q^{14}+2q^{12}-6q^{10}-
\\
-8q^{8}-2q^{6}+7q^{4}-8-8q^{-2}+q^{-4}+4q^{-6}-4q^{-10}-2q^{-12}+q^{-16}-q^{-18}\big)+
\\
+A^{8}\big(q^{32}+q^{26}+2q^{24}-2q^{20}+3q^{18}+7q^{16}-6q^{12}-2q^{10}+9q^{8}+7q^{6}-3q^{4}-6q^{2}+5q^{-2}+3q^{-4}-q^{-6}-q^{-8}+q^{-12}\big)+
\\
+A^{10}\big(-q^{32}-3q^{24}+3q^{20}+q^{18}-5q^{16}-4q^{14}+3q^{12}+5q^{10}-q^{8}-4q^{6}-2q^{4}+2q^{2}+1-q^{-4}\big)+
\\
+A^{12}\big(q^{30}-q^{28}+q^{24}+q^{22}-q^{20}-2q^{18}+q^{16}+2q^{14}-q^{10}-q^{8}+q^{6}\big)
\end{array}
$$

$$\nonumber]
\begin{array}{c}
h_{[4]} =\ \ \ \big(q^{8}+q^{4}+q^{2}+1+q^{-2}+q^{-4}+q^{-6}+2q^{-8}+q^{-10}+q^{-12}+q^{-16}+q^{-18}+q^{-20}+q^{-22}+q^{-32}\big)+
\\
+A^{2}\big(-q^{22}-q^{18}-q^{16}-2q^{12}-q^{10}-2q^{8}-4q^{6}-3q^{4}-4q^{2}-3-3q^{-2}-5q^{-4}-5q^{-6}-6q^{-8}-
\\
-3q^{-10}-2q^{-12}-2q^{-14}-2q^{-16}-5q^{-18}-3q^{-20}-2q^{-22}-q^{-24}+q^{-26}-q^{-28}-q^{-30}-q^{-32}-q^{-34}\big)+
\\
+A^{4}\big(q^{34}+q^{30}+q^{26}+q^{24}+q^{22}+4q^{20}+3q^{18}+4q^{16}+4q^{14}+q^{12}+6q^{10}+7q^{8}+10q^{6}+9q^{4}+4q^{2}+
\\
+4+5q^{-2}+10q^{-4}+13q^{-6}+8q^{-8}+6q^{-10}-q^{-12}+2q^{-14}+5q^{-16}+6q^{-18}+7q^{-20}+2q^{-22}+2q^{-30}+q^{-32}+q^{-34}\big)+
\\
+A^{6}\big(-q^{44}-q^{38}-2q^{34}-2q^{32}-3q^{30}-5q^{28}-q^{26}-2q^{24}-2q^{22}-7q^{20}-13q^{18}-8q^{16}-4q^{14}+2q^{12}-3q^{10}-16q^{8}-18q^{6}-16q^{4}-
\\
-2q^{2}+5-2q^{-2}-9q^{-4}-19q^{-6}-13q^{-8}-4q^{-10}+2q^{-12}+4q^{-14}-4q^{-16}-6q^{-18}-6q^{-20}-3q^{-22}+q^{-24}-q^{-30}-q^{-32}\big)+
\\
+A^{8}\big(q^{52}-q^{50}+q^{48}+q^{46}+3q^{42}+q^{40}+3q^{38}+2q^{36}-2q^{34}+5q^{32}+6q^{30}+12q^{28}+6q^{26}-
\\
\end{array}
$$

\end{footnotesize}

\subsection*{\fbox{Knot $6_3$}\hspace{1cm} {\large $(a_1b_1|a_2b_2) = (2,-1|1,-2)$}}

\begin{footnotesize}

$$
h_{[1]} =\ \ \ (-q^{2}+1-q^{-2})A^{-2}+(q^{4}-q^{2}+3-q^{-2}+q^{-4})+(-q^{2}+1-q^{-2})A^{2}
$$

$$
\begin{array}{c}
h_{[2]} =\ \ \ q^{2}-1+2q^{-4}-q^{-6}-q^{-8}+q^{-10})A^{-4}+(-q^{8}+q^{4}-4q^{2}-1+4q^{-2}-3q^{-4}-3q^{-6}+2q^{-8}-q^{-12})A^{-2}+
\\
+(q^{12}-q^{10}+q^{8}+4q^{6}-3q^{4}-q^{2}+9-q^{-2}-3q^{-4}+4q^{-6}+q^{-8}-q^{-10}+q^{-12})+
\\
+(-q^{12}+2q^{8}-3q^{6}-3q^{4}+4q^{2}-1-4q^{-2}+q^{-4}-q^{-8})A^{2}+(q^{10}-q^{8}-q^{6}+2q^{4}-1+q^{-2})A^{4}
\Big)
\end{array}
$$

$$
\begin{array}{c}
h_{[3]} =\ \ \
A^{-6}\big(-1+q^{-2}-q^{-6}-q^{-8}+q^{-10}+2q^{-12}-q^{-14}-2q^{-16}+q^{-20}+q^{-22}-q^{-24}\big)+
\\
+A^{-4}\big(q^{10}+4q^{2}+1-3q^{-2}-2q^{-4}+5q^{-6}+6q^{-8}-2q^{-10}-6q^{-12}+4q^{-16}+3q^{-18}-2q^{-20}-q^{-22}+q^{-26}\big)+
\\
+A^{-2}\big(-q^{18}-q^{12}-4q^{10}-2q^{8}+4q^{6}-11q^{2}-9+4q^{-2}+11q^{-4}-5q^{-6}-
\\
-13q^{-8}-5q^{-10}+6q^{-12}+4q^{-14}-4q^{-16}-4q^{-18}+q^{-22}-q^{-26}\big)+
\\
+\big(q^{24}-q^{22}+q^{20}+2q^{18}+2q^{16}-3q^{14}+9q^{10}+10q^{8}-6q^{6}-9q^{4}+5q^{2}+23+5q^{-2}-
\\
-9q^{-4}-6q^{-6}+10q^{-8}+9q^{-10}-3q^{-14}+2q^{-16}+2q^{-18}+q^{-20}-q^{-22}+q^{-24}\big)+
\\
+A^{2}\big(-q^{26}+q^{22}-4q^{18}-4q^{16}+4q^{14}+6q^{12}-5q^{10}-13q^{8}-5q^{6}+11q^{4}+
\\
+4q^{2}-9-11q^{-2}+4q^{-6}-2q^{-8}-4q^{-10}-q^{-12}-q^{-18}\big)+
\\
+A^{4}\big(q^{26}-q^{22}-2q^{20}+3q^{18}+4q^{16}-6q^{12}-2q^{10}+6q^{8}+5q^{6}-2q^{4}-3q^{2}+1+4q^{-2}+q^{-10}\big)+
\\
+A^{6}\big(-q^{24}+q^{22}+q^{20}-2q^{16}-q^{14}+2q^{12}+q^{10}-q^{8}-q^{6}+q^{2}-1\big)
\end{array}
$$

$$\nonumber
\begin{array}{c}
h_{[4]} =\ \ \ A^{-8}\big(q^{-4}-q^{-6}+q^{-10}+q^{-14}-2q^{-16}-q^{-18}+q^{-20}+q^{-22}+
%\\
%+
3q^{-24}-2q^{-26}-2q^{-28}-q^{-30}+\\
+3q^{-34}-q^{-40}-q^{-42}+q^{-44}\big)%+
%\\
+A^{-6}\big(-q^{10}-q^{4}-4+3q^{-4}-9q^{-10}-4q^{-12}+4q^{-14}+6q^{-16}+8q^{-18}-7q^{-20}-9q^{-22}-
\\
-5q^{-24}+q^{-26}+10q^{-28}+2q^{-30}-q^{-32}-4q^{-34}-4q^{-36}+2q^{-38}+q^{-40}+q^{-42}-q^{-46}\big)+
\\
+A^{-4}\big(q^{22}+q^{18}+2q^{14}+4q^{12}+q^{10}+q^{8}-3q^{6}+4q^{4}+15q^{2}+8+3q^{-2}-
\\
-16q^{-4}-9q^{-6}+12q^{-8}+21q^{-10}+23q^{-12}-8q^{-14}-20q^{-16}-12q^{-18}+3q^{-20}+
\\
+23q^{-22}+10q^{-24}-9q^{-28}-8q^{-30}+4q^{-32}+4q^{-34}+4q^{-36}-2q^{-40}+q^{-46}\big)+
\\
+A^{-2}\big(-q^{32}-2q^{26}-q^{24}-4q^{22}-q^{20}+2q^{18}-4q^{16}-8q^{14}-18q^{12}-7q^{10}+11q^{8}+
\\
+10q^{6}-2q^{4}-39q^{2}-37-7q^{-2}+22q^{-4}+36q^{-6}-10q^{-8}-38q^{-10}-38q^{-12}-12q^{-14}+25q^{-16}+16q^{-18}-
\\
\end{array}
$$

\end{footnotesize}

\subsection*{\fbox{Knot $7_3$} \hspace{1cm}{\large $(a_1b_1|a_2b_2) = (1,-1|1,5)$}}

\begin{footnotesize}

$$
h_{[1]} =\ \ \ A^{-6}\Big((-q^{2}-q^{-2})A^{-2}+(q^{4}-q^{2}+2-q^{-2}+q^{-4})+(q^{4}-q^{2}+1-q^{-2}+q^{-4})A^{2}\Big)
$$

$$
\begin{array}{c}
h_{[2]} =\ \ \ A^{-12}q^{-24}\Big((q^{18}+q^{12}+q^{10}+q^{6})A^{-4}+
\\
+(-q^{24}+q^{20}-2q^{18}-2q^{16}+q^{14}-q^{12}-2q^{10}-q^{6}-q^{4})A^{-2}+
\\
+(-2q^{26}+3q^{22}-2q^{20}-3q^{18}+4q^{16}-2q^{12}+2q^{10}-q^{6}+q^{4})+
\\
+(q^{32}+q^{30}-2q^{28}+4q^{24}-q^{22}-3q^{20}+3q^{18}+q^{16}-2q^{14}+2q^{12}-q^{8}+q^{6})A^{2}+
\\
+(q^{32}-q^{30}-q^{28}+2q^{26}-2q^{22}+q^{20}+q^{18}-q^{16}+q^{14}-q^{10}+q^{8})A^{4}
\Big)
\end{array}
$$

$$
\begin{array}{c}
h_{[3]} =\ \ \ A^{-24}\big(-q^{-24}-q^{-32}-q^{-34}-q^{-36}-q^{-40}-q^{-42}-q^{-44}-q^{-48}\big)+
\\
+A^{-22}\big(q^{-14}-q^{-20}+2q^{-22}+2q^{-24}+2q^{-26}-q^{-28}+q^{-30}+2q^{-32}+
\\
+4q^{-34}+q^{-36}+2q^{-38}+q^{-40}+3q^{-42}+q^{-44}+2q^{-46}+q^{-48}+q^{-50}\big)+
\\
+A^{-20}\big(q^{-8}+2q^{-10}-2q^{-14}-2q^{-16}+3q^{-18}+4q^{-20}-5q^{-24}-2q^{-26}+2q^{-28}+
\\
+2q^{-30}-3q^{-32}-3q^{-34}-2q^{-36}-2q^{-40}-q^{-42}-q^{-44}-q^{-46}-q^{-48}-q^{-50}\big)+
\\
+A^{-18}\big(-q^{2}-1-2q^{-2}+q^{-4}+2q^{-6}-6q^{-10}-6q^{-12}+q^{-14}+7q^{-16}-8q^{-20}-
\\
-7q^{-22}+5q^{-24}+4q^{-26}-2q^{-28}-5q^{-30}+2q^{-34}-q^{-38}+q^{-40}-q^{-46}+q^{-48}\big)+
\\
+A^{-16}\big(q^{8}-q^{4}-3q^{2}+1+5q^{-2}+4q^{-4}-4q^{-6}-7q^{-8}+10q^{-12}+5q^{-14}-5q^{-16}-8q^{-18}+
\\
+3q^{-20}+7q^{-22}+q^{-24}-5q^{-26}-q^{-28}+2q^{-30}+q^{-32}-q^{-34}+q^{-36}-q^{-42}+q^{-44}\big)+
\\
+A^{-14}\big(q^{12}+q^{10}-3q^{6}-q^{4}+3q^{2}+5-q^{-2}-6q^{-4}-3q^{-6}+5q^{-8}+6q^{-10}-q^{-12}-
\\
-6q^{-14}-q^{-16}+5q^{-18}+3q^{-20}-2q^{-22}-2q^{-24}+q^{-26}+q^{-28}+q^{-32}-q^{-38}+q^{-40}\big)+
\\
+A^{-12}\big(q^{12}-q^{10}-q^{8}+2q^{4}+q^{2}-2-2q^{-2}+q^{-4}+2q^{-6}+q^{-8}-
\\
-2q^{-10}-q^{-12}+q^{-14}+q^{-16}-q^{-20}+q^{-28}-q^{-34}+q^{-36}\big)
\end{array}
$$

$$
\begin{array}{c}
h_{[4]} =\ \ \ A^{-32}\big(q^{-52}+q^{-62}+q^{-64}+q^{-66}+q^{-68}+
\\
+q^{-72}+q^{-74}+2q^{-76}+q^{-78}+q^{-80}+q^{-82}+q^{-84}+q^{-86}+q^{-88}+q^{-92}\big)+
\\
+A^{-30}\big(-q^{-38}+q^{-46}-2q^{-48}-2q^{-50}-2q^{-52}-2q^{-54}+q^{-56}-q^{-58}-2q^{-60}-4q^{-62}-5q^{-64}-3q^{-66}-3q^{-68}-
\\
-2q^{-70}-4q^{-72}-5q^{-74}-4q^{-76}-5q^{-78}-3q^{-80}-3q^{-82}-4q^{-84}-2q^{-86}-3q^{-88}-2q^{-90}-q^{-92}-q^{-94}\big)+
\\
+A^{-28}\big(-q^{-28}-q^{-30}-2q^{-32}+2q^{-36}+q^{-38}+q^{-40}-4q^{-42}-5q^{-44}-q^{-46}+q^{-48}+
\\
+7q^{-50}+2q^{-52}-q^{-54}-3q^{-56}-2q^{-58}+5q^{-60}+6q^{-62}+6q^{-64}+4q^{-66}+q^{-68}+5q^{-70}+
\\
+4q^{-72}+7q^{-74}+6q^{-76}+4q^{-78}+5q^{-80}+3q^{-82}+3q^{-84}+4q^{-86}+2q^{-88}+3q^{-90}+q^{-92}+q^{-94}\big)+
\\
+A^{-26}\big(q^{-14}+q^{-16}+2q^{-18}+2q^{-20}-q^{-22}-q^{-24}-2q^{-26}+3q^{-28}+9q^{-30}+8q^{-32}+6q^{-34}-
\\
\end{array}
$$

\end{footnotesize}

\subsection*{\fbox{Knot $7_5$} \hspace{1cm}{\large $(a_1b_1|a_2b_2) = (-2,1|-1,-4)$}}

\begin{footnotesize}

$$
h_{[1]} =\ \ \ A^6\Big((q^{4}-q^{2}+2-q^{-2}+q^{-4})A^{-2}+(q^{4}-2q^{2}+2-2q^{-2}+q^{-4})+(-q^{2}+1-q^{-2})A^{2}
\Big)
$$

$$
\begin{array}{c}
h_{[2]} =\ \ \ A^{12}q^{24}\Big((q^{-8}-q^{-10}+q^{-12}+2q^{-14}-2q^{-16}+q^{-18}
+3q^{-20}-3q^{-22}+3q^{-26}-q^{-28}-q^{-30}+q^{-32})A^{-4}+
\\+(q^{-6}-2q^{-8}+3q^{-12}-5q^{-14}+6q^{-18}-5q^{-20}-3q^{-22}
+6q^{-24}-3q^{-28}+q^{-30}+q^{-32})A^{-2}+
\\+(q^{-4}-2q^{-6}+4q^{-10}-5q^{-12}-q^{-14}+7q^{-16}-5q^{-18}-5q^{-20}+5q^{-22}-3q^{-26})+
\\+(-q^{-4}+2q^{-8}-3q^{-10}-q^{-12}+5q^{-14}-2q^{-16}-3q^{-18}+3q^{-20}+q^{-22}-q^{-24})A^{2}+
\\+(q^{-6}-q^{-8}+2q^{-12}-q^{-14}-q^{-16}+q^{-18})A^{4}
\Big)
\end{array}
$$

$$
\begin{array}{c}
h_{[3]} =\ \ \
A^{12}\big(q^{36}-q^{34}+q^{32}+q^{30}+q^{28}-q^{26}+q^{24}+2q^{22}-2q^{18}+2q^{16}+
\\
+3q^{14}+q^{12}-5q^{10}+4q^{6}+3q^{4}-3q^{2}-3+q^{-2}+3q^{-4}-q^{-8}-q^{-10}+q^{-12}\big)+
\\
+A^{14}\big(q^{40}-2q^{38}+q^{34}-4q^{30}+4q^{26}-q^{24}-9q^{22}+q^{20}+10q^{18}+4q^{16}-
\\
-13q^{14}-7q^{12}+9q^{10}+13q^{8}-3q^{6}-11q^{4}-3q^{2}+8+5q^{-2}-q^{-4}-4q^{-6}+q^{-10}+q^{-12}\big)+
\\
+A^{16}\big(q^{44}-2q^{42}+2q^{38}+q^{36}-4q^{34}-q^{32}+8q^{30}+q^{28}-12q^{26}-5q^{24}+14q^{22}+
\\
+11q^{20}-15q^{18}-19q^{16}+5q^{14}+19q^{12}+3q^{10}-16q^{8}-11q^{6}+5q^{4}+8q^{2}+1-5q^{-2}-2q^{-4}+q^{-8}\big)+
\\
+A^{18}\big(q^{48}-2q^{46}+2q^{42}+q^{40}-4q^{38}-3q^{36}+8q^{34}+3q^{32}-11q^{30}-9q^{28}+12q^{26}+
\\
+16q^{24}-8q^{22}-20q^{20}+18q^{16}+10q^{14}-10q^{12}-10q^{10}+q^{8}+6q^{6}+3q^{4}-2q^{2}-1-q^{-2}\big)+
\\
+A^{20}\big(-q^{50}+q^{46}-2q^{42}-2q^{40}+6q^{38}+3q^{36}-6q^{34}-7q^{32}+7q^{30}+
\\
+13q^{28}-2q^{26}-14q^{24}-3q^{22}+10q^{20}+9q^{18}-4q^{16}-6q^{14}-q^{12}+3q^{10}+2q^{8}\big)+
\\
+A^{22}\big(q^{50}-q^{44}+3q^{42}+q^{40}-3q^{38}-4q^{36}+3q^{34}+6q^{32}-
\\
-q^{30}-7q^{28}-2q^{26}+4q^{24}+4q^{22}-2q^{20}-2q^{18}-q^{16}+q^{14}\big)+
\\
+A^{24}\big(-q^{48}+q^{46}-q^{42}-q^{40}+q^{38}+2q^{36}-q^{34}-2q^{32}+q^{28}+q^{26}-q^{24}\big)
\end{array}
$$

$$
\begin{array}{c}
h_{[4]} =\ \ \ A^{16}\big(q^{64}-q^{62}+q^{60}+q^{58}+2q^{54}-q^{52}+q^{50}+2q^{48}-q^{46}+3q^{44}-q^{42}+
\\
+3q^{40}+2q^{38}-3q^{36}+q^{34}-2q^{32}+5q^{30}+5q^{28}-3q^{26}-q^{24}-7q^{22}+3q^{20}+8q^{18}+2q^{16}+
\\
+q^{14}-9q^{12}-3q^{10}+3q^{8}+4q^{6}+6q^{4}-4q^{2}-3-2q^{-2}+4q^{-6}-q^{-12}-q^{-14}+q^{-16}\big)+
\\
+A^{18}\big(q^{70}-2q^{68}+q^{64}-2q^{62}+q^{60}-4q^{58}+2q^{54}-6q^{52}-q^{50}-4q^{48}+
\\
+6q^{46}+8q^{44}-9q^{42}-9q^{40}-12q^{38}+12q^{36}+24q^{34}+2q^{32}-12q^{30}-30q^{28}-
\\
-4q^{26}+27q^{24}+24q^{22}+11q^{20}-26q^{18}-26q^{16}-3q^{14}+16q^{12}+28q^{10}+4q^{8}-
\\
-11q^{6}-15q^{4}-8q^{2}+9+8q^{-2}+5q^{-4}-q^{-6}-5q^{-8}-q^{-10}+q^{-14}+q^{-16}\big)+
\\
+A^{20}\big(q^{76}-2q^{74}+2q^{70}-q^{68}+2q^{66}-4q^{64}+3q^{62}+6q^{60}-4q^{58}-3q^{56}-
\\
\end{array}
$$

\end{footnotesize}

\subsection*{\fbox{Knot $8_2$} \hspace{1cm}{\large $(a_1b_1|a_2b_2) = (1,-1|1,-5)$}}

\begin{footnotesize}

$$
h_{[1]} =\ \ \ A^4\Big((q^{4}+1+q^{-4})A^{-2}+(-q^{6}+q^{4}-2q^{2}+1-2q^{-2}+q^{-4}-q^{-6})+(q^{4}-q^{2}+1-q^{-2}+q^{-4})A^{2}\Big)
$$

$$
\begin{array}{c}
h_{[2]} =\ \ \ A^8q^{16}\Big((q^{-4}+q^{-8}+q^{-10}+q^{-12}+q^{-14}+q^{-16}+q^{-20}+q^{-22}+q^{-28})A^{-4}+
\\
+(-q^{2}-2q^{-4}-q^{-6}-q^{-8}-2q^{-10}-q^{-12}-2q^{-14}-3q^{-16}-3q^{-22}-q^{-24}+q^{-26}-q^{-28}-q^{-30})A^{-2}+
\\
+(q^{6}-q^{4}+q^{2}+2-q^{-2}+q^{-4}+2q^{-6}+3q^{-10}+q^{-12}-2q^{-14}+3q^{-16}+4q^{-18}-3q^{-20}+4q^{-24}-q^{-28}+q^{-30})+
\\
+(-q^{6}+q^{2}-2-2q^{-6}+2q^{-8}-4q^{-12}+q^{-14}+3q^{-16}-3q^{-18}-2q^{-20}+2q^{-22}-q^{-26})A^{2}+
\\
+(q^{4}-q^{2}+q^{-2}-q^{-4}+q^{-6}+q^{-8}-2q^{-10}+2q^{-14}-q^{-16}-q^{-18}+q^{-20})A^{4}
\Big)
\end{array}
$$

$$\nonumber
\begin{array}{c}
h_{[3]}
=A^{6}\big(q^{24}+q^{20}+q^{18}+2q^{16}+q^{14}+2q^{12}+q^{10}+2q^{8}+2q^{6}+2q^{4}+q^{2}+2+q^{-2}+2q^{-4}+q^{-6}+q^{-8}+q^{-12}+
\\
+q^{-14}+q^{-16}+q^{-24}\big)+A^{8}\big(-q^{34}-q^{30}-3q^{26}-q^{24}-3q^{22}-q^{20}-5q^{18}-4q^{16}-6q^{14}-3q^{12}-5q^{10}-4q^{8}-
\\
-7q^{6}-4q^{4}-5q^{2}-2-5q^{-2}-4q^{-4}-5q^{-6}-q^{-8}-q^{-10}-q^{-12}-4q^{-14}-2q^{-16}-q^{-18}+q^{-20}-q^{-22}-q^{-24}-q^{-26}\big)+
\\
+A^{10}\big(q^{42}+q^{36}+2q^{34}+2q^{32}+2q^{28}+4q^{26}+6q^{24}+3q^{22}+2q^{20}+4q^{18}+8q^{16}+8q^{14}+5q^{12}+3q^{10}+6q^{8}+
\\
+8q^{6}+9q^{4}+4q^{2}+1+2q^{-2}+7q^{-4}+8q^{-6}+3q^{-8}-2q^{-10}+4q^{-14}+5q^{-16}+q^{-18}-q^{-20}+q^{-24}+q^{-26}\big)+
\\
+A^{12}\big(-q^{48}+q^{46}-q^{44}-q^{42}-2q^{40}+q^{38}-q^{36}-3q^{34}-5q^{32}-q^{30}+q^{28}-q^{26}-9q^{24}-6q^{22}-2q^{20}+3q^{18}-6q^{16}-
\\
+9q^{14}-8q^{12}+q^{10}+2q^{8}-3q^{6}-11q^{4}-8q^{2}-1+7q^{-2}-q^{-4}-8q^{-6}-8q^{-8}+q^{-10}+4q^{-12}-4q^{-16}-2q^{-18}+q^{-22}-q^{-24}\big)+
\\
+A^{14}\big(q^{50}+2q^{42}+2q^{40}-q^{36}+q^{34}+4q^{32}+5q^{30}-q^{28}-3q^{26}+9q^{22}+5q^{20}-3q^{18}-6q^{16}+
\\
+3q^{14}+10q^{12}+6q^{10}-5q^{8}-7q^{6}+10q^{2}+6-3q^{-2}-6q^{-4}+5q^{-8}+3q^{-10}-q^{-12}-q^{-14}+q^{-18}\big)+
\\
+A^{16}\big(-q^{50}+q^{44}-2q^{42}-q^{40}-q^{38}+q^{36}+q^{34}-3q^{30}-3q^{28}+2q^{26}+4q^{24}-q^{22}-6q^{20}-
\\
-3q^{18}+5q^{16}+5q^{14}-q^{12}-7q^{10}-3q^{8}+4q^{6}+5q^{4}-q^{2}-4-2q^{-2}+2q^{-4}+q^{-6}-q^{-10}\big)+
\\
+A^{18}\big(q^{48}-q^{46}+q^{40}-q^{32}+q^{28}+q^{26}-q^{24}-2q^{22}+q^{20}+2q^{18}+q^{16}-2q^{14}-2q^{12}+q^{10}+2q^{8}-q^{4}-q^{2}+1\big)
\end{array}
$$

$$
\begin{array}{c}
h_{[4]}
=A^{8}\big(q^{40}+q^{36}+q^{34}+2q^{32}+2q^{30}+2q^{28}+2q^{26}+3q^{24}+3q^{22}+4q^{20}+3q^{18}+
\\
+4q^{16}+3q^{14}+4q^{12}+4q^{10}+4q^{8}+3q^{6}+3q^{4}+3q^{2}+4+3q^{-2}+3q^{-4}+2q^{-6}+2q^{-8}+
\\
+2q^{-10}+2q^{-12}+2q^{-14}+2q^{-16}+q^{-18}+q^{-20}+q^{-24}+q^{-26}+q^{-28}+q^{-30}+q^{-40}\big)+
\\
+A^{10}\big(-q^{54}-q^{50}-q^{48}-q^{46}-3q^{44}-2q^{42}-3q^{40}-4q^{38}-5q^{36}-8q^{34}-8q^{32}-9q^{30}-
\\
-9q^{28}-10q^{26}-13q^{24}-13q^{22}-14q^{20}-13q^{18}-13q^{16}-15q^{14}-15q^{12}-15q^{10}-13q^{8}-11q^{6}-
\\
-12q^{4}-12q^{2}-13-12q^{-2}-9q^{-4}-8q^{-6}-6q^{-8}-7q^{-10}-8q^{-12}-7q^{-14}-7q^{-16}-3q^{-18}-2q^{-20}-
\\
-2q^{-22}-2q^{-24}-5q^{-26}-3q^{-28}-2q^{-30}-q^{-32}+q^{-34}-q^{-36}-q^{-38}-q^{-40}-q^{-42}\big)+
\\
+A^{12}\big(q^{66}+q^{62}+2q^{58}+2q^{56}+2q^{54}+3q^{52}+2q^{50}+6q^{48}+8q^{46}+7q^{44}+9q^{42}+6q^{40}+
\\
\end{array}
$$

\end{footnotesize}

\subsection*{\fbox{Knot $8_5$} \hspace{1cm}{\large $(a_1b_1|a_2b_2) = (-1,3|-1,3)$}}

\begin{footnotesize}

$$
h_{[1]} =\ \ \ A^{-4}\Big((q^{4}-q^{2}+2-q^{-2}+q^{-4})A^{-2}+(-q^{6}+q^{4}-3q^{2}+1-3q^{-2}+q^{-4}-q^{-6})+(q^{4}+2+q^{-4})A^{2}
\Big)
$$

$$
\begin{array}{c}
h_{[2]} =\ \ \ A^{-8}q^{-16}\Big((q^{20}-q^{18}-q^{16}+3q^{14}+q^{12}-3q^{10}+2q^{8}+2q^{6}-2q^{4}+q^{2}+1-q^{-2}+q^{-4})A^{-4}+
\\
+(-q^{26}+2q^{22}-3q^{20}-5q^{18}+3q^{16}-8q^{12}-q^{10}+2q^{8}-5q^{6}-q^{4}+q^{2}-3-q^{-6})A^{-2}+
\\
+(q^{30}-q^{28}+5q^{24}+q^{22}-3q^{20}+8q^{18}+7q^{16}-q^{14}+5q^{12}+8q^{10}+4q^{6}+4q^{4}-q^{2}+2+2q^{-2}-q^{-4}+q^{-6})+
\\
+(-q^{30}-q^{28}+q^{26}-2q^{24}-5q^{22}-q^{20}-2q^{18}-7q^{16}-4q^{14}-3q^{12}-6q^{10}-3q^{8}-q^{6}-3q^{4}-q^{2}-q^{-2})A^{2}+
\\
+(q^{28}+2q^{22}+2q^{20}+3q^{16}+2q^{14}+q^{12}+2q^{10}+2q^{8}+q^{4})A^{4}
\Big)
\end{array}
$$

$$
\begin{array}{c}
h_{[3]}
=A^{-18}\big(1-q^{-2}-q^{-4}+3q^{-8}+2q^{-10}-2q^{-12}-4q^{-14}+q^{-16}+5q^{-18}+4q^{-20}-
\\
-4q^{-22}-3q^{-24}+q^{-26}+5q^{-28}-2q^{-32}+2q^{-36}+q^{-44}-q^{-46}+q^{-48}\big)+
\\
+A^{-16}\big(-q^{10}+q^{6}+2q^{4}-3q^{2}-6-3q^{-2}+6q^{-4}+5q^{-6}-6q^{-8}-16q^{-10}-6q^{-12}+6q^{-14}+8q^{-16}-10q^{-18}-
\\
-16q^{-20}-6q^{-22}+8q^{-24}+q^{-26}-8q^{-28}-8q^{-30}+q^{-32}-2q^{-36}-3q^{-38}-2q^{-42}-q^{-46}-q^{-50}\big)+
\\
+A^{-14}\big(q^{18}-q^{14}-q^{12}+4q^{10}+7q^{8}+2q^{6}-6q^{4}-2q^{2}+13+22q^{-2}+8q^{-4}-5q^{-6}-q^{-8}+
\\
+23q^{-10}+30q^{-12}+15q^{-14}-4q^{-16}+5q^{-18}+21q^{-20}+23q^{-22}+4q^{-24}+q^{-26}+6q^{-28}+
\\
+12q^{-30}+5q^{-32}+3q^{-34}+3q^{-36}+3q^{-38}+2q^{-40}+2q^{-42}+q^{-44}+q^{-46}+q^{-50}\big)+
\\
+A^{-12}\big(-q^{24}+q^{22}-2q^{18}-5q^{16}-q^{14}+3q^{12}-14q^{8}-17q^{6}-9q^{4}+4q^{2}-11-27q^{-2}-36q^{-4}-
\\
-18q^{-6}-11q^{-8}-17q^{-10}-38q^{-12}-32q^{-14}-23q^{-16}-9q^{-18}-20q^{-20}-19q^{-22}-19q^{-24}-7q^{-26}-
\\
-9q^{-28}-6q^{-30}-7q^{-32}-4q^{-34}-5q^{-36}-q^{-38}-2q^{-40}-q^{-42}-2q^{-44}+q^{-46}-q^{-48}\big)+
\\
+A^{-10}\big(q^{26}+q^{24}-q^{20}+2q^{18}+7q^{16}+7q^{14}+2q^{12}+q^{10}+10q^{8}+20q^{6}+20q^{4}+13q^{2}+
\\
+13+23q^{-2}+33q^{-4}+31q^{-6}+25q^{-8}+22q^{-10}+27q^{-12}+29q^{-14}+25q^{-16}+18q^{-18}+15q^{-20}+
\\
+13q^{-22}+11q^{-24}+8q^{-26}+7q^{-28}+4q^{-30}+3q^{-32}+2q^{-34}+2q^{-36}+q^{-38}+q^{-42}\big)+
\\
+A^{-8}\big(-q^{26}-q^{24}-q^{22}+q^{20}-2q^{18}-4q^{16}-7q^{14}-3q^{12}-4q^{10}-6q^{8}-
\\
-14q^{6}-12q^{4}-14q^{2}-10-18q^{-2}-16q^{-4}-21q^{-6}-15q^{-8}-18q^{-10}-13q^{-12}-16q^{-14}-
\\
-10q^{-16}-12q^{-18}-6q^{-20}-7q^{-22}-2q^{-24}-4q^{-26}-q^{-28}-2q^{-30}-q^{-34}\big)+
\\
+A^{-6}\big(q^{24}+2q^{16}+2q^{14}+2q^{12}+3q^{8}+4q^{6}+5q^{4}+2q^{2}+5+4q^{-2}+6q^{-4}+
\\
+4q^{-6}+5q^{-8}+4q^{-10}+5q^{-12}+2q^{-14}+3q^{-16}+2q^{-18}+2q^{-20}+q^{-24}\big)
\end{array}
$$

$$
\begin{array}{c}
h_{[4]}
=A^{-24}\big(q^{-8}-q^{-10}-q^{-12}+4q^{-18}+q^{-20}-q^{-22}-3q^{-24}-6q^{-26}+3q^{-28}+
\\
+5q^{-30}+8q^{-32}+3q^{-34}-8q^{-36}-7q^{-38}-5q^{-40}+8q^{-42}+13q^{-44}+3q^{-46}-
\\
-3q^{-48}-12q^{-50}-3q^{-52}+8q^{-54}+7q^{-56}+4q^{-58}-6q^{-60}-4q^{-62}+2q^{-64}+
\\
+2q^{-66}+4q^{-68}-2q^{-70}-q^{-72}+q^{-74}+2q^{-78}-q^{-80}+q^{-84}-q^{-86}+q^{-88}\big)+
\\
+A^{-22}\big(-q^{6}+q^{2}+1+2q^{-2}-4q^{-4}-6q^{-6}-4q^{-8}+12q^{-12}+7q^{-14}-3q^{-16}-17q^{-18}-26q^{-20}-
\\
-6q^{-22}+11q^{-24}+24q^{-26}+11q^{-28}-25q^{-30}-38q^{-32}-33q^{-34}+5q^{-36}+33q^{-38}+17q^{-40}-13q^{-42}-
\\
-45q^{-44}-32q^{-46}+6q^{-48}+21q^{-50}+15q^{-52}-17q^{-54}-27q^{-56}-11q^{-58}+2q^{-60}+11q^{-62}-3q^{-64}-
\\
-10q^{-66}-6q^{-68}-4q^{-70}+3q^{-72}-2q^{-74}-3q^{-76}-q^{-78}-3q^{-80}+q^{-82}-q^{-84}-q^{-86}-q^{-90}\big)+
\\
\end{array}
$$

$$
\begin{array}{c}
+A^{-12}\big(q^{42}+q^{40}+2q^{38}+q^{32}+4q^{30}+11q^{28}+11q^{26}+10q^{24}+8q^{22}+6q^{20}+21q^{18}+
\\
+30q^{16}+44q^{14}+46q^{12}+41q^{10}+45q^{8}+54q^{6}+76q^{4}+99q^{2}+105+112q^{-2}+102q^{-4}+114q^{-6}+
\\
+129q^{-8}+146q^{-10}+159q^{-12}+150q^{-14}+138q^{-16}+137q^{-18}+132q^{-20}+144q^{-22}+131q^{-24}+
\\
+118q^{-26}+98q^{-28}+84q^{-30}+79q^{-32}+70q^{-34}+59q^{-36}+47q^{-38}+32q^{-40}+30q^{-42}+20q^{-44}+
\\
+19q^{-46}+14q^{-48}+9q^{-50}+8q^{-52}+5q^{-54}+3q^{-56}+3q^{-58}+q^{-60}+2q^{-62}+q^{-66}\big)+
\\
+A^{-10}\big(-q^{42}-q^{40}-q^{38}-q^{36}+q^{34}-2q^{32}-4q^{30}-6q^{28}-9q^{26}-5q^{24}-
\\
-6q^{22}-8q^{20}-13q^{18}-22q^{16}-23q^{14}-26q^{12}-25q^{10}-28q^{8}-37q^{6}-42q^{4}-
\\
-52q^{2}-53-53q^{-2}-58q^{-4}-57q^{-6}-66q^{-8}-69q^{-10}-67q^{-12}-67q^{-14}-59q^{-16}-
\\
-59q^{-18}-58q^{-20}-53q^{-22}-50q^{-24}-40q^{-26}-35q^{-28}-29q^{-30}-24q^{-32}-21q^{-34}-
\\
-14q^{-36}-12q^{-38}-8q^{-40}-5q^{-42}-5q^{-44}-2q^{-46}-2q^{-48}-2q^{-50}-q^{-54}\big)+
\\
+A^{-8}\big(q^{40}+2q^{30}+2q^{28}+2q^{26}+2q^{24}+3q^{20}+4q^{18}+7q^{16}+6q^{14}+
\\
+5q^{12}+6q^{10}+7q^{8}+10q^{6}+12q^{4}+10q^{2}+13+10q^{-2}+13q^{-4}+14q^{-6}+
\\
+14q^{-8}+14q^{-10}+13q^{-12}+12q^{-14}+13q^{-16}+10q^{-18}+11q^{-20}+8q^{-22}+
\\
+9q^{-24}+6q^{-26}+5q^{-28}+4q^{-30}+3q^{-32}+2q^{-34}+2q^{-36}+q^{-40}\big)
\end{array}
$$

\end{footnotesize}

\subsection*{\fbox{Knot $8_7$} \hspace{1cm}{\large $(a_1b_1|a_2b_2) = (-2,1|-1,4)$}}

\begin{footnotesize}

$$
\begin{array}{c}
h_{[1]} =\ \ \ A^{-2}\Big((-q^{4}+q^{2}-2+q^{-2}-q^{-4})A^{-2}+
\\
+(q^{6}-q^{4}+3q^{2}-2+3q^{-2}-q^{-4}+q^{-6})+(-q^{4}+q^{2}-1+q^{-2}-q^{-4})A^{2}
\end{array}
$$

$$
\begin{array}{c}
h_{[2]} =\ \ \ A^{-4}q^{-8}\Big((q^{14}-q^{12}+3q^{8}-q^{6}-2q^{4}+4q^{2}-2q^{-2}+2q^{-4}-q^{-8}+q^{-10})A^{-4}+
\\
+(-q^{20}+q^{16}-4q^{14}-2q^{12}+5q^{10}-4q^{8}-8q^{6}+5q^{4}-7+2q^{-2}-3q^{-6}+q^{-8}-q^{-12})A^{-2}+
\\
+(q^{24}-q^{22}+q^{20}+4q^{18}-3q^{16}-2q^{14}+10q^{12}-7q^{8}+8q^{6}+5q^{4}-5q^{2}+5+2q^{-2}-2q^{-4}+3q^{-6}+q^{-8}-q^{-10}+q^{-12})+
\\
+(-q^{24}+2q^{20}-3q^{18}-3q^{16}+5q^{14}-6q^{10}+2q^{8}+2q^{6}-4q^{4}+q^{2}-3q^{-2}+q^{-4}-q^{-8})A^{2}+
\\
+(q^{22}-q^{20}-q^{18}+2q^{16}-2q^{12}+q^{10}+q^{8}-q^{6}+q^{4}-1+q^{-2})A^{4}
\Big)
\end{array}
$$

$$
\begin{array}{c}
h_{[3]} =\ \ \ A^{-12}\big(-q^{6}+q^{4}-1-2q^{-2}+q^{-4}+2q^{-6}-q^{-8}-5q^{-10}-q^{-12}+3q^{-14}+3q^{-16}-
\\
-5q^{-18}-3q^{-20}+q^{-22}+3q^{-24}-q^{-26}-2q^{-28}+q^{-32}-q^{-34}+q^{-40}-q^{-42}\big)+
\\
+A^{-10}\big(q^{16}+4q^{8}+2q^{6}-2q^{4}-3q^{2}+7+11q^{-2}+3q^{-4}-11q^{-6}-2q^{-8}+14q^{-10}+17q^{-12}-5q^{-14}-
\\
-10q^{-16}+2q^{-18}+14q^{-20}+4q^{-22}-4q^{-24}-2q^{-26}+4q^{-28}+q^{-30}+q^{-32}+2q^{-36}-q^{-38}+q^{-44}\big)+
\\
+A^{-8}\big(-q^{24}-q^{18}-4q^{16}-2q^{14}+3q^{12}+q^{10}-11q^{8}-13q^{6}+14q^{2}-1-
\\
-25q^{-2}-24q^{-4}+7q^{-6}+19q^{-8}-7q^{-10}-30q^{-12}-14q^{-14}+9q^{-16}+7q^{-18}-
\\
-12q^{-20}-11q^{-22}-3q^{-24}+q^{-26}-2q^{-28}-2q^{-30}-q^{-32}-2q^{-34}-2q^{-36}-q^{-44}\big)+
\\
+A^{-6}\big(q^{30}-q^{28}+q^{26}+2q^{24}+2q^{22}-3q^{20}+8q^{16}+11q^{14}-5q^{12}-11q^{10}+q^{8}+
\\
+27q^{6}+16q^{4}-10q^{2}-21+10q^{-2}+31q^{-4}+17q^{-6}-15q^{-8}-8q^{-10}+12q^{-12}+22q^{-14}+
\\
+2q^{-16}+7q^{-22}+4q^{-24}+5q^{-26}+2q^{-28}+q^{-30}-q^{-32}+3q^{-34}+q^{-36}+q^{-38}-q^{-40}+q^{-42}\big)+
\\
+A^{-4}\big(-q^{32}+q^{28}-4q^{24}-4q^{22}+4q^{20}+7q^{18}-4q^{16}-14q^{14}-7q^{12}+
\\
+13q^{10}+12q^{8}-8q^{6}-21q^{4}-6q^{2}+13+10q^{-2}-11q^{-4}-14q^{-6}-3q^{-8}+7q^{-10}-
\\
-q^{-12}-5q^{-14}-6q^{-16}-3q^{-18}-2q^{-20}-3q^{-26}-3q^{-28}-q^{-36}\big)+
\\
+A^{-2}\big(q^{32}-q^{28}-2q^{26}+3q^{24}+4q^{22}-7q^{18}-3q^{16}+6q^{14}+8q^{12}-
\\
-2q^{10}-8q^{8}-3q^{6}+7q^{4}+6q^{2}-2-5q^{-2}+4q^{-6}+2q^{-8}-q^{-14}+q^{-16}+q^{-18}+
\\
+3q^{-20}-q^{-22}+q^{-28}\big)+\big(-q^{30}+q^{28}+q^{26}-2q^{22}-q^{20}+2q^{18}+2q^{16}-
\\
-q^{14}-2q^{12}-q^{10}+2q^{8}+q^{6}-q^{4}-q^{2}+q^{-2}-q^{-10}+q^{-16}-q^{-18}\big)
\end{array}
$$

$$
\begin{array}{c}
h_{[4]} =\ \ \ A^{-16}\big(q^{4}-q^{2}+q^{-2}+3q^{-6}-2q^{-8}-2q^{-10}+2q^{-12}+2q^{-14}+11q^{-16}-5q^{-20}-7q^{-22}-
\\
-2q^{-24}+22q^{-26}+16q^{-28}+6q^{-30}-13q^{-32}-21q^{-34}+14q^{-36}+19q^{-38}+23q^{-40}+3q^{-42}-18q^{-44}-3q^{-46}+
\\
+4q^{-48}+16q^{-50}+9q^{-52}-5q^{-54}-2q^{-58}+5q^{-60}+3q^{-62}-q^{-64}+2q^{-66}-q^{-68}+q^{-70}+q^{-72}-q^{-74}+q^{-76}\big)+
\\
+A^{-14}\big(-q^{18}-q^{12}-5q^{8}-2q^{6}+2q^{4}-q^{2}-18q^{-2}-19q^{-4}-6q^{-6}+5q^{-8}+22q^{-10}-
\\
-16q^{-12}-50q^{-14}-63q^{-16}-37q^{-18}+39q^{-20}+35q^{-22}-12q^{-24}-90q^{-26}-122q^{-28}-33q^{-30}+
\\
+24q^{-32}+44q^{-34}-19q^{-36}-99q^{-38}-76q^{-40}-36q^{-42}+18q^{-44}+14q^{-46}-30q^{-48}-35q^{-50}-35q^{-52}-
\\
-8q^{-54}+2q^{-56}-8q^{-58}-5q^{-60}-12q^{-62}-5q^{-64}-q^{-66}-4q^{-68}-2q^{-72}-q^{-74}-q^{-78}\big)+
\\
+A^{-12}\big(q^{30}+q^{26}+2q^{22}+5q^{20}+2q^{18}+3q^{16}-2q^{14}+5q^{12}+23q^{10}+22q^{8}+23q^{6}-11q^{4}-15q^{2}+21+66q^{-2}+
\\
+116q^{-4}+62q^{-6}-16q^{-8}-52q^{-10}-2q^{-12}+168q^{-14}+220q^{-16}+163q^{-18}+q^{-20}-92q^{-22}+35q^{-24}+186q^{-26}+
\\
+274q^{-28}+180q^{-30}+7q^{-32}-14q^{-34}+42q^{-36}+162q^{-38}+169q^{-40}+86q^{-42}+32q^{-44}+7q^{-46}+49q^{-48}+69q^{-50}+
\end{array}
$$

$$
\begin{array}{c}
+49q^{-52}+35q^{-54}+8q^{-56}+15q^{-58}+16q^{-60}+12q^{-62}+12q^{-64}+4q^{-66}+4q^{-68}+3q^{-70}+q^{-72}+2q^{-74}+q^{-78}\big)+
\\
+A^{-10}\big(-q^{40}-2q^{34}-q^{32}-5q^{30}-2q^{28}+2q^{26}-6q^{24}-9q^{22}-26q^{20}-20q^{18}+3q^{16}+7q^{14}+
\\
+2q^{12}-65q^{10}-101q^{8}-80q^{6}-18q^{4}+70q^{2}+5-124q^{-2}-249q^{-4}-252q^{-6}-48q^{-8}+72q^{-10}+45q^{-12}-
\\
-182q^{-14}-399q^{-16}-335q^{-18}-151q^{-20}+37q^{-22}-21q^{-24}-241q^{-26}-336q^{-28}-303q^{-30}-121q^{-32}-
\\
-40q^{-34}-103q^{-36}-158q^{-38}-198q^{-40}-126q^{-42}-63q^{-44}-54q^{-46}-51q^{-48}-72q^{-50}-51q^{-52}-
\\
-32q^{-54}-25q^{-56}-12q^{-58}-18q^{-60}-10q^{-62}-7q^{-64}-7q^{-66}-q^{-68}-3q^{-70}-q^{-72}-q^{-76}\big)+
\\
+A^{-8}\big(q^{48}-q^{46}+q^{44}+2q^{42}+3q^{38}-2q^{36}+3q^{34}+13q^{32}+8q^{30}+11q^{28}-12q^{26}-6q^{24}+26q^{22}+
\\
+50q^{20}+74q^{18}+11q^{16}-39q^{14}-34q^{12}+34q^{10}+186q^{8}+182q^{6}+92q^{4}-56q^{2}-112+83q^{-2}+267q^{-4}+
\\
\end{array}
$$

\end{footnotesize}

\subsection*{\fbox{Knot $8_9$} \hspace{1cm}{\large $(a_1b_1|a_2b_2) = (3,-1|1,-3)$}}

\begin{footnotesize}

$$
\begin{array}{c}
h_{[1]} =\ \ \ (q^{4}-q^{2}+2-q^{-2}+q^{-4})A^{-2}+(-q^{6}+q^{4}-3q^{2}+3-3q^{-2}+q^{-4}-q^{-6})+(q^{4}-q^{2}+2-q^{-2}+q^{-4})A^{2}
\end{array}
$$

$$
\begin{array}{c}
h_{[2]} =\ \ \
(q^{8}-q^{6}+2q^{2}-2+4q^{-4}-2q^{-6}-q^{-8}+3q^{-10}-q^{-14}+q^{-16})A^{-4}+
\\
+(-q^{14}+q^{10}-3q^{8}+4q^{4}-7q^{2}-2+8q^{-2}-6q^{-4}-6q^{-6}+5q^{-8}-q^{-10}-4q^{-12}+q^{-14}-q^{-18})A^{-2}+
\\
+(q^{18}-q^{16}+q^{14}+3q^{12}-3q^{10}+q^{8}+9q^{6}-7q^{4}-3q^{2}+15-
\\
-3q^{-2}-7q^{-4}+9q^{-6}+q^{-8}-3q^{-10}+3q^{-12}+q^{-14}-q^{-16}+q^{-18})+
\\
+(-q^{18}+q^{14}-4q^{12}-q^{10}+5q^{8}-6q^{6}-6q^{4}+8q^{2}-2-7q^{-2}+4q^{-4}-3q^{-8}+q^{-10}-q^{-14})A^{2}+
\\
+(q^{16}-q^{14}+3q^{10}-q^{8}-2q^{6}+4q^{4}-2+2q^{-2}-q^{-6}+q^{-8})A^{4}
\end{array}
$$

$$
\begin{array}{c}
h_{[3]} =\ \ \ A^{-6}\big(q^{12}-q^{10}+q^{4}-q^{2}+2q^{-2}+q^{-4}-3q^{-6}-q^{-8}+3q^{-10}+5q^{-12}-
\\
-3q^{-14}-3q^{-16}+q^{-18}+5q^{-20}+q^{-22}-2q^{-24}-q^{-26}+2q^{-28}+q^{-30}-q^{-34}+q^{-36}\big)+
\\
+A^{-4}\big(-q^{22}+q^{16}-2q^{14}-5q^{6}+6q^{2}+1-14q^{-2}-7q^{-4}+9q^{-6}+13q^{-8}-11q^{-10}-17q^{-12}-
\\
-3q^{-14}+12q^{-16}+2q^{-18}-9q^{-20}-9q^{-22}+2q^{-24}+2q^{-26}-q^{-28}-4q^{-30}-q^{-38}\big)+
\\
+A^{-2}\big(q^{30}+2q^{22}+q^{20}+2q^{16}+5q^{14}-6q^{10}+2q^{8}+20q^{6}+10q^{4}-18q^{2}-16+21q^{-2}+
\\
+35q^{-4}-2q^{-6}-27q^{-8}-4q^{-10}+26q^{-12}+19q^{-14}-7q^{-16}-9q^{-18}+5q^{-20}+11q^{-22}+
\\
+3q^{-24}-2q^{-26}+q^{-28}+3q^{-30}+q^{-32}+q^{-38}\big)+\big(-q^{36}+q^{34}-q^{32}-q^{30}-2q^{28}+
\\
+q^{26}-3q^{24}-4q^{22}-2q^{20}+5q^{18}-5q^{16}-16q^{14}-12q^{12}+19q^{10}+13q^{8}-24q^{6}-40q^{4}+
\\
+4q^{2}+37+4q^{-2}-40q^{-4}-24q^{-6}+13q^{-8}+19q^{-10}-12q^{-12}-16q^{-14}-5q^{-16}+5q^{-18}-
\\
-2q^{-20}-4q^{-22}-3q^{-24}+q^{-26}-2q^{-28}-q^{-30}-q^{-32}+q^{-34}-q^{-36}\big)%+
%\\
+A^{2}\big(q^{38}+q^{32}+3q^{30}+q^{28}-2q^{26}+3q^{24}+\\+11q^{22}+5q^{20}-9q^{18}%-
%\\
-7q^{16}+19q^{14}+26q^{12}-4q^{10}-27q^{8}-2q^{6}+35q^{4}+21q^{2}-16-18q^{-2}+10q^{-4}+
\\
+20q^{-6}+2q^{-8}-6q^{-10}+5q^{-14}+2q^{-16}+q^{-20}+2q^{-22}+q^{-30}\big)%+
%\\
+A^{4}\big(-q^{38}-4q^{30}-q^{28}+2q^{26}+2q^{24}-9q^{22}-\\-9q^{20}+2q^{18}+12q^{16}-3q^{14}%-
%\\
-17q^{12}-11q^{10}+13q^{8}+9q^{6}-7q^{4}-14q^{2}+1+6q^{-2}-5q^{-6}-2q^{-14}+q^{-16}-q^{-22}\big)%+
%\\
+A^{6}\big(q^{36}-q^{34}+\\+q^{30}+2q^{28}-q^{26}-2q^{24}+q^{22}+5q^{20}+q^{18}-3q^{16}%-
%\\
-3q^{14}+5q^{12}+3q^{10}-q^{8}-3q^{6}+q^{4}+2q^{2}-q^{-2}+q^{-4}-q^{-10}+q^{-12}\big)
\end{array}
$$

$$
\begin{array}{c}
h_{[4]} =\ \ \ A^{-8}\big(q^{16}-q^{14}-q^{8}+2q^{6}-q^{4}+q^{2}+1-
\\
-2q^{-2}+q^{-4}-3q^{-6}+2q^{-8}+4q^{-10}+q^{-14}-7q^{-16}-q^{-18}+5q^{-20}+4q^{-22}+
\\
+7q^{-24}-7q^{-26}-5q^{-28}+3q^{-32}+10q^{-34}-2q^{-38}-3q^{-40}-2q^{-42}+6q^{-44}+
\\
+q^{-46}+2q^{-48}-q^{-50}-2q^{-52}+2q^{-54}+q^{-58}-q^{-62}+q^{-64}\big)+
\\
+A^{-6}\big(-q^{30}+2q^{22}-2q^{20}-q^{16}-4q^{14}+3q^{12}-2q^{10}+4q^{8}+3q^{6}-8q^{4}-5q^{2}-
\\
-12+7q^{-2}+18q^{-4}+3q^{-6}-8q^{-8}-35q^{-10}-13q^{-12}+18q^{-14}+24q^{-16}+19q^{-18}-31q^{-20}-
\\
-37q^{-22}-15q^{-24}+9q^{-26}+33q^{-28}+2q^{-30}-17q^{-32}-25q^{-34}-18q^{-36}+10q^{-38}+
\\
+6q^{-40}+4q^{-42}-7q^{-44}-13q^{-46}-q^{-48}-2q^{-50}+2q^{-52}-4q^{-56}-q^{-60}-q^{-66}\big)+
\\
\end{array}
$$

\end{footnotesize}

\subsection*{\fbox{Knot $8_{10}$}\hspace{1cm} {\large $(a_1b_1|a_2b_2) = (-2,2|-1,3)$}}

\begin{footnotesize}

$$
\begin{array}{c}
h_{[1]} =\ \ \ A^{-2}\Big((-q^{4}+q^{2}-3+q^{-2}-q^{-4})A^{-2}+
\\
+(q^{6}-q^{4}+4q^{2}-2+4q^{-2}-q^{-4}+q^{-6})+(-q^{4}+q^{2}-2+q^{-2}-q^{-4})A^{2}
\Big)
\end{array}
$$

$$
\begin{array}{c}
h_{[2]} =\ \ \ A^{-4}q^{-8}\Big((q^{14}-q^{12}+4q^{8}-q^{6}-3q^{4}+7q^{2}+1-3q^{-2}+3q^{-4}+q^{-6}-q^{-8}+q^{-10})A^{-4}+
\\
+(-q^{20}+q^{16}-5q^{14}-3q^{12}+6q^{10}-7q^{8}-13q^{6}+6q^{4}-2q^{2}-13+q^{-2}-5q^{-6}-q^{-12})A^{-2}+
\\
+(q^{24}-q^{22}+q^{20}+5q^{18}-3q^{16}-2q^{14}+15q^{12}+q^{10}-8q^{8}+
\\
+15q^{6}+10q^{4}-6q^{2}+10+6q^{-2}-2q^{-4}+4q^{-6}+2q^{-8}-q^{-10}+q^{-12})+
\\
+(-q^{24}+2q^{20}-4q^{18}-4q^{16}+6q^{14}-2q^{12}-10q^{10}+3q^{8}+q^{6}-9q^{4}-5q^{-2}-q^{-8})A^{2}+
\\
+(q^{22}-q^{20}-q^{18}+3q^{16}-3q^{12}+3q^{10}+q^{8}-2q^{6}+2q^{4}+q^{2}-1+q^{-2})A^{4}
\Big)
\end{array}
$$

$$
\begin{array}{c}
h_{[3]} =\ \ \ A^{-12}\big(-q^{6}+q^{4}-1-3q^{-2}+q^{-4}+3q^{-6}-q^{-8}-9q^{-10}-4q^{-12}+5q^{-14}+7q^{-16}-11q^{-18}-
\\
-9q^{-20}-q^{-22}+7q^{-24}-2q^{-26}-6q^{-28}-2q^{-30}+2q^{-32}-q^{-34}-q^{-36}-q^{-38}+q^{-40}-q^{-42}\big)+
\\
+A^{-10}\big(q^{16}+5q^{8}+3q^{6}-2q^{4}-4q^{2}+11+19q^{-2}+8q^{-4}-16q^{-6}-q^{-8}+29q^{-10}+38q^{-12}-3q^{-14}-
\\
-13q^{-16}+9q^{-18}+36q^{-20}+15q^{-22}-3q^{-24}-q^{-26}+13q^{-28}+8q^{-30}+3q^{-32}+4q^{-36}+q^{-38}+q^{-40}+q^{-44}\big)+
\\
+A^{-8}\big(-q^{24}-q^{18}-5q^{16}-3q^{14}+3q^{12}+q^{10}-16q^{8}-22q^{6}-4q^{4}+18q^{2}-7-48q^{-2}-
\\
-50q^{-4}+4q^{-6}+21q^{-8}-29q^{-10}-70q^{-12}-39q^{-14}+4q^{-16}-4q^{-18}-40q^{-20}-33q^{-22}-
\\
-12q^{-24}-4q^{-26}-13q^{-28}-12q^{-30}-5q^{-32}-3q^{-34}-4q^{-36}-2q^{-38}-q^{-40}-q^{-44}\big)+
\\
+A^{-6}\big(q^{30}-q^{28}+q^{26}+2q^{24}+3q^{22}-3q^{20}+10q^{16}+17q^{14}-3q^{12}-
\\
-13q^{10}+4q^{8}+48q^{6}+32q^{4}-9q^{2}-26+35q^{-2}+69q^{-4}+41q^{-6}-15q^{-8}+
\\
+9q^{-10}+48q^{-12}+56q^{-14}+13q^{-16}+15q^{-18}+21q^{-20}+28q^{-22}+11q^{-24}+
\\
+11q^{-26}+10q^{-28}+8q^{-30}+q^{-32}+4q^{-34}+2q^{-36}+2q^{-38}-q^{-40}+q^{-42}\big)+
\\
+A^{-4}\big(-q^{32}+q^{28}-5q^{24}-5q^{22}+4q^{20}+8q^{18}-7q^{16}-22q^{14}-12q^{12}+17q^{10}+
\\
+13q^{8}-21q^{6}-42q^{4}-12q^{2}+17+2q^{-2}-36q^{-4}-35q^{-6}-9q^{-8}-21q^{-12}-23q^{-14}-
\\
-15q^{-16}-8q^{-18}-13q^{-20}-10q^{-22}-4q^{-24}-4q^{-26}-5q^{-28}-2q^{-30}-q^{-32}-q^{-36}\big)+
\\
+A^{-2}\big(q^{32}-q^{28}-2q^{26}+4q^{24}+5q^{22}-9q^{18}-2q^{16}+11q^{14}+13q^{12}-5q^{10}-11q^{8}+
\\
+18q^{4}+9q^{2}-3-5q^{-2}+10q^{-4}+10q^{-6}+4q^{-8}+q^{-10}+7q^{-12}+5q^{-14}+3q^{-16}+q^{-18}+5q^{-20}+
\\
+q^{-22}+q^{-24}+q^{-28}\big)+\big(-q^{30}+q^{28}+q^{26}-3q^{22}-q^{20}+3q^{18}+3q^{16}-3q^{14}-4q^{12}+
\\
+5q^{8}-q^{6}-3q^{4}-2q^{2}+2-2q^{-4}-q^{-6}+q^{-8}-q^{-10}-q^{-12}-q^{-14}+q^{-16}-q^{-18}\big)
\end{array}
$$

$$
\begin{array}{c}
h_{[4]} =\ \ \ A^{-16}\big(q^{4}-q^{2}+q^{-2}+3q^{-6}-2q^{-8}-2q^{-10}+2q^{-12}+2q^{-14}+
\\
+11q^{-16}-5q^{-20}-7q^{-22}-2q^{-24}+22q^{-26}+16q^{-28}+6q^{-30}-13q^{-32}-21q^{-34}+
\\
+14q^{-36}+19q^{-38}+23q^{-40}+3q^{-42}-18q^{-44}-3q^{-46}+4q^{-48}+16q^{-50}+9q^{-52}-
\\
-5q^{-54}-2q^{-58}+5q^{-60}+3q^{-62}-q^{-64}+2q^{-66}-q^{-68}+q^{-70}+q^{-72}-q^{-74}+q^{-76}\big)+
\\
+A^{-14}\big(-q^{18}-q^{12}-5q^{8}-2q^{6}+2q^{4}-q^{2}-18q^{-2}-19q^{-4}-6q^{-6}+5q^{-8}+22q^{-10}-16q^{-12}-
\\
-50q^{-14}-63q^{-16}-37q^{-18}+39q^{-20}+35q^{-22}-12q^{-24}-90q^{-26}-122q^{-28}-33q^{-30}+24q^{-32}+
\\
+44q^{-34}-19q^{-36}-99q^{-38}-76q^{-40}-36q^{-42}+18q^{-44}+14q^{-46}-30q^{-48}-35q^{-50}-35q^{-52}-
\\
-8q^{-54}+2q^{-56}-8q^{-58}-5q^{-60}-12q^{-62}-5q^{-64}-q^{-66}-4q^{-68}-2q^{-72}-q^{-74}-q^{-78}\big)+
\\
\end{array}
$$

$$
\begin{array}{c}
+33q^{8}+110q^{6}+91q^{4}+3q^{2}-33-33q^{-2}+55q^{-4}+106q^{-6}+76q^{-8}+41q^{-10}-9q^{-12}+
\\
+26q^{-14}+60q^{-16}+67q^{-18}+64q^{-20}+28q^{-22}+30q^{-24}+35q^{-26}+31q^{-28}+39q^{-30}+
\\
+20q^{-32}+23q^{-34}+16q^{-36}+9q^{-38}+11q^{-40}+6q^{-42}+6q^{-44}+4q^{-46}+q^{-48}+2q^{-50}+q^{-54}\big)+
\\
+A^{-2}\big(-q^{54}+q^{50}+q^{48}+2q^{46}-5q^{44}-5q^{42}-q^{40}+3q^{38}+14q^{36}+q^{34}-10q^{32}-17q^{30}-
\\
-15q^{28}+16q^{26}+19q^{24}+14q^{22}-11q^{20}-37q^{18}-15q^{16}+27q^{12}+19q^{10}-19q^{8}-24q^{6}-
\\
-30q^{4}+q^{2}+17-2q^{-2}-8q^{-4}-28q^{-6}-15q^{-8}-2q^{-10}-7q^{-12}-2q^{-14}-16q^{-16}-11q^{-18}-
\\
-7q^{-20}-11q^{-22}-3q^{-24}-8q^{-26}-4q^{-28}-2q^{-30}-6q^{-32}-q^{-34}-2q^{-36}-q^{-38}-q^{-42}\big)+
\\
+\big(q^{52}-q^{50}-q^{48}+4q^{42}-2q^{38}-3q^{36}-4q^{34}+6q^{32}+4q^{30}+3q^{28}-3q^{26}-9q^{24}+
\\
+q^{22}+2q^{20}+8q^{18}+3q^{16}-7q^{14}-q^{12}-3q^{10}+5q^{8}+5q^{6}-2q^{4}+q^{2}-3+2q^{-2}+3q^{-4}-
\\
-q^{-6}+3q^{-8}-q^{-10}+2q^{-12}+q^{-14}-q^{-16}+2q^{-18}+q^{-22}+q^{-24}-q^{-26}+q^{-28}\big)
\end{array}
$$

\end{footnotesize}

\subsection*{\fbox{Knot $8_{16}$}\hspace{1cm} {\large $(a_1b_1|a_2b_2) = (1,-1|1,-2|1,-2)$}}

\begin{footnotesize}

$$
\begin{array}{c}
h_{[1]} =\ \ \ A^2\Big((-q^{4}+2q^{2}-2+2q^{-2}-q^{-4})A^{-2}+
\\
+(q^{6}-2q^{4}+4q^{2}-4+4q^{-2}-2q^{-4}+q^{-6})+(-q^{4}+2q^{2}-3+2q^{-2}-q^{-4})A^{2}
\Big)
\end{array}
$$

$$
\begin{array}{c}
h_{[2]} =\ \ \ A^4q^{8}\Big((q^{2}-2+q^{-2}+3q^{-4}-5q^{-6}+2q^{-8}+5q^{-10}-6q^{-12}-q^{-14}+5q^{-16}-2q^{-18}-2q^{-20}+q^{-22})A^{-4}+
\\
+(-q^{8}+q^{6}+q^{4}-6q^{2}+4+4q^{-2}-12q^{-4}+6q^{-6}+10q^{-8}-14q^{-10}-
\\
-q^{-12}+13q^{-14}-4q^{-16}-5q^{-18}+4q^{-20}+q^{-22}-q^{-24})A^{-2}+
\\
+(q^{12}-2q^{10}+2q^{8}+4q^{6}-8q^{4}+6q^{2}+9-18q^{-2}+8q^{-4}+18q^{-6}-
\\
-21q^{-8}-4q^{-10}+19q^{-12}-7q^{-14}-8q^{-16}+6q^{-18}-2q^{-22}+q^{-24})+
\\
+(-q^{12}+q^{10}+q^{8}-6q^{6}+4q^{4}+6q^{2}-16+4q^{-2}+16q^{-4}-16q^{-6}-
\\
-6q^{-8}+14q^{-10}-2q^{-12}-6q^{-14}+3q^{-16}+q^{-18}-q^{-20})A^{2}+
\\
+(q^{10}-2q^{8}+q^{6}+4q^{4}-7q^{2}+1+9q^{-2}-7q^{-4}-3q^{-6}+6q^{-8}-q^{-10}-2q^{-12}+q^{-14})A^{4}
\Big)
\end{array}
$$

$$
\begin{array}{c}
h_{[3]} =\ \ \ \big(-q^{18}+2q^{16}-q^{14}-2q^{12}+q^{10}+3q^{8}-q^{6}-6q^{4}+4q^{2}+8-3q^{-2}-11q^{-4}+q^{-6}+
\\
+13q^{-8}+3q^{-10}-12q^{-12}-7q^{-14}+6q^{-16}+9q^{-18}-2q^{-20}-6q^{-22}-q^{-24}+2q^{-26}+2q^{-28}-q^{-30}\big)+
\\
+A^{2}\big(q^{28}-q^{26}+q^{22}+4q^{20}-4q^{18}-3q^{16}+7q^{14}+4q^{12}-10q^{10}-9q^{8}+
\\
+17q^{6}+14q^{4}-19q^{2}-27+13q^{-2}+35q^{-4}+2q^{-6}-35q^{-8}-17q^{-10}+22q^{-12}+26q^{-14}-
\\
-6q^{-16}-20q^{-18}-5q^{-20}+9q^{-22}+7q^{-24}-3q^{-26}-3q^{-28}-q^{-30}+q^{-32}\big)+
\\
+A^{4}\big(-q^{36}+q^{34}-2q^{30}-3q^{28}+2q^{26}+4q^{24}-9q^{22}-6q^{20}+13q^{18}+
\\
+3q^{16}-22q^{14}-14q^{12}+38q^{10}+28q^{8}-40q^{6}-51q^{4}+29q^{2}+71+4q^{-2}-
\\
-69q^{-4}-33q^{-6}+42q^{-8}+49q^{-10}-12q^{-12}-38q^{-14}-8q^{-16}+20q^{-18}+
\\
+13q^{-20}-6q^{-22}-7q^{-24}+q^{-26}+3q^{-28}+q^{-30}-q^{-32}\big)+
\\
+A^{6}\big(q^{42}-2q^{40}+2q^{38}+2q^{36}-4q^{32}+5q^{30}+8q^{28}-7q^{26}-
\\
-6q^{24}+22q^{22}+11q^{20}-30q^{18}-28q^{16}+52q^{14}+53q^{12}-47q^{10}-88q^{8}+
\\
+20q^{6}+103q^{4}+24q^{2}-90-59q^{-2}+45q^{-4}+68q^{-6}-9q^{-8}-46q^{-10}-14q^{-12}+
\\
+22q^{-14}+13q^{-16}-8q^{-18}-8q^{-20}+2q^{-22}+2q^{-24}-2q^{-28}+q^{-30}\big)+
\\
=A^{8}\big(-q^{44}+q^{42}-2q^{38}-2q^{36}+3q^{34}+2q^{32}-12q^{30}-6q^{28}+
\\
+20q^{26}+11q^{24}-34q^{22}-38q^{20}+42q^{18}+65q^{16}-31q^{14}-92q^{12}-
\\
-7q^{10}+96q^{8}+47q^{6}-67q^{4}-66q^{2}+21+60q^{-2}+9q^{-4}-33q^{-6}-18q^{-8}+
\\
+12q^{-10}+14q^{-12}-2q^{-14}-6q^{-16}+2q^{-20}+q^{-22}-q^{-24}\big)+
\\
+A^{10}\big(q^{44}-q^{42}+q^{38}+3q^{36}-5q^{34}-3q^{32}+11q^{30}+12q^{28}-17q^{26}-
\\
-27q^{24}+19q^{22}+49q^{20}-7q^{18}-59q^{16}-20q^{14}+57q^{12}+40q^{10}-31q^{8}-46q^{6}+
\\
+5q^{4}+33q^{2}+10-16q^{-2}-11q^{-4}+4q^{-6}+7q^{-8}-q^{-10}-2q^{-12}-q^{-14}+q^{-16}\big)+
\\
+A^{12}\big(-q^{42}+2q^{40}-q^{38}-2q^{36}+q^{34}+5q^{32}-2q^{30}-12q^{28}+3q^{26}+18q^{24}+q^{22}-
\\
-20q^{20}-12q^{18}+20q^{16}+15q^{14}-10q^{12}-15q^{10}+q^{8}+11q^{6}+2q^{4}-5q^{2}-2+q^{-2}+2q^{-4}-q^{-6}\big)
\end{array}
$$

$$
\begin{array}{c}
h_{[4]} =\ \ \ \big(q^{28}-2q^{26}+q^{24}+2q^{22}-2q^{20}+q^{18}-4q^{16}+4q^{14}+4q^{12}-
\\
-7q^{10}-5q^{6}+11q^{4}+10q^{2}-13-9q^{-2}-11q^{-4}+19q^{-6}+25q^{-8}-7q^{-10}-18q^{-12}-
\\
-31q^{-14}+8q^{-16}+32q^{-18}+16q^{-20}-34q^{-24}-16q^{-26}+7q^{-28}+17q^{-30}+20q^{-32}-
\\
-8q^{-34}-11q^{-36}-8q^{-38}-q^{-40}+9q^{-42}+2q^{-44}+q^{-46}-2q^{-48}-2q^{-50}+q^{-52}\big)+
\\
+A^{2}\big(-q^{42}+q^{40}-2q^{36}+q^{34}-4q^{32}+5q^{30}+q^{28}-9q^{26}+5q^{24}-2q^{22}+12q^{20}-4q^{18}-
\\
-26q^{16}+9q^{14}+13q^{12}+36q^{10}-2q^{8}-63q^{6}-25q^{4}+12q^{2}+84+53q^{-2}-58q^{-4}-81q^{-6}-66q^{-8}+
\\
-60q^{-10}+112q^{-12}+43q^{-14}-32q^{-16}-114q^{-18}-49q^{-20}+37q^{-22}+71q^{-24}+61q^{-26}-28q^{-28}-50q^{-30}-
\\
-35q^{-32}-q^{-34}+35q^{-36}+16q^{-38}+2q^{-40}-11q^{-42}-12q^{-44}+q^{-46}+2q^{-48}+3q^{-50}+q^{-52}-q^{-54}\big)+
\\
\end{array}
$$

\end{footnotesize}

\subsection*{\fbox{Knot $8_{17}$}\hspace{1cm} {\large $(a_1b_1|a_2b_2) = (2,-1|1,-1|1,-2)$}}

\begin{footnotesize}

$$
\begin{array}{c}
h_{[1]} =\ \ \ (q^{4}-2q^{2}+3-2q^{-2}+q^{-4})A^{-2}+
\\
+(-q^{6}+2q^{4}-4q^{2}+5-4q^{-2}+2q^{-4}-q^{-6})+(q^{4}-2q^{2}+3-2q^{-2}+q^{-4})A^{2}
\end{array}
$$

$$
\begin{array}{c}
h_{[2]} =\ \ \ (q^{8}-2q^{6}+5q^{2}-6-2q^{-2}+10q^{-4}-5q^{-6}-5q^{-8}+6q^{-10}-2q^{-14}+q^{-16})A^{-4}+
\\
+(-q^{14}+q^{12}+2q^{10}-6q^{8}+2q^{6}+11q^{4}-15q^{2}-4+22q^{-2}-
\\
-10q^{-4}-13q^{-6}+13q^{-8}+q^{-10}-7q^{-12}+2q^{-14}+q^{-16}-q^{-18})A^{-2}+
\\
+(q^{18}-2q^{16}+q^{14}+5q^{12}-9q^{10}+q^{8}+18q^{6}-19q^{4}-10q^{2}+31-
\\
-10q^{-2}-19q^{-4}+18q^{-6}+q^{-8}-9q^{-10}+5q^{-12}+q^{-14}-2q^{-16}+q^{-18})+
\\
+(-q^{18}+q^{16}+2q^{14}-7q^{12}+q^{10}+13q^{8}-13q^{6}-10q^{4}+22q^{2}-
\\
-4-15q^{-2}+11q^{-4}+2q^{-6}-6q^{-8}+2q^{-10}+q^{-12}-q^{-14})A^{2}+
\\
+(q^{16}-2q^{14}+6q^{10}-5q^{8}-5q^{6}+10q^{4}-2q^{2}-6+5q^{-2}-2q^{-6}+q^{-8})A^{4}
\end{array}
$$

$$
\begin{array}{c}
h_{[3]} =\ \ \ A^{-6}\big(q^{12}-2q^{10}+2q^{6}+2q^{4}-5q^{2}-5+10q^{-2}+8q^{-4}-12q^{-6}-14q^{-8}+9q^{-10}+23q^{-12}-
\\
-6q^{-14}-21q^{-16}-3q^{-18}+16q^{-20}+8q^{-22}-9q^{-24}-6q^{-26}+3q^{-28}+3q^{-30}-2q^{-34}+q^{-36}\big)+
\\
+A^{-4}\big(-q^{22}+q^{20}+q^{18}-5q^{14}+2q^{12}+9q^{10}-19q^{6}-5q^{4}+33q^{2}+18-42q^{-2}-
\\
-39q^{-4}+33q^{-6}+62q^{-8}-15q^{-10}-62q^{-12}-14q^{-14}+46q^{-16}+27q^{-18}-22q^{-20}-
\\
-25q^{-22}+5q^{-24}+14q^{-26}+2q^{-28}-7q^{-30}-q^{-32}+q^{-34}+q^{-36}-q^{-38}\big)+
\\
+A^{-2}\big(q^{30}-q^{28}-q^{26}+q^{24}+4q^{22}-2q^{20}-9q^{18}+4q^{16}+18q^{14}-
\\
-2q^{12}-37q^{10}-12q^{8}+62q^{6}+39q^{4}-69q^{2}-80+48q^{-2}+116q^{-4}-7q^{-6}-
\\
-107q^{-8}-37q^{-10}+73q^{-12}+56q^{-14}-34q^{-16}-45q^{-18}+6q^{-20}+27q^{-22}+
\\
\end{array}
$$

$$
\begin{array}{c}
h_{[4]} =\ \ \ A^{-8}\big(q^{16}-2q^{14}+2q^{10}-q^{8}+3q^{6}-7q^{4}-q^{2}+8+3q^{-2}+
\\
+6q^{-4}-24q^{-6}-11q^{-8}+19q^{-10}+23q^{-12}+23q^{-14}-43q^{-16}-42q^{-18}+q^{-20}+38q^{-22}+
\\
+66q^{-24}-18q^{-26}-53q^{-28}-41q^{-30}+3q^{-32}+63q^{-34}+22q^{-36}-9q^{-38}-33q^{-40}-
\\
-24q^{-42}+18q^{-44}+12q^{-46}+10q^{-48}-5q^{-50}-10q^{-52}+2q^{-54}+3q^{-58}-2q^{-62}+q^{-64}\big)+
\\
+A^{-6}\big(-q^{30}+q^{28}+q^{26}-q^{24}+q^{22}-5q^{20}+4q^{18}+7q^{16}-5q^{14}-2q^{12}-20q^{10}+
\\
+12q^{8}+37q^{6}+7q^{4}-20q^{2}-87-12q^{-2}+97q^{-4}+98q^{-6}+27q^{-8}-172q^{-10}-148q^{-12}+
\\
+30q^{-14}+175q^{-16}+204q^{-18}-63q^{-20}-204q^{-22}-154q^{-24}+22q^{-26}+208q^{-28}+
\\
+108q^{-30}-27q^{-32}-124q^{-34}-100q^{-36}+38q^{-38}+63q^{-40}+51q^{-42}-9q^{-44}-
\\
\end{array}
$$

$$
\begin{array}{c}
+8q^{-44}+3q^{-46}-6q^{-48}+4q^{-50}-2q^{-52}+2q^{-54}+q^{-56}-2q^{-58}+q^{-60}\big)+
\\
+A^{2}\big(-q^{64}+q^{62}+q^{60}-3q^{58}-3q^{54}+6q^{52}+9q^{50}-13q^{48}-12q^{46}-17q^{44}+
\\
+29q^{42}+62q^{40}-5q^{38}-65q^{36}-133q^{34}+q^{32}+203q^{30}+196q^{28}+11q^{26}-376q^{24}-
\\
-358q^{22}+77q^{20}+493q^{18}+551q^{16}-142q^{14}-673q^{12}-571q^{10}+95q^{8}+793q^{6}+
\\
+525q^{4}-143q^{2}-650-493q^{-2}+204q^{-4}+457q^{-6}+310q^{-8}-124q^{-10}-336q^{-12}-
\\
-113q^{-14}+63q^{-16}+166q^{-18}+57q^{-20}-64q^{-22}-42q^{-24}-27q^{-26}+30q^{-28}+20q^{-30}-
\\
-8q^{-32}-9q^{-36}+6q^{-38}+3q^{-40}-4q^{-42}+q^{-44}-2q^{-46}+q^{-48}+q^{-50}-q^{-52}\big)+
\\
+A^{4}\big(q^{66}-q^{64}+3q^{58}+5q^{56}-8q^{54}-8q^{52}-9q^{50}+19q^{48}+
\\
+43q^{46}-q^{44}-41q^{42}-95q^{40}-14q^{38}+129q^{36}+145q^{34}+49q^{32}-230q^{30}-
\\
-268q^{28}-19q^{26}+282q^{24}+421q^{22}+9q^{20}-378q^{18}-433q^{16}-67q^{14}+
\\
+461q^{12}+390q^{10}+28q^{8}-363q^{6}-364q^{4}+54q^{2}+245+229q^{-2}-32q^{-4}-
\\
-190q^{-6}-76q^{-8}+14q^{-10}+95q^{-12}+35q^{-14}-32q^{-16}-19q^{-18}-18q^{-20}+
\\
+17q^{-22}+8q^{-24}-5q^{-26}-5q^{-30}+4q^{-32}+q^{-34}-q^{-36}-q^{-40}+q^{-42}\big)+
\\
+A^{6}\big(-q^{66}+q^{64}+q^{62}-2q^{60}-q^{58}-6q^{56}+7q^{54}+15q^{52}-11q^{48}-
\\
-44q^{46}-9q^{44}+51q^{42}+63q^{40}+38q^{38}-100q^{36}-124q^{34}-27q^{32}+108q^{30}+
\\
+208q^{28}+22q^{26}-154q^{24}-204q^{22}-63q^{20}+204q^{18}+175q^{16}+30q^{14}-148q^{12}-
\\
-172q^{10}+27q^{8}+98q^{6}+97q^{4}-12q^{2}-87-20q^{-2}+7q^{-4}+37q^{-6}+12q^{-8}-20q^{-10}-
\\
-2q^{-12}-5q^{-14}+7q^{-16}+4q^{-18}-5q^{-20}+q^{-22}-q^{-24}+q^{-26}+q^{-28}-q^{-30}\big)+
\\
+A^{8}\big(q^{64}-2q^{62}+3q^{58}+2q^{54}-10q^{52}-5q^{50}+10q^{48}+12q^{46}+18q^{44}-
\\
-24q^{42}-33q^{40}-9q^{38}+22q^{36}+63q^{34}+3q^{32}-41q^{30}-53q^{28}-18q^{26}+
\\
+66q^{24}+38q^{22}+q^{20}-42q^{18}-43q^{16}+23q^{14}+23q^{12}+19q^{10}-11q^{8}-
\\
-24q^{6}+6q^{4}+3q^{2}+8-q^{-2}-7q^{-4}+3q^{-6}-q^{-8}+2q^{-10}-2q^{-14}+q^{-16}\big)
\end{array}
$$

\end{footnotesize}

\subsection*{\fbox{Knot $8_{18}$}\hspace{1cm} {\large $(a_1b_1|a_2b_2) = (1,-1|1,-1|1,-1|1,-1)$}}

\begin{footnotesize}

$$
\begin{array}{c}
h_{[1]} =\ \ \ (q^{4}-3q^{2}+3-3q^{-2}+q^{-4})A^{-2}+
\\
+(-q^{6}+3q^{4}-4q^{2}+7-4q^{-2}+3q^{-4}-q^{-6})+(q^{4}-3q^{2}+3-3q^{-2}+q^{-4})A^{2}
\end{array}
$$

$$
\begin{array}{c}
h_{[2]} =\ \ \ (q^{8}-3q^{6}+8q^{2}-9-3q^{-2}+15q^{-4}-7q^{-6}-8q^{-8}+9q^{-10}-3q^{-14}+q^{-16})A^{-4}+
\\
+(-q^{14}+2q^{12}+3q^{10}-9q^{8}+3q^{6}+16q^{4}-23q^{2}-8+31q^{-2}-15q^{-4}-
\\
-20q^{-6}+19q^{-8}+2q^{-10}-10q^{-12}+3q^{-14}+2q^{-16}-q^{-18})A^{-2}+
\\
+(q^{18}-3q^{16}+q^{14}+7q^{12}-14q^{10}+2q^{8}+27q^{6}-26q^{4}-13q^{2}+
\\
+47-13q^{-2}-26q^{-4}+27q^{-6}+2q^{-8}-14q^{-10}+7q^{-12}+q^{-14}-3q^{-16}+q^{-18})+
\\
+(-q^{18}+2q^{16}+3q^{14}-10q^{12}+2q^{10}+19q^{8}-20q^{6}-15q^{4}+31q^{2}-
\\
-8-23q^{-2}+16q^{-4}+3q^{-6}-9q^{-8}+3q^{-10}+2q^{-12}-q^{-14})A^{2}+
\\
+(q^{16}-3q^{14}+9q^{10}-8q^{8}-7q^{6}+15q^{4}-3q^{2}-9+8q^{-2}-3q^{-6}+q^{-8})A^{4}
\end{array}
$$

$$
\begin{array}{c}
h_{[3]} =\ \ \ A^{-6}\big(q^{12}-3q^{10}+5q^{6}+3q^{4}-9q^{2}-11+20q^{-2}+16q^{-4}-21q^{-6}-31q^{-8}+17q^{-10}+42q^{-12}-
\\
-6q^{-14}-42q^{-16}-7q^{-18}+28q^{-20}+17q^{-22}-17q^{-24}-11q^{-26}+4q^{-28}+6q^{-30}-3q^{-34}+q^{-36}\big)+
\\
+A^{-4}\big(-q^{22}+2q^{20}+2q^{18}-2q^{16}-9q^{14}+3q^{12}+19q^{10}-2q^{8}-36q^{6}-14q^{4}+64q^{2}+
\\
+37-69q^{-2}-79q^{-4}+59q^{-6}+113q^{-8}-14q^{-10}-114q^{-12}-27q^{-14}+79q^{-16}+55q^{-18}-
\\
-39q^{-20}-44q^{-22}+6q^{-24}+27q^{-26}+4q^{-28}-11q^{-30}-3q^{-32}+2q^{-34}+2q^{-36}-q^{-38}\big)+
\\
+A^{-2}\big(q^{30}-2q^{28}-2q^{26}+3q^{24}+7q^{22}-4q^{20}-18q^{18}+9q^{16}+36q^{14}-70q^{10}-26q^{8}+
\\
+107q^{6}+75q^{4}-122q^{2}-151+71q^{-2}+200q^{-4}-10q^{-6}-190q^{-8}-73q^{-10}+126q^{-12}+98q^{-14}-59q^{-16}-
\\
-82q^{-18}+11q^{-20}+48q^{-22}+8q^{-24}-24q^{-26}-6q^{-28}+8q^{-30}+4q^{-32}-2q^{-34}-2q^{-36}+q^{-38}\big)+
\\
\end{array}
$$

$$
\begin{array}{c}
h_{[4]} =\ \ \ A^{-8}\big(q^{16}-3q^{14}+5q^{10}+3q^{6}-16q^{4}-4q^{2}+19+14q^{-2}+13q^{-4}-57q^{-6}-
\\
-36q^{-8}+38q^{-10}+67q^{-12}+63q^{-14}-90q^{-16}-114q^{-18}-18q^{-20}+92q^{-22}+160q^{-24}-
\\
-17q^{-26}-123q^{-28}-113q^{-30}-q^{-32}+140q^{-34}+65q^{-36}-15q^{-38}-78q^{-40}-57q^{-42}+
\\
+34q^{-44}+31q^{-46}+23q^{-48}-10q^{-50}-21q^{-52}+q^{-54}+q^{-56}+6q^{-58}-3q^{-62}+q^{-64}\big)+
\\
+A^{-6}\big(-q^{30}+2q^{28}+2q^{26}-3q^{24}-2q^{22}-9q^{20}+10q^{18}+19q^{16}-6q^{14}-15q^{12}-
\\
-50q^{10}+22q^{8}+92q^{6}+41q^{4}-43q^{2}-206-63q^{-2}+200q^{-4}+255q^{-6}+97q^{-8}-356q^{-10}-
\\
-375q^{-12}+2q^{-14}+381q^{-16}+480q^{-18}-70q^{-20}-441q^{-22}-385q^{-24}+12q^{-26}+440q^{-28}+
\\
+274q^{-30}-32q^{-32}-273q^{-34}-231q^{-36}+62q^{-38}+143q^{-40}+118q^{-42}-14q^{-44}-92q^{-46}-
\\
\end{array}
$$

\end{footnotesize}

\subsection*{\fbox{Knot $8_{19}$}\hspace{1cm} {\large $(a_1b_1|a_2b_2) = (1,3|1,3)=(1,1|1,1|1,1|1,1)$}}

\begin{footnotesize}

$$
h_{[1]} =\ \ \ A^{-8}\Big(A^{-2}+(-q^{4}-q^{2}-1-q^{-2}-q^{-4})+(q^{6}+q^{2}+1+q^{-2}+q^{-6})A^{2}
\Big)
$$

$$
\begin{array}{c}
h_{[2]} =\ \ \ A^{-16}q^{-32}\Big(q^{16}A^{-4}+(-q^{26}-q^{24}-q^{22}-2q^{20}-q^{18}-q^{16}-2q^{14}-q^{12})A^{-2}+
\\
+(q^{34}+q^{32}+2q^{30}+2q^{28}+3q^{26}+3q^{24}+4q^{22}+4q^{20}+5q^{18}+3q^{16}+3q^{14}+2q^{12}+2q^{10})+
\\
+(-q^{40}-q^{38}-q^{36}-2q^{34}-2q^{32}-3q^{30}-4q^{28}-4q^{26}-5q^{24}-5q^{22}-4q^{20}-5q^{18}-5q^{16}-3q^{14}-2q^{12}-2q^{10}-q^{8})A^{2}+
\\
+(q^{44}+q^{38}+q^{36}+q^{34}+2q^{32}+q^{30}+2q^{28}+2q^{26}+2q^{24}+2q^{22}+3q^{20}+q^{18}+2q^{16}+2q^{14}+q^{12}+q^{8})A^{4}
\Big)
\end{array}
$$

$$
\begin{array}{c}
h_{[3]} =\ \ \ A^{-30}q^{-48}+A^{-28}\big(-q^{-32}-q^{-34}-q^{-36}-q^{-38}-2q^{-40}-2q^{-42}-q^{-44}-q^{-46}-2q^{-48}-2q^{-50}-q^{-52}\big)+
\\
+A^{-26}\big(q^{-18}+q^{-20}+2q^{-22}+2q^{-24}+4q^{-26}+4q^{-28}+5q^{-30}+5q^{-32}+7q^{-34}+
\\
+7q^{-36}+8q^{-38}+8q^{-40}+9q^{-42}+7q^{-44}+6q^{-46}+4q^{-48}+5q^{-50}+3q^{-52}+2q^{-54}\big)+
\\
+A^{-24}\big(-q^{-6}-q^{-8}-2q^{-10}-3q^{-12}-4q^{-14}-5q^{-16}-7q^{-18}-8q^{-20}-10q^{-22}-
\\
-12q^{-24}-14q^{-26}-16q^{-28}-19q^{-30}-19q^{-32}-20q^{-34}-20q^{-36}-20q^{-38}-19q^{-40}-
\\
-19q^{-42}-16q^{-44}-13q^{-46}-10q^{-48}-7q^{-50}-5q^{-52}-4q^{-54}-q^{-56}\big)+
\\
+A^{-22}\big(q^{4}+q^{2}+2+2q^{-2}+4q^{-4}+4q^{-6}+6q^{-8}+7q^{-10}+10q^{-12}+11q^{-14}+14q^{-16}+
\\
+16q^{-18}+20q^{-20}+21q^{-22}+24q^{-24}+24q^{-26}++28q^{-28}+28q^{-30}+31q^{-32}+29q^{-34}+30q^{-36}+
\\
+26q^{-38}+25q^{-40}+21q^{-42}+20q^{-44}+15q^{-46}+12q^{-48}+7q^{-50}+6q^{-52}+3q^{-54}+2q^{-56}\big)+
\\
+A^{-20}\big(-q^{12}-q^{10}-q^{8}-q^{6}-2q^{4}-3q^{2}-3-4q^{-2}-6q^{-4}-7q^{-6}-8q^{-8}-9q^{-10}-12q^{-12}-
\\
-13q^{-14}-14q^{-16}-15q^{-18}-18q^{-20}-20q^{-22}-21q^{-24}-21q^{-26}-22q^{-28}-22q^{-30}-22q^{-32}-21q^{-34}-
\\
-21q^{-36}-19q^{-38}-16q^{-40}-13q^{-42}-12q^{-44}-10q^{-46}-7q^{-48}-4q^{-50}-3q^{-52}-2q^{-54}-q^{-56}\big)+
\\
+A^{-18}\big(q^{18}+q^{10}+q^{8}+q^{6}+q^{4}+2q^{2}+2+2q^{-2}+2q^{-4}+4q^{-6}+3q^{-8}+4q^{-10}+
\\
+4q^{-12}+6q^{-14}+5q^{-16}+6q^{-18}+5q^{-20}+7q^{-22}++7q^{-24}+7q^{-26}+6q^{-28}+8q^{-30}+
\\
+6q^{-32}+6q^{-34}+5q^{-36}+6q^{-38}+4q^{-40}+4q^{-42}+2q^{-44}+3q^{-46}+2q^{-48}+q^{-50}+q^{-54}\big)
\end{array}
$$

$$
\begin{array}{c}
h_{[4]} =\ \ \ A^{-40}q^{-96}+
\\
+A^{-38}\big(-q^{-74}-q^{-76}-q^{-78}-q^{-80}-q^{-82}-2q^{-84}-2q^{-86}-2q^{-88}-q^{-90}-q^{-92}-2q^{-94}-2q^{-96}-2q^{-98}-q^{-100}\big)+
\\
+A^{-36}\big(q^{-54}+q^{-56}+2q^{-58}+2q^{-60}+3q^{-62}+4q^{-64}+6q^{-66}+6q^{-68}+
\\
+7q^{-70}+7q^{-72}+9q^{-74}+10q^{-76}+12q^{-78}+11q^{-80}+12q^{-82}+12q^{-84}+13q^{-86}+
\\
+11q^{-88}+10q^{-90}+7q^{-92}+7q^{-94}+6q^{-96}+6q^{-98}+3q^{-100}+2q^{-102}\big)+
\\
+A^{-34}\big(-q^{-36}-q^{-38}-2q^{-40}-3q^{-42}-4q^{-44}-6q^{-46}-8q^{-48}-10q^{-50}-
\\
-12q^{-52}-15q^{-54}-18q^{-56}-21q^{-58}-25q^{-60}-28q^{-62}-32q^{-64}-37q^{-66}-40q^{-68}-
\\
-43q^{-70}-45q^{-72}-46q^{-74}-47q^{-76}-47q^{-78}-46q^{-80}-44q^{-82}-42q^{-84}-39q^{-86}-
\\
\end{array}
$$

\end{footnotesize}

\subsection*{\fbox{Knot $8_{20}$}\hspace{1cm} {\large $(a_1b_1|a_2b_2) = (-1,-3|-1,3)$}}

\begin{footnotesize}

$$
h_{[1]} =\ \ \ A^2\Big((-q^{2}+1-q^{-2})A^{-2}+(q^{4}+2+q^{-4})+(-q^{2}-q^{-2})A^{2}
\Big)
$$

$$
\begin{array}{c}
h_{[2]} =\ \ \ A^4q^{8}\Big((q^{-2}-q^{-6}+q^{-8}+q^{-10}-q^{-12})A^{-4}+(-q^{4}-q^{2}-q^{-2}-3q^{-4}-q^{-6}+q^{-8}-q^{-10}-q^{-12})A^{-2}+
\\
+(q^{8}+q^{4}+3q^{2}+3+q^{-2}+3q^{-4}+3q^{-6}+2q^{-8}+q^{-10}+q^{-12}+q^{-16})+
\\
+(-q^{8}-q^{6}-2q^{2}-3-2q^{-2}-2q^{-4}-2q^{-6}-q^{-8}-q^{-10}-q^{-12})A^{2}+(q^{6}+1+q^{-2}+q^{-6})A^{4}
\Big)
\end{array}%
$$

\end{footnotesize}

\subsection*{\fbox{Knot $8_{21}$}\hspace{1cm} {\large $(a_1b_1|a_2b_2) = (-2,2|-1,-3)$}}

\begin{footnotesize}

$$
h_{[1]} =\ \ \ A^4\Big((2q^{2}-1+2q^{-2})A^{-2}+(-q^{4}+q^{2}-3+q^{-2}-q^{-4})+(q^{2}-1+q^{-2})A^{2}
\Big)
$$

$$
\begin{array}{c}
h_{[2]} =\ \ \ A^8q^{16}\Big((3q^{-8}-q^{-10}+5q^{-14}-2q^{-18}+3q^{-20}+q^{-22})A^{-4}+
\\
+(-q^{-2}-q^{-4}+2q^{-6}-4q^{-8}-6q^{-10}+4q^{-12}-2q^{-14}-8q^{-16}+q^{-20}-2q^{-22}-q^{-24})A^{-2}+
\\
+(q^{2}-1+4q^{-4}-q^{-6}-3q^{-8}+8q^{-10}+3q^{-12}-3q^{-14}+3q^{-16}+3q^{-18}+q^{-22})+
\\
+(-q^{2}+2q^{-2}-2q^{-4}-3q^{-6}+3q^{-8}-4q^{-12}-q^{-18})A^{2}+
\\
+(1-q^{-2}-q^{-4}+2q^{-6}-q^{-10}+q^{-12})A^{4} \Big)
\end{array}%
$$

\end{footnotesize}

\subsection*{\fbox{Knot $10_{139}$}\hspace{1cm} {\large $(a_1b_1|a_2b_2) = (2,3|1,4)$}}

\begin{footnotesize}

$$
h_{[1]} =\ \ \ A^{-10}\Big((q^{2}-1+q^{-2})A^{-2}+(-q^{6}-q^{4}-2-q^{-4}-q^{-6})+(q^{8}+q^{4}+q^{2}+q^{-2}+q^{-4}+q^{-8})A^{2}
\Big)
$$

$$
\begin{array}{c}
h_{[2]} =\ \ \ A^{-20}q^{-40}\Big((q^{28}-q^{26}-q^{24}+2q^{22}-q^{18}+q^{16})A^{-4}+
\\
+(-q^{38}-q^{36}-2q^{28}-2q^{26}+q^{24}-q^{22}-3q^{20}-q^{18}-q^{14}-q^{12})A^{-2}+
\\
+(q^{46}+q^{44}+2q^{42}+q^{40}+q^{38}+2q^{36}+4q^{34}+3q^{32}+q^{30}+3
\\
+q^{28}+6q^{26}+3q^{24}+4q^{22}+4q^{20}+4q^{18}+3q^{16}+2q^{14}+q^{12}+2q^{10})+
\\
+(-q^{52}-q^{50}-q^{48}-2q^{46}-q^{44}-2q^{42}-5q^{40}-3q^{38}-q^{36}-4q^{34}-6q^{32}-4q^{30}-
\\
L-4q^{28}-6q^{26}-5q^{24}-6q^{22}-5q^{20}-3q^{18}-5q^{16}-3q^{14}-q^{12}-2q^{10}-q^{8})A^{2}+
\\
+(q^{56}+q^{50}+q^{48}+q^{46}+2q^{44}+q^{40}+3q^{38}+2q^{36}+3q^{32}+3q^{30}+
\\
+q^{28}+3q^{26}+3q^{24}+q^{22}+4q^{20}+q^{18}+q^{16}+2q^{14}+q^{12}+q^{8})A^{4}
\Big)
\end{array}%
$$

\end{footnotesize}

\end{document}